\documentclass[aps,prl,reprint,twocolumn,superscriptaddress,floatfix,nofootinbib,longbibliography]{revtex4-2}
\usepackage{graphicx,amsmath,amsfonts,amssymb,amsthm,xr}
\usepackage{epsfig,amsmath,amssymb,color,dsfont,upgreek,physics}
\usepackage{mathrsfs}
\usepackage{mathtools}
\usepackage{bbold}
\usepackage{float}
\usepackage[caption=false]{subfig}

\usepackage[bookmarks=true,colorlinks,linkcolor=OrangeRed,urlcolor=NavyBlue,citecolor=RoyalBlue]{hyperref}
\usepackage[dvipsnames]{xcolor}
\usepackage{orcidlink}
\usepackage[capitalise]{cleveref}

\usepackage{graphicx}
\usepackage{dcolumn}
\usepackage{bm}
\usepackage{color}

\newcommand{\Zt}{\mathbb{Z}_2}
\newcommand{\rr}{\mathbf{r}}

\begin{document}

\preprint{}

\title{String Breaking Dynamics and Glueball Formation in a $2+1$D Lattice Gauge Theory}

\author{Kaidi Xu${}^{\orcidlink{0000-0003-2184-0829}}$}
\thanks{These authors contributed equally to this work.}
\affiliation{Max Planck Institute of Quantum Optics, 85748 Garching, Germany}
\affiliation{Munich Center for Quantum Science and Technology (MCQST), 80799 Munich, Germany}

\author{Umberto Borla${}^{\orcidlink{0000-0002-4224-5335}}$}
\thanks{These authors contributed equally to this work.}
\affiliation{Max Planck Institute of Quantum Optics, 85748 Garching, Germany}
\affiliation{Munich Center for Quantum Science and Technology (MCQST), 80799 Munich, Germany}
\affiliation{Racah Institute of Physics, The Hebrew University of Jerusalem, Givat Ram, Jerusalem 91904, Israel}

\author{Sergej Moroz${}^{\orcidlink{0000-0002-4615-2507}}$}
\affiliation{Department of Engineering and Physics, Karlstad University, Karlstad, Sweden}
\affiliation{Nordita, Stockholm University and KTH Royal Institute of Technology, 10691 Stockholm, Sweden}

\author{Jad C.~Halimeh${}^{\orcidlink{0000-0002-0659-7990}}$}
\email{jad.halimeh@physik.lmu.de}
\affiliation{Max Planck Institute of Quantum Optics, 85748 Garching, Germany}
\affiliation{Department of Physics and Arnold Sommerfeld Center for Theoretical Physics (ASC), Ludwig Maximilian University of Munich, 80333 Munich, Germany}
\affiliation{Munich Center for Quantum Science and Technology (MCQST), 80799 Munich, Germany}

\date{\today}

\begin{abstract}
With the advent of advanced quantum processors capable of probing lattice gauge theories (LGTs) in higher spatial dimensions, it is crucial to understand string dynamics in such models to guide upcoming experiments and to make connections to high-energy physics (HEP). Using tensor network methods, we study the far-from-equilibrium quench dynamics of electric flux strings between two static charges in the $2+1$D $\mathbb{Z}_2$ LGT with dynamical matter. We calculate the probabilities of finding the time-evolved wave function in string configurations of the same length as the initial string. At resonances determined by the the electric field strength and the mass, we identify various string breaking processes accompanied with matter creation. Away from resonance strings exhibit intriguing confined dynamics which, for strong electric fields, we fully characterize through effective perturbative models. Starting in maximal-length strings, we find that the wave function enters a dynamical regime where it splits into shorter strings and disconnected loops, with the latter bearing qualitative resemblance to glueballs in quantum chromodynamics (QCD). Our findings can be probed on state-of-the-art superconducting-qubit and trapped-ion quantum processors.
\end{abstract}

\maketitle

\textbf{\emph{Introduction.---}}
String breaking is a paradigmatic phenomenon with roots in QCD: Pull a quark-antiquark pair sufficiently apart and the flux string between them becomes so expensive that it breaks through the creation of more quark-antiquark pairs \cite{Weinberg_book,Gattringer_book,Zee_book}. Understanding the real-time dynamics of string breaking in QCD from a first-principles approach remains an outstanding challenge in HEP \cite{Ellis_book}. Given the difficulties associated with a numerical or experimental realization of $3+1$D QCD, approaching this problem with simpler models would be worthwhile. Indeed, string breaking is not specific to QCD, but also appears in LGTs with a confined phase, and which are amenable to investigation with current state-of-the-art numerical and experimental techniques \cite{Rothe_book,Kogut1975}.

\begin{figure}[t!]
    \centering
    \includegraphics[width=\linewidth]{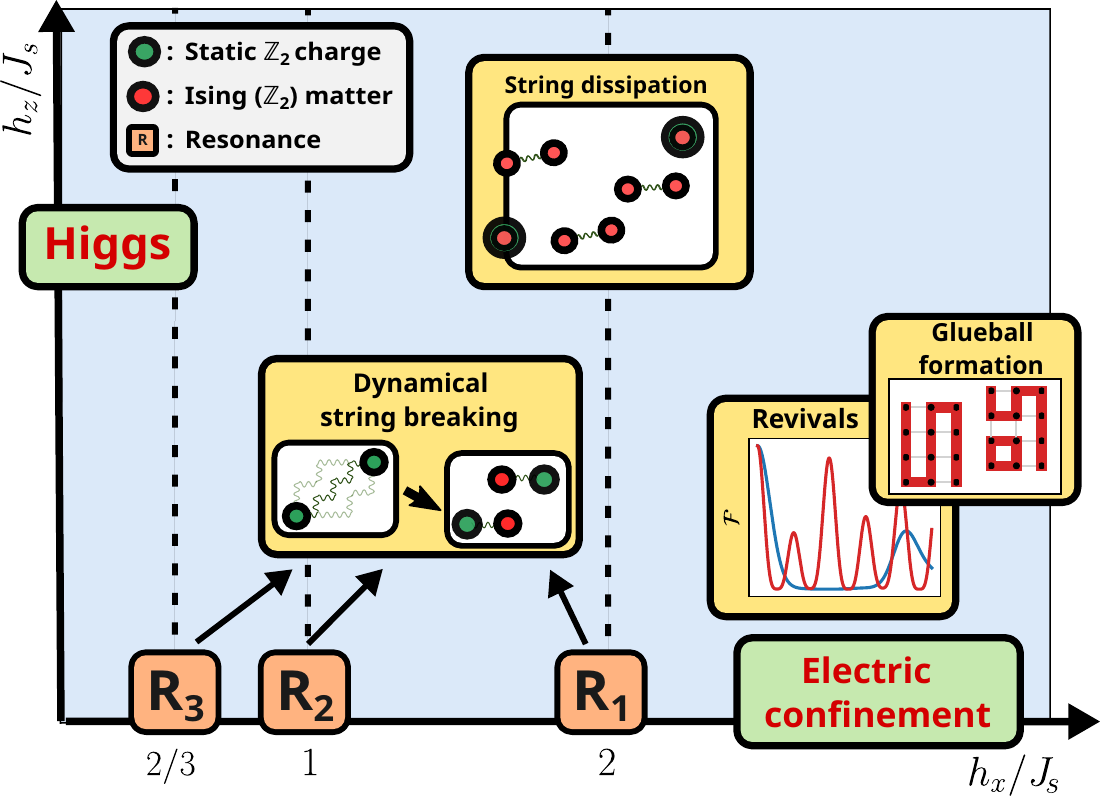}
    \caption{Dynamical quantum phase diagram of the model \cref{eq:2DFS} in the confined/Higgs regimes. Deep in the confined phase one can either observe rich unbroken string dynamics, including quantum revivals and loop nucleation, or string breaking in correspondence of certain mesonic resonances. As the Higgs phase is approached, strings dissipate, and mesons whose mobility is initially restricted spread across the lattice.}
    \label{fig:fig_1}
\end{figure}

Originally developed to enable nonperturbative calculations of QCD for better insights into the nature of quark confinement \cite{Wilson1974}, LGTs have also proven to be a powerful framework to study out-of-equilibrium phenomena using tensor network methods and quantum simulation~\cite{Dalmonte_review, Zohar_review, Aidelsburger2011, Zohar_NewReview, klco2021standard, Bauer_review, dimeglio2023quantum, Cheng_review, Halimeh_review, Cohen:2021imf, Lee:2024jnt, Turro:2024pxu,bauer2025efficientusequantumcomputers}. Their relevance extends beyond HEP, as they also provide a powerful framework for modeling emergent gauge structures in condensed matter systems, including quantum spin liquids, frustrated magnets, and, more speculatively, high-$T_\mathrm{c}$ superconductivity \cite{Wegner1971,Kogut_review,wen2004quantum,Savary2016,Calzetta_book}, as well as give rise to ergodicity-breaking mechanisms central to quantum many-body physics, such as disorder-free localization \cite{Smith2017,Brenes2018,smith2017absence,karpov2021disorder,Sous2021,Chakraborty2022,Halimeh2021enhancing}, quantum many-body scarring \cite{Surace2020,Desaules2022weak,Desaules2022prominent,aramthottil2022scar}, Hilbert-space fragmentation \cite{Desaules2024ergodicitybreaking,desaules2024massassistedlocaldeconfinementconfined,jeyaretnam2025hilbertspacefragmentationorigin,ciavarella2025generichilbertspacefragmentation}, and nonstabilizerness \cite{Tarabunga2023many,hartse2024stabilizerscars,Smith2025nonstabilizerness,Falcao2025Nonstabilizerness,Esposito2025magic}. The last decade has seen an explosion of quantum simulation experiments on both digital and analog platforms observing different features of LGTs both in and out of equilibrium~\cite{Martinez2016,Klco2018,Goerg2019,Schweizer2019,Mil2020,Yang2020,Wang2021,Su2022,Zhou2022,Wang2023,Zhang2023,Ciavarella2024quantum,Ciavarella:2024lsp,Farrell:2023fgd,Farrell:2024fit,zhu2024probingfalsevacuumdecay,Ciavarella:2021nmj,Ciavarella:2023mfc,Ciavarella:2021lel,Gustafson:2023kvd,Gustafson:2024kym,Lamm:2024jnl,Farrell:2022wyt,Farrell:2022vyh,Li:2024lrl,Zemlevskiy:2024vxt,Lewis:2019wfx,Atas:2021ext,ARahman:2022tkr,Atas:2022dqm,Mendicelli:2022ntz,Kavaki:2024ijd,Than:2024zaj,Angelides2025first,gyawali2024observationdisorderfreelocalizationefficient,schuhmacher2025observationhadronscatteringlattice,davoudi2025quantumcomputationhadronscattering}. In parallel to this impressive quantum simulation effort, tensor networks have been part and parcel of investigations of real-time dynamics of LGTs \cite{Pichler2016,Chanda2020confinement,Notarnicola2020real,Rigobello2021entanglement,vandamme2022dqpt,su2024particlecollider,Belyansky2024high,Calajo2024digital,Calajo2025QMBS,cataldi2025disorderfreelocalizationfragmentationnonabelian}, including in $2+1$D \cite{osborne2024quantummanybodyscarring21d,Budde2024qmbs,osborne2023disorderfreelocalization21dlattice}, in most cases surpassing state-of-the-art quantum simulators \cite{Banuls_review,magnifico2024tensornetworkslatticegauge}. 

Recently, several quantum simulation experiments of LGTs have appeared that observe string breaking in one \cite{de2024observationstringbreakingdynamicsquantum,liu2024stringbreakingmechanismlattice,alexandrou2025realizingstringbreakingdynamics} and two \cite{cochran2024visualizingdynamicschargesstrings,gonzalezcuadra2024observationstringbreaking2,crippa2024analysisconfinementstring2} spatial dimensions. This has been accompanied by tensor network studies of the string roughening transition \cite{xu2025tensornetworkstudyrougheningtransition,dimarcantonio2025rougheningdynamicselectricflux} and string breaking \cite{borla2025stringbreaking21dmathbbz2} in $2+1$D. The current huge interest in string breaking in $2+1$D LGTs is well justified. In two spatial dimensions, the string has an additional transverse mode of vibration, and the dynamics of the string becomes generally richer. Importantly, in two spatial dimensions, truly far-from-equilibrium string configurations can be prepared that fundamentally depart from the ground-state string configuration of the model \cite{borla2025stringbreaking21dmathbbz2}. Furthermore, studying string breaking in $2+1$D is a natural next step toward the ultimate goal of probing its real-time dynamics in $3+1$D QCD.

Whereas Refs.~\cite{cochran2024visualizingdynamicschargesstrings,gonzalezcuadra2024observationstringbreaking2} are restricted to short-time near-equilibrium dynamics of string breaking in $2+1$D due to small system sizes and device noise, Refs.~\cite{xu2025tensornetworkstudyrougheningtransition,dimarcantonio2025rougheningdynamicselectricflux} are concerned with the roughening transition, and do not address string breaking. In fact, the pure $\mathbb{Z}_2$ LGT considered in Ref.~\cite{dimarcantonio2025rougheningdynamicselectricflux} cannot exhibit string breaking due to the absence of dynamical matter. On the other hand, Ref.~\cite{borla2025stringbreaking21dmathbbz2} maps out the ground-state phase diagram of the $2+1$D $\mathbb{Z}_2$ LGT in the presence of static charges, but does not probe any dynamical features. Thus, a study of truly far-from-equilibrium string dynamics and breaking in a $2+1$D LGT is still missing from both the numerical and quantum simulation points of view. 

In this Letter, we address this by using tensor networks \cite{Uli_review,Paeckel_review,Montangero_book,tenpy2024} with the time-dependent variational principle (TDVP) \cite{Haegeman2011,Haegeman2013,Haegeman2016} for time evolution calculation to simulate the far-from-equilibrium quench dynamics of strings in a $2+1$D $\mathbb{Z}_2$ LGT. We find that in several relevant setups, consisting of various string initial states, string breaking can occur if certain resonance conditions are satisfied (see \cref{fig:fig_1}). Away from the resonance condition, strings that are arbitrarily far from equilibrium exhibit confined oscillations that we are able to fully characterize through effective perturbative models in the strong electric coupling limit. We also find that long strings can decrease their length by spontaneously nucleating disconnected electric loops, which we argue are qualitatively similar to glueballs from QCD \cite{greensite2011introduction}.

\textbf{\emph{Model.---}} We consider a $2+1$D $\mathbb{Z}_2$ LGT with Ising matter on a square lattice \cite{Fradkin1979,Trebst2007breakdown,Vidal2009low-energy,Wu2012phase}, described by the Hamiltonian
 
\begin{align}\nonumber
\hat{H}=&-J_s\sum_\rr \hat\tau^z_\rr-J_p\sum_{\rr^*} \hat B_{\rr^*}\\
&-h_z \sum_{\rr,\eta} \hat\tau^x_\rr \hat\sigma^z_{\rr,\eta} \hat\tau^x_{\rr+\eta}-h_x\sum_{\rr,\eta} \hat\sigma^x_{\rr,\eta}.
\label{eq:2DFS}
\end{align}
Here, $\hat{\tau}^{z}_\mathbf{r}$ represents the particle-number operator for Ising matter on site $\mathbf{r}$ and $\hat\sigma^{x(z)}_{\rr,\eta}$ is the $\Zt$ electric (gauge) field operator on the link emanating from site $\mathbf{r}$ in the direction $\eta$, with coupling strength $h_x$. $\hat B_{\rr^*}=\prod_{b\in \square_{\rr^*}} \hat\sigma^z_b$ is the four-body plaquette operator, with the index $\rr^*$ labeling the sites of a dual lattice formed by the centers of the plaquettes. The three-body term in Hamiltonian~\eqref{eq:2DFS} describes gauge-matter interactions with strength $h_z$. We also define the star (or vertex) operator $\hat A_{\rr}=\prod_{\eta \in +_\rr}\hat\sigma^x_{\rr, \eta}$ as the product of $\hat\sigma^x$ on the four links connecting at the vertex $\rr$.

\begin{figure}[t!]
    \includegraphics[width=\linewidth]{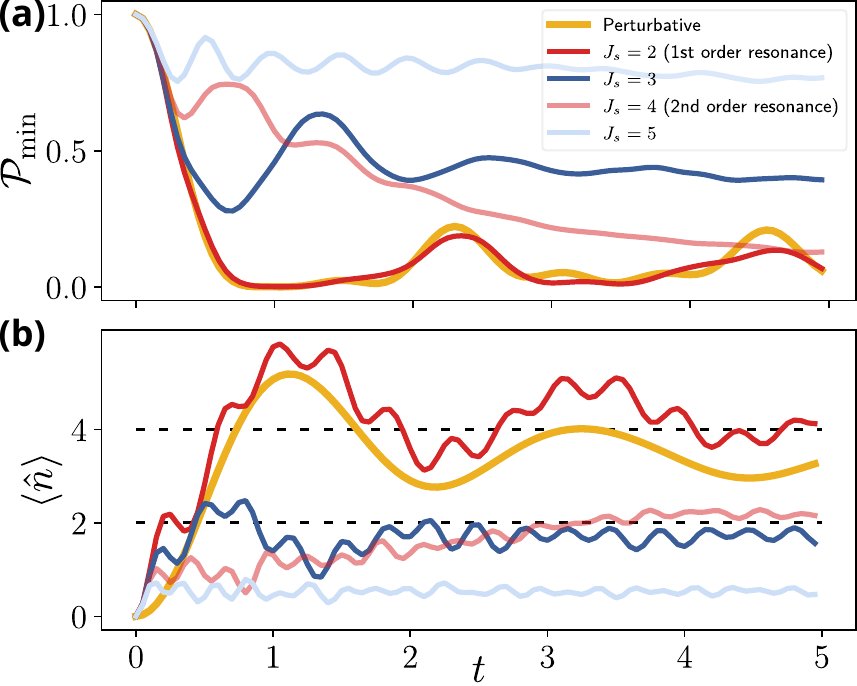}
    \caption{Minimal string probability (top) and total particle number (bottom) as a function of time for the initial state corresponding to an L-shaped string. We take $h_x=4$, $h_z=1$ and $J_p=1$, which places the system in the confined regime. For the first two resonances $J_s^{\text{1}}=2$ and $J_s^{\text{2}}=4$, the string probability drops to very small values at timescales $t^*_1\approx 0.8$ and $t_2^*\approx 5$ respectively. String breaking coincides with the creation of one and two mesons (pairs of $\Zt$ charges) respectively. In the off-resonance case, matter fluctuations are still present, but the minimal string probability maintains a finite value at large timescales. The black curves show perturbative results valid in the regime where $h_x \rightarrow \infty$ while maintaining the resonance conditions $h_x=2J_s$. To match the numerical results, we choose $J_p=h_z=1$ in the effective model.}
    \label{fig:resonances}
\end{figure}

Equation \eqref{eq:2DFS} is invariant under local gauge transformations $\hat{G}_\rr = \hat A_{\rr}\hat\tau^z_{\rr}$,
which relate $\Zt$ electric lines, where $\hat{\sigma}^x=-1$, emanating from a site to its total $\Zt$ charge. Physical states satisfy Gauss's law: $\hat{G}_\rr|\psi\rangle=Q_\rr |\psi \rangle$, where $Q_\rr={\pm} 1$ denotes the absence or presence of a static background $\Zt$ charge on that particular site. The Hilbert space separates into disconnected sectors, each determined by a different distribution of background charges. In order to satisfy Gauss's law, a basis of physical gauge-invariant states must be formed by $\Zt$ charges, either dynamical or static, connected by electric strings. In the following, we will study the time evolution of individual string configurations, i.e., the case where there are two static charges at sites $\rr_1$ and $\rr_2$ ($Q_{\rr_1}=Q_{\rr_2}=-1$, $Q_{\rr\neq\rr_1,\rr_2}=+1$).

In this work, we want to study how initial configurations consisting of strings connecting the two static charges evolve in time. Since a string consists of flipped electric links (where $\hat{\sigma}^x=-1$ locally), its energy cost compared to the vacuum is determined by its length $l$ through $E_{\text{str}}=2 l h_x$. If the string connects two dynamical charges, the additional cost $4J_s$ related to matter creation on top of the vacuum must be included.
The $J_p$ (plaquette) term in the Hamiltonian provides kinetic energy to the strings by allowing transitions between different string configurations, with the effect of delocalizing the initial product state. The $\Zt$ charge fluctuations, on the other hand, provide a mechanism for string breaking by pair creation.

\textbf{\emph{String breaking dynamics.---}}
We first analyze the possibility of string breaking triggered by the formation of a pair of $\Zt$ charges. In order for broken-string states to be dynamically accessible, it is typically necessary to find resonance conditions for which the cost of matter creation precisely compensates the gain in electric energy associated with the broken string. Consider the case where a string of length $l$ is split into two pieces through the formation of a meson of length $d$ anywhere along the path of the string. When matter fluctuations can be ignored, i.e., deep in the confined phase $h_x \gg h_z$, the condition for which broken and unbroken strings have equal energy is $2h_xl=4J_s + 2(l-d)h_x$, i.e., $2 J_s=d\, h_x$. While length-one mesons, corresponding to the first resonance $h_x^{(1)}=2J_s$, can be directly created through the $h_z$ term in the Hamiltonian, longer mesons can only be the result of higher-order processes. As a consequence, they are energetically suppressed and are expected to cause string breaking at larger timescales $t_{\text{br}}\approx h_x^{d-1}/(l \,h_z^d)$. 

Deep in the confining phase, the Hilbert space is fragmented into a number of sectors labeled by the length of the strings, which are separated from each other by a large gap $\Delta E \approx 2h_x$. In this scenario, the only relevant fluctuations are those which preserve the length of the string. These are triggered by first-order resonance processes caused by the action of $B_{\rr^*}$ on plaquettes where exactly two of four electric links are flipped. We note that fourth-order processes involving $h_z$ can also generate a plaquette term perturbatively, but these are heavily suppressed and do not meaningfully contribute to the dynamics. 

Let us consider the example of strings of minimal length $l$ given by the Manhattan distance between the two static charges: $l= \Delta x+\Delta y$ and different shapes. Unbroken strings oscillate between different minimal length configurations contained within the patch spanned by the two static charges. The presence or absence of the string in the time-evolved state is therefore detected by the total overlap   

\begin{align}\label{eq:string_prob}
    \mathcal{P}_{\text{min}}(t) = \sum_{\gamma \in \mathcal{S}} \lvert\langle \psi_{\gamma}| \psi(t)\rangle\rvert^2, 
\end{align}
where $\psi_\gamma$ is the product state string configuration with $\hat{\sigma}^x = -1$ on the links belonging to the path $\gamma$, and $\mathcal{S}$ is the set of all possible string configurations in the patch between the two static charges that are of the same length as the initial strength.
$\mathcal{P}_{\text{min}}$ is expected to approach unity in the strong confinement regime, and only drops if the string is broken. When this happens, we expect to observe matter creation within the patch, in the form of mesonic pairs connected by an anti-string. For lower values of the electric coupling, on the other hand, deviations from this resonance condition in the absence of pair production may be used to characterize the ``floppiness'' of the string, i.e., the amount of transverse fluctuations.

\begin{figure}[t!]
    \centering
    \includegraphics[width=0.99\linewidth]{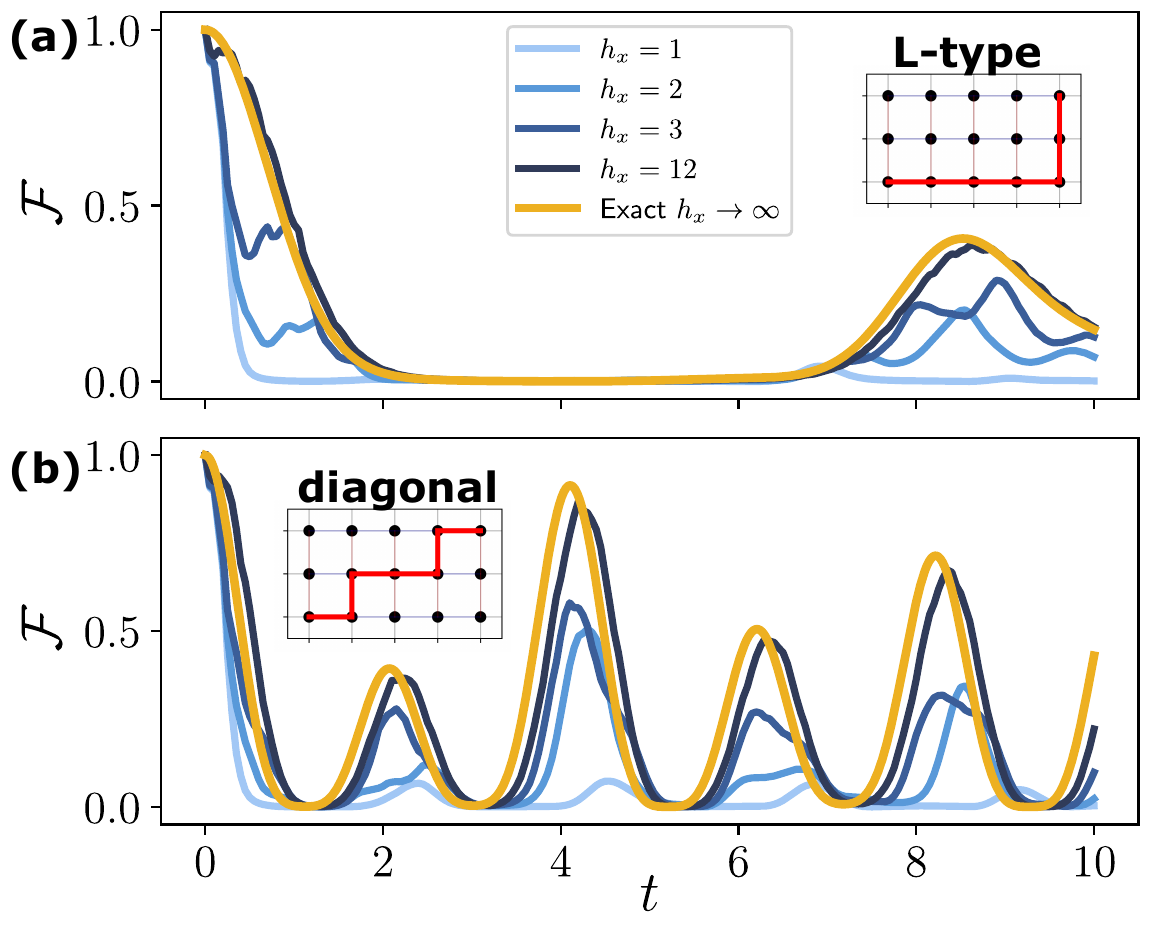}
    \caption{Comparison between perturbative analytical results for the fidelity $\mathcal{F}(t)$ of different initial states, obtained using the free-fermion picture, and numerical TDVP results. We see how deep in the confined regime the perturbative results becomes exact. As the electric coupling $h_x$ is lowered, deviations are increasingly manifest. The parameters of the quench Hamiltonian are $J_p=1$, $J_s=15$, and $h_z=1$.}
    \label{fig:ff_deviation}
\end{figure}

In \cref{fig:resonances}, we plot $\mathcal{P}_{\text{min}}$ 
as a function of evolution time together with the total number of particles in the system. We tune $J_s$ to resonant and off-resonant values. We observe distinctive behavior when the resonance condition is met. In particular, we find that at the first-order resonance $h_x=2J_s$ string breaking occurs at very short timescales, and is associated with the rapid creation of four particles, i.e., of two $\Zt$ mesons. For this resonance, exact results can be obtained through degenerate perturbation theory in the limit $h_x\rightarrow \infty$ \cite{SM}. These match remarkably well the numerical results obtained at lower electric coupling, with only small deviations caused by larger matter fluctuations. We are also able to observe string breaking at larger timescales in correspondence of the second-order resonance $h_x=J_s$, associated with the production of a single meson of length two.

It is also interesting to note that in the case of the first-order resonance, the string very quickly breaks and the concomitant matter creation occurs along the initial string configuration and then spreads within the patch, while everywhere else the matter sites are almost completely empty. In the case of the second-order resonance, the string breaking occurs significantly later in time, as expected, but the concomitant matter creation is also restricted within the patch. This can clearly be seen in snapshot animations of our tensor network simulations \cite{videos}.

\textbf{\emph{Off-resonance string dynamics.---}} 
We now turn our attention to the time evolution of unbroken strings away from resonance. As a general setup, we consider the deeply confined phase $h_x \gg J_p, h_z$, where the dynamics can be understood perturbatively as explained above. We then study how time-dependent observables deviate from this picture as the electric coupling is decreased.

\begin{figure}[t!]
\captionsetup[subfigure]{labelformat=empty}
    \subfloat[\label{subfig:loop_prob}]{}
    \subfloat[\label{subfig:loop_fids}]{}
    \subfloat[\label{subfig:loop_configs}]{}
    \includegraphics[width=\linewidth]{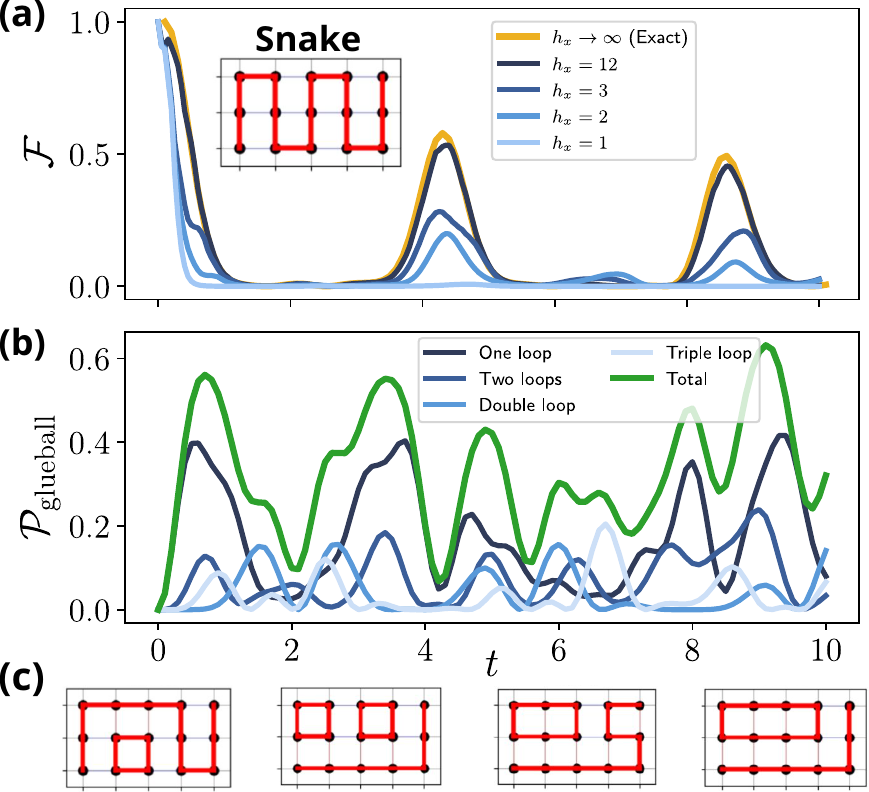}
\caption{(a) Fidelity of the ``snake'' initial state for different values of the electric field $h_x$. Similarly to the minimal length strings case, we see excellent agreement with the perturbative results at large $h_x$. The parameters for the quench in are $J_p=1$ and $J_s=15$. (b) Perturbative ($h_x\rightarrow \infty$) results for the probability for an initial ``snake'' string state to nucleate one or more electric loops, resolved by size and number of the loops.  (c) Examples of configurations containing simple or extended electric loops.}
    \label{fig:snake_probs}
\end{figure}

As a first example, we take strings of minimal length $l$.  In this case, nonzero matrix elements only occur between configurations differing by the application of a plaquette operator at a corner. This effective description is equivalent to one-dimensional free fermions (or hardcore bosons) on a chain of length $l$ and filling $N$ equal to the vertical displacement of the charges \cite{borla2025stringbreaking21dmathbbz2}. This effective model can be used to predict string dynamics in the limit of very strong electric field. To this end, we define the fidelity $\mathcal{F}(t) = |\langle \psi_0|\psi(t) \rangle|^2$, which can be computed analytically in the free-fermion picture and directly compared with the results obtained from our tensor network simulations. As displayed in \cref{fig:ff_deviation}, deep in the confined regime the time-evolution of $\mathcal{F}$ matches perfectly the perturbative description. For a variety of initial states, we observe robust quasiperiodic revivals whose amplitude and position in time strongly depend on the starting configuration. As the electric coupling is decreased, deviations from the integrable limit are evident. We note, however, that robust peaks of moderate amplitude persist well beyond the regime where perturbation theory is expected to be applicable, and only disappear as $h_x \approx h_z$ where the matter fluctuations are very strong.

\textbf{\textit{Dynamical glueball formation.---}} Next, we analyze a case where the initial state is a long $l=14$ ``snake'' shaped string, as shown in the inset of \cref{subfig:loop_prob}. This is a very far-from-equilibrium string configuration, beyond what any current experiment or numerical simulation has probed. A crucial difference compared to the minimal strings is that such configuration can resonate with states that include a shorter string \textit{and} one or more loops of various sizes, shown in Figs.~\ref{subfig:loop_fids} and~\ref{subfig:loop_configs}. At any generic time $t$, therefore, there is a finite probability for the nucleation of electric loops, which can be interpreted as a peculiar form of string breaking that does not involve matter creation. Similarly, a short string can increase its size by absorbing a loop. The effective Hamiltonian operating in this sector of the Hilbert space, for this particular setup, can be obtained through degenerate perturbation theory \cite{SM}. This allows us to extract exact results for the loop creation probability, which are shown in Fig.~\ref{subfig:loop_fids}. We note that the total probability for a string to nucleate a loop is sizable and exhibits irregularly spaced peaks, which persist up to very large timescales.

We argue that these loops are qualitatively analogous to glueballs from QCD. Indeed, the latter are color-singlet bound states of gluons and can be viewed as closed loops of flux or excited configurations of the gluon field without any valence quarks \cite{Mathieu2009glueballs}. The loops we observe forming dynamically in our simulations are gauge-invariant and do not carry static charges. They are closed self-contained excitations of the gauge field, much like glueballs in QCD. 

\textbf{\textit{Summary and outlook.---}}
We have analyzed in detail the quench dynamics of strings in the confined phase of a paradigmatic $\mathbb{Z}_2$ lattice gauge theory in two spatial dimensions using a combination of tensor network methods and perturbative techniques. By tuning the parameters of the model to resonant values, we observe clear signatures of string breaking, signaled by concomitant matter pair creation and by drastically reduced weights of string configurations in the time-evolved state. 

Away from resonance, arbitrarily long strings oscillate between a large number of configurations. These include states which contain shorter strings and disconnected loops, realizing a different form of partial string breaking without matter creation. We argue that the loops forming dynamically in our quench protocol are qualitative analogs of glueballs from QCD. Like the latter, they are gauge-invariant closed-loop self-contained excitations of the gauge field.

The realistic setup that we consider, together with the intrinsically two-dimensional character of the string configurations, suggest that our results can be probed in the near future on state-of-the-art superconducting-qubit and trapped-ion quantum devices. Given our ability to tune the Hamiltonian parameters to speed up the dynamics, the physics outlined above, including the dynamical formation of glueballs, occurs within accessible evolution times on current quantum devices.

\medskip

\footnotesize{\begin{acknowledgments}
K.X., U.B., and J.C.H.~acknowledge funding by the Max Planck Society, the Deutsche Forschungsgemeinschaft (DFG, German Research Foundation) under Germany’s Excellence Strategy – EXC-2111 – 390814868, and the European Research Council (ERC) under the European Union’s Horizon Europe research and innovation program (Grant Agreement No.~101165667)—ERC Starting Grant QuSiGauge. This work is part of the Quantum Computing for High-Energy Physics (QC4HEP) working group. S.M. is supported by Vetenskapsr{\aa}det (grant number 2021-03685) and Nordita.
\end{acknowledgments}}
\normalsize

\bibliography{biblio}

\begin{thebibliography}{121}%
\makeatletter
\providecommand \@ifxundefined [1]{%
 \@ifx{#1\undefined}
}%
\providecommand \@ifnum [1]{%
 \ifnum #1\expandafter \@firstoftwo
 \else \expandafter \@secondoftwo
 \fi
}%
\providecommand \@ifx [1]{%
 \ifx #1\expandafter \@firstoftwo
 \else \expandafter \@secondoftwo
 \fi
}%
\providecommand \natexlab [1]{#1}%
\providecommand \enquote  [1]{``#1''}%
\providecommand \bibnamefont  [1]{#1}%
\providecommand \bibfnamefont [1]{#1}%
\providecommand \citenamefont [1]{#1}%
\providecommand \href@noop [0]{\@secondoftwo}%
\providecommand \href [0]{\begingroup \@sanitize@url \@href}%
\providecommand \@href[1]{\@@startlink{#1}\@@href}%
\providecommand \@@href[1]{\endgroup#1\@@endlink}%
\providecommand \@sanitize@url [0]{\catcode `\\12\catcode `\$12\catcode
  `\&12\catcode `\#12\catcode `\^12\catcode `\_12\catcode `\%12\relax}%
\providecommand \@@startlink[1]{}%
\providecommand \@@endlink[0]{}%
\providecommand \url  [0]{\begingroup\@sanitize@url \@url }%
\providecommand \@url [1]{\endgroup\@href {#1}{\urlprefix }}%
\providecommand \urlprefix  [0]{URL }%
\providecommand \Eprint [0]{\href }%
\providecommand \doibase [0]{https://doi.org/}%
\providecommand \selectlanguage [0]{\@gobble}%
\providecommand \bibinfo  [0]{\@secondoftwo}%
\providecommand \bibfield  [0]{\@secondoftwo}%
\providecommand \translation [1]{[#1]}%
\providecommand \BibitemOpen [0]{}%
\providecommand \bibitemStop [0]{}%
\providecommand \bibitemNoStop [0]{.\EOS\space}%
\providecommand \EOS [0]{\spacefactor3000\relax}%
\providecommand \BibitemShut  [1]{\csname bibitem#1\endcsname}%
\let\auto@bib@innerbib\@empty
\bibitem [{\citenamefont {Weinberg}(1995)}]{Weinberg_book}%
  \BibitemOpen
  \bibfield  {author} {\bibinfo {author} {\bibfnamefont {S.}~\bibnamefont
  {Weinberg}},\ }\href {https://books.google.de/books?id=doeDB3\_WLvwC} {\emph
  {\bibinfo {title} {The Quantum Theory of Fields}}},\ Vol. 2: Modern
  Applications\ (\bibinfo  {publisher} {Cambridge University Press},\ \bibinfo
  {year} {1995})\BibitemShut {NoStop}%
\bibitem [{\citenamefont {Gattringer}\ and\ \citenamefont
  {Lang}(2009)}]{Gattringer_book}%
  \BibitemOpen
  \bibfield  {author} {\bibinfo {author} {\bibfnamefont {C.}~\bibnamefont
  {Gattringer}}\ and\ \bibinfo {author} {\bibfnamefont {C.}~\bibnamefont
  {Lang}},\ }\href {https://books.google.de/books?id=l2hZKnlYDxoC} {\emph
  {\bibinfo {title} {Quantum Chromodynamics on the Lattice: An Introductory
  Presentation}}},\ Lecture Notes in Physics\ (\bibinfo  {publisher} {Springer
  Berlin Heidelberg},\ \bibinfo {year} {2009})\BibitemShut {NoStop}%
\bibitem [{\citenamefont {Zee}(2003)}]{Zee_book}%
  \BibitemOpen
  \bibfield  {author} {\bibinfo {author} {\bibfnamefont {A.}~\bibnamefont
  {Zee}},\ }\href {https://books.google.de/books?id=85G9QgAACAAJ} {\emph
  {\bibinfo {title} {Quantum Field Theory in a Nutshell}}}\ (\bibinfo
  {publisher} {Princeton University Press},\ \bibinfo {year}
  {2003})\BibitemShut {NoStop}%
\bibitem [{\citenamefont {Ellis}\ \emph {et~al.}(2003)\citenamefont {Ellis},
  \citenamefont {Stirling},\ and\ \citenamefont {Webber}}]{Ellis_book}%
  \BibitemOpen
  \bibfield  {author} {\bibinfo {author} {\bibfnamefont {R.}~\bibnamefont
  {Ellis}}, \bibinfo {author} {\bibfnamefont {W.}~\bibnamefont {Stirling}},\
  and\ \bibinfo {author} {\bibfnamefont {B.}~\bibnamefont {Webber}},\ }\href
  {https://books.google.de/books?id=TqrPVoS6s0UC} {\emph {\bibinfo {title} {QCD
  and Collider Physics}}},\ Cambridge Monographs on Particle Physics, Nuclear
  Physics and Cosmology\ (\bibinfo  {publisher} {Cambridge University Press},\
  \bibinfo {year} {2003})\BibitemShut {NoStop}%
\bibitem [{\citenamefont {Rothe}(2005)}]{Rothe_book}%
  \BibitemOpen
  \bibfield  {author} {\bibinfo {author} {\bibfnamefont {H.}~\bibnamefont
  {Rothe}},\ }\href {https://books.google.de/books?id=U1hBLG-\_WxAC} {\emph
  {\bibinfo {title} {Lattice Gauge Theories: An Introduction}}},\ EBSCO ebook
  academic collection\ (\bibinfo  {publisher} {World Scientific},\ \bibinfo
  {year} {2005})\BibitemShut {NoStop}%
\bibitem [{\citenamefont {Kogut}\ and\ \citenamefont
  {Susskind}(1975)}]{Kogut1975}%
  \BibitemOpen
  \bibfield  {author} {\bibinfo {author} {\bibfnamefont {J.}~\bibnamefont
  {Kogut}}\ and\ \bibinfo {author} {\bibfnamefont {L.}~\bibnamefont
  {Susskind}},\ }\bibfield  {title} {\bibinfo {title} {Hamiltonian formulation
  of wilson's lattice gauge theories},\ }\href
  {https://doi.org/10.1103/PhysRevD.11.395} {\bibfield  {journal} {\bibinfo
  {journal} {Phys. Rev. D}\ }\textbf {\bibinfo {volume} {11}},\ \bibinfo
  {pages} {395} (\bibinfo {year} {1975})}\BibitemShut {NoStop}%
\bibitem [{\citenamefont {Wilson}(1974)}]{Wilson1974}%
  \BibitemOpen
  \bibfield  {author} {\bibinfo {author} {\bibfnamefont {K.~G.}\ \bibnamefont
  {Wilson}},\ }\bibfield  {title} {\bibinfo {title} {Confinement of quarks},\
  }\href {https://doi.org/10.1103/PhysRevD.10.2445} {\bibfield  {journal}
  {\bibinfo  {journal} {Phys. Rev. D}\ }\textbf {\bibinfo {volume} {10}},\
  \bibinfo {pages} {2445} (\bibinfo {year} {1974})}\BibitemShut {NoStop}%
\bibitem [{\citenamefont {Dalmonte}\ and\ \citenamefont
  {Montangero}(2016)}]{Dalmonte_review}%
  \BibitemOpen
  \bibfield  {author} {\bibinfo {author} {\bibfnamefont {M.}~\bibnamefont
  {Dalmonte}}\ and\ \bibinfo {author} {\bibfnamefont {S.}~\bibnamefont
  {Montangero}},\ }\bibfield  {title} {\bibinfo {title} {Lattice gauge theory
  simulations in the quantum information era},\ }\href
  {https://doi.org/10.1080/00107514.2016.1151199} {\bibfield  {journal}
  {\bibinfo  {journal} {Contemporary Physics}\ }\textbf {\bibinfo {volume}
  {57}},\ \bibinfo {pages} {388} (\bibinfo {year} {2016})},\ \Eprint
  {https://arxiv.org/abs/https://doi.org/10.1080/00107514.2016.1151199}
  {https://doi.org/10.1080/00107514.2016.1151199} \BibitemShut {NoStop}%
\bibitem [{\citenamefont {Zohar}\ \emph {et~al.}(2015)\citenamefont {Zohar},
  \citenamefont {Cirac},\ and\ \citenamefont {Reznik}}]{Zohar_review}%
  \BibitemOpen
  \bibfield  {author} {\bibinfo {author} {\bibfnamefont {E.}~\bibnamefont
  {Zohar}}, \bibinfo {author} {\bibfnamefont {J.~I.}\ \bibnamefont {Cirac}},\
  and\ \bibinfo {author} {\bibfnamefont {B.}~\bibnamefont {Reznik}},\
  }\bibfield  {title} {\bibinfo {title} {Quantum simulations of lattice gauge
  theories using ultracold atoms in optical lattices},\ }\href
  {https://doi.org/10.1088/0034-4885/79/1/014401} {\bibfield  {journal}
  {\bibinfo  {journal} {Rep. Prog. Phys.}\ }\textbf {\bibinfo {volume} {79}},\
  \bibinfo {pages} {014401} (\bibinfo {year} {2015})}\BibitemShut {NoStop}%
\bibitem [{\citenamefont {Aidelsburger}\ \emph {et~al.}(2011)\citenamefont
  {Aidelsburger}, \citenamefont {Atala}, \citenamefont {Nascimbene},
  \citenamefont {Trotzky}, \citenamefont {Chen},\ and\ \citenamefont
  {Bloch}}]{Aidelsburger2011}%
  \BibitemOpen
  \bibfield  {author} {\bibinfo {author} {\bibfnamefont {M.}~\bibnamefont
  {Aidelsburger}}, \bibinfo {author} {\bibfnamefont {M.}~\bibnamefont {Atala}},
  \bibinfo {author} {\bibfnamefont {S.}~\bibnamefont {Nascimbene}}, \bibinfo
  {author} {\bibfnamefont {S.}~\bibnamefont {Trotzky}}, \bibinfo {author}
  {\bibfnamefont {Y.-.~A.}\ \bibnamefont {Chen}},\ and\ \bibinfo {author}
  {\bibfnamefont {I.}~\bibnamefont {Bloch}},\ }\bibfield  {title} {\bibinfo
  {title} {Experimental realization of strong effective magnetic fields in an
  optical lattice},\ }\href {https://doi.org/10.1103/PhysRevLett.107.255301}
  {\bibfield  {journal} {\bibinfo  {journal} {Phys. Rev. Lett.}\ }\textbf
  {\bibinfo {volume} {107}},\ \bibinfo {pages} {255301} (\bibinfo {year}
  {2011})}\BibitemShut {NoStop}%
\bibitem [{\citenamefont {{Zohar}}(2022)}]{Zohar_NewReview}%
  \BibitemOpen
  \bibfield  {author} {\bibinfo {author} {\bibfnamefont {E.}~\bibnamefont
  {{Zohar}}},\ }\bibfield  {title} {\bibinfo {title} {{Quantum simulation of
  lattice gauge theories in more than one space
  dimension{\textemdash}requirements, challenges and methods}},\ }\href
  {https://doi.org/10.1098/rsta.2021.0069} {\bibfield  {journal} {\bibinfo
  {journal} {Philos. Trans. Royal Soc. A}\ }\textbf {\bibinfo {volume} {380}},\
  \bibinfo {eid} {20210069} (\bibinfo {year} {2022})},\ \Eprint
  {https://arxiv.org/abs/2106.04609} {arXiv:2106.04609 [quant-ph]} \BibitemShut
  {NoStop}%
\bibitem [{\citenamefont {Klco}\ \emph {et~al.}(2022)\citenamefont {Klco},
  \citenamefont {Roggero},\ and\ \citenamefont {Savage}}]{klco2021standard}%
  \BibitemOpen
  \bibfield  {author} {\bibinfo {author} {\bibfnamefont {N.}~\bibnamefont
  {Klco}}, \bibinfo {author} {\bibfnamefont {A.}~\bibnamefont {Roggero}},\ and\
  \bibinfo {author} {\bibfnamefont {M.~J.}\ \bibnamefont {Savage}},\ }\bibfield
   {title} {\bibinfo {title} {Standard model physics and the digital quantum
  revolution: thoughts about the interface},\ }\href
  {https://doi.org/10.1088/1361-6633/ac58a4} {\bibfield  {journal} {\bibinfo
  {journal} {Reports on Progress in Physics}\ }\textbf {\bibinfo {volume}
  {85}},\ \bibinfo {pages} {064301} (\bibinfo {year} {2022})}\BibitemShut
  {NoStop}%
\bibitem [{\citenamefont {Bauer}\ \emph {et~al.}(2023)\citenamefont {Bauer},
  \citenamefont {Davoudi}, \citenamefont {Balantekin}, \citenamefont
  {Bhattacharya}, \citenamefont {Carena}, \citenamefont {de~Jong},
  \citenamefont {Draper}, \citenamefont {El-Khadra}, \citenamefont {Gemelke},
  \citenamefont {Hanada}, \citenamefont {Kharzeev}, \citenamefont {Lamm},
  \citenamefont {Li}, \citenamefont {Liu}, \citenamefont {Lukin}, \citenamefont
  {Meurice}, \citenamefont {Monroe}, \citenamefont {Nachman}, \citenamefont
  {Pagano}, \citenamefont {Preskill}, \citenamefont {Rinaldi}, \citenamefont
  {Roggero}, \citenamefont {Santiago}, \citenamefont {Savage}, \citenamefont
  {Siddiqi}, \citenamefont {Siopsis}, \citenamefont {Van~Zanten}, \citenamefont
  {Wiebe}, \citenamefont {Yamauchi}, \citenamefont {Yeter-Aydeniz},\ and\
  \citenamefont {Zorzetti}}]{Bauer_review}%
  \BibitemOpen
  \bibfield  {author} {\bibinfo {author} {\bibfnamefont {C.~W.}\ \bibnamefont
  {Bauer}}, \bibinfo {author} {\bibfnamefont {Z.}~\bibnamefont {Davoudi}},
  \bibinfo {author} {\bibfnamefont {A.~B.}\ \bibnamefont {Balantekin}},
  \bibinfo {author} {\bibfnamefont {T.}~\bibnamefont {Bhattacharya}}, \bibinfo
  {author} {\bibfnamefont {M.}~\bibnamefont {Carena}}, \bibinfo {author}
  {\bibfnamefont {W.~A.}\ \bibnamefont {de~Jong}}, \bibinfo {author}
  {\bibfnamefont {P.}~\bibnamefont {Draper}}, \bibinfo {author} {\bibfnamefont
  {A.}~\bibnamefont {El-Khadra}}, \bibinfo {author} {\bibfnamefont
  {N.}~\bibnamefont {Gemelke}}, \bibinfo {author} {\bibfnamefont
  {M.}~\bibnamefont {Hanada}}, \bibinfo {author} {\bibfnamefont
  {D.}~\bibnamefont {Kharzeev}}, \bibinfo {author} {\bibfnamefont
  {H.}~\bibnamefont {Lamm}}, \bibinfo {author} {\bibfnamefont {Y.-Y.}\
  \bibnamefont {Li}}, \bibinfo {author} {\bibfnamefont {J.}~\bibnamefont
  {Liu}}, \bibinfo {author} {\bibfnamefont {M.}~\bibnamefont {Lukin}}, \bibinfo
  {author} {\bibfnamefont {Y.}~\bibnamefont {Meurice}}, \bibinfo {author}
  {\bibfnamefont {C.}~\bibnamefont {Monroe}}, \bibinfo {author} {\bibfnamefont
  {B.}~\bibnamefont {Nachman}}, \bibinfo {author} {\bibfnamefont
  {G.}~\bibnamefont {Pagano}}, \bibinfo {author} {\bibfnamefont
  {J.}~\bibnamefont {Preskill}}, \bibinfo {author} {\bibfnamefont
  {E.}~\bibnamefont {Rinaldi}}, \bibinfo {author} {\bibfnamefont
  {A.}~\bibnamefont {Roggero}}, \bibinfo {author} {\bibfnamefont {D.~I.}\
  \bibnamefont {Santiago}}, \bibinfo {author} {\bibfnamefont {M.~J.}\
  \bibnamefont {Savage}}, \bibinfo {author} {\bibfnamefont {I.}~\bibnamefont
  {Siddiqi}}, \bibinfo {author} {\bibfnamefont {G.}~\bibnamefont {Siopsis}},
  \bibinfo {author} {\bibfnamefont {D.}~\bibnamefont {Van~Zanten}}, \bibinfo
  {author} {\bibfnamefont {N.}~\bibnamefont {Wiebe}}, \bibinfo {author}
  {\bibfnamefont {Y.}~\bibnamefont {Yamauchi}}, \bibinfo {author}
  {\bibfnamefont {K.}~\bibnamefont {Yeter-Aydeniz}},\ and\ \bibinfo {author}
  {\bibfnamefont {S.}~\bibnamefont {Zorzetti}},\ }\bibfield  {title} {\bibinfo
  {title} {Quantum simulation for high-energy physics},\ }\href
  {https://doi.org/10.1103/PRXQuantum.4.027001} {\bibfield  {journal} {\bibinfo
   {journal} {PRX Quantum}\ }\textbf {\bibinfo {volume} {4}},\ \bibinfo {pages}
  {027001} (\bibinfo {year} {2023})}\BibitemShut {NoStop}%
\bibitem [{\citenamefont {Di~Meglio}\ \emph {et~al.}(2024)\citenamefont
  {Di~Meglio}, \citenamefont {Jansen}, \citenamefont {Tavernelli},
  \citenamefont {Alexandrou}, \citenamefont {Arunachalam}, \citenamefont
  {Bauer}, \citenamefont {Borras}, \citenamefont {Carrazza}, \citenamefont
  {Crippa}, \citenamefont {Croft}, \citenamefont {de~Putter}, \citenamefont
  {Delgado}, \citenamefont {Dunjko}, \citenamefont {Egger}, \citenamefont
  {Fern\'andez-Combarro}, \citenamefont {Fuchs}, \citenamefont {Funcke},
  \citenamefont {Gonz\'alez-Cuadra}, \citenamefont {Grossi}, \citenamefont
  {Halimeh}, \citenamefont {Holmes}, \citenamefont {K\"uhn}, \citenamefont
  {Lacroix}, \citenamefont {Lewis}, \citenamefont {Lucchesi}, \citenamefont
  {Martinez}, \citenamefont {Meloni}, \citenamefont {Mezzacapo}, \citenamefont
  {Montangero}, \citenamefont {Nagano}, \citenamefont {Pascuzzi}, \citenamefont
  {Radescu}, \citenamefont {Ortega}, \citenamefont {Roggero}, \citenamefont
  {Schuhmacher}, \citenamefont {Seixas}, \citenamefont {Silvi}, \citenamefont
  {Spentzouris}, \citenamefont {Tacchino}, \citenamefont {Temme}, \citenamefont
  {Terashi}, \citenamefont {Tura}, \citenamefont {T\"uys\"uz}, \citenamefont
  {Vallecorsa}, \citenamefont {Wiese}, \citenamefont {Yoo},\ and\ \citenamefont
  {Zhang}}]{dimeglio2023quantum}%
  \BibitemOpen
  \bibfield  {author} {\bibinfo {author} {\bibfnamefont {A.}~\bibnamefont
  {Di~Meglio}}, \bibinfo {author} {\bibfnamefont {K.}~\bibnamefont {Jansen}},
  \bibinfo {author} {\bibfnamefont {I.}~\bibnamefont {Tavernelli}}, \bibinfo
  {author} {\bibfnamefont {C.}~\bibnamefont {Alexandrou}}, \bibinfo {author}
  {\bibfnamefont {S.}~\bibnamefont {Arunachalam}}, \bibinfo {author}
  {\bibfnamefont {C.~W.}\ \bibnamefont {Bauer}}, \bibinfo {author}
  {\bibfnamefont {K.}~\bibnamefont {Borras}}, \bibinfo {author} {\bibfnamefont
  {S.}~\bibnamefont {Carrazza}}, \bibinfo {author} {\bibfnamefont
  {A.}~\bibnamefont {Crippa}}, \bibinfo {author} {\bibfnamefont
  {V.}~\bibnamefont {Croft}}, \bibinfo {author} {\bibfnamefont
  {R.}~\bibnamefont {de~Putter}}, \bibinfo {author} {\bibfnamefont
  {A.}~\bibnamefont {Delgado}}, \bibinfo {author} {\bibfnamefont
  {V.}~\bibnamefont {Dunjko}}, \bibinfo {author} {\bibfnamefont {D.~J.}\
  \bibnamefont {Egger}}, \bibinfo {author} {\bibfnamefont {E.}~\bibnamefont
  {Fern\'andez-Combarro}}, \bibinfo {author} {\bibfnamefont {E.}~\bibnamefont
  {Fuchs}}, \bibinfo {author} {\bibfnamefont {L.}~\bibnamefont {Funcke}},
  \bibinfo {author} {\bibfnamefont {D.}~\bibnamefont {Gonz\'alez-Cuadra}},
  \bibinfo {author} {\bibfnamefont {M.}~\bibnamefont {Grossi}}, \bibinfo
  {author} {\bibfnamefont {J.~C.}\ \bibnamefont {Halimeh}}, \bibinfo {author}
  {\bibfnamefont {Z.}~\bibnamefont {Holmes}}, \bibinfo {author} {\bibfnamefont
  {S.}~\bibnamefont {K\"uhn}}, \bibinfo {author} {\bibfnamefont
  {D.}~\bibnamefont {Lacroix}}, \bibinfo {author} {\bibfnamefont
  {R.}~\bibnamefont {Lewis}}, \bibinfo {author} {\bibfnamefont
  {D.}~\bibnamefont {Lucchesi}}, \bibinfo {author} {\bibfnamefont {M.~L.}\
  \bibnamefont {Martinez}}, \bibinfo {author} {\bibfnamefont {F.}~\bibnamefont
  {Meloni}}, \bibinfo {author} {\bibfnamefont {A.}~\bibnamefont {Mezzacapo}},
  \bibinfo {author} {\bibfnamefont {S.}~\bibnamefont {Montangero}}, \bibinfo
  {author} {\bibfnamefont {L.}~\bibnamefont {Nagano}}, \bibinfo {author}
  {\bibfnamefont {V.~R.}\ \bibnamefont {Pascuzzi}}, \bibinfo {author}
  {\bibfnamefont {V.}~\bibnamefont {Radescu}}, \bibinfo {author} {\bibfnamefont
  {E.~R.}\ \bibnamefont {Ortega}}, \bibinfo {author} {\bibfnamefont
  {A.}~\bibnamefont {Roggero}}, \bibinfo {author} {\bibfnamefont
  {J.}~\bibnamefont {Schuhmacher}}, \bibinfo {author} {\bibfnamefont
  {J.}~\bibnamefont {Seixas}}, \bibinfo {author} {\bibfnamefont
  {P.}~\bibnamefont {Silvi}}, \bibinfo {author} {\bibfnamefont
  {P.}~\bibnamefont {Spentzouris}}, \bibinfo {author} {\bibfnamefont
  {F.}~\bibnamefont {Tacchino}}, \bibinfo {author} {\bibfnamefont
  {K.}~\bibnamefont {Temme}}, \bibinfo {author} {\bibfnamefont
  {K.}~\bibnamefont {Terashi}}, \bibinfo {author} {\bibfnamefont
  {J.}~\bibnamefont {Tura}}, \bibinfo {author} {\bibfnamefont {C.}~\bibnamefont
  {T\"uys\"uz}}, \bibinfo {author} {\bibfnamefont {S.}~\bibnamefont
  {Vallecorsa}}, \bibinfo {author} {\bibfnamefont {U.-J.}\ \bibnamefont
  {Wiese}}, \bibinfo {author} {\bibfnamefont {S.}~\bibnamefont {Yoo}},\ and\
  \bibinfo {author} {\bibfnamefont {J.}~\bibnamefont {Zhang}},\ }\bibfield
  {title} {\bibinfo {title} {Quantum computing for high-energy physics: State
  of the art and challenges},\ }\href
  {https://doi.org/10.1103/PRXQuantum.5.037001} {\bibfield  {journal} {\bibinfo
   {journal} {PRX Quantum}\ }\textbf {\bibinfo {volume} {5}},\ \bibinfo {pages}
  {037001} (\bibinfo {year} {2024})}\BibitemShut {NoStop}%
\bibitem [{\citenamefont {Cheng}\ and\ \citenamefont
  {Zhai}(2024)}]{Cheng_review}%
  \BibitemOpen
  \bibfield  {author} {\bibinfo {author} {\bibfnamefont {Y.}~\bibnamefont
  {Cheng}}\ and\ \bibinfo {author} {\bibfnamefont {H.}~\bibnamefont {Zhai}},\
  }\bibfield  {title} {\bibinfo {title} {Emergent u(1) lattice gauge theory in
  rydberg atom arrays},\ }\href {https://doi.org/10.1038/s42254-024-00749-6}
  {\bibfield  {journal} {\bibinfo  {journal} {Nature Reviews Physics}\ }\textbf
  {\bibinfo {volume} {6}},\ \bibinfo {pages} {566} (\bibinfo {year}
  {2024})}\BibitemShut {NoStop}%
\bibitem [{\citenamefont {Halimeh}\ \emph {et~al.}(2025)\citenamefont
  {Halimeh}, \citenamefont {Aidelsburger}, \citenamefont {Grusdt},
  \citenamefont {Hauke},\ and\ \citenamefont {Yang}}]{Halimeh_review}%
  \BibitemOpen
  \bibfield  {author} {\bibinfo {author} {\bibfnamefont {J.~C.}\ \bibnamefont
  {Halimeh}}, \bibinfo {author} {\bibfnamefont {M.}~\bibnamefont
  {Aidelsburger}}, \bibinfo {author} {\bibfnamefont {F.}~\bibnamefont
  {Grusdt}}, \bibinfo {author} {\bibfnamefont {P.}~\bibnamefont {Hauke}},\ and\
  \bibinfo {author} {\bibfnamefont {B.}~\bibnamefont {Yang}},\ }\bibfield
  {title} {\bibinfo {title} {Cold-atom quantum simulators of gauge theories},\
  }\bibfield  {journal} {\bibinfo  {journal} {Nature Physics}\ }\href
  {https://doi.org/10.1038/s41567-024-02721-8} {10.1038/s41567-024-02721-8}
  (\bibinfo {year} {2025})\BibitemShut {NoStop}%
\bibitem [{\citenamefont {Cohen}\ \emph {et~al.}(2021)\citenamefont {Cohen},
  \citenamefont {Lamm}, \citenamefont {Lawrence},\ and\ \citenamefont
  {Yamauchi}}]{Cohen:2021imf}%
  \BibitemOpen
  \bibfield  {author} {\bibinfo {author} {\bibfnamefont {T.~D.}\ \bibnamefont
  {Cohen}}, \bibinfo {author} {\bibfnamefont {H.}~\bibnamefont {Lamm}},
  \bibinfo {author} {\bibfnamefont {S.}~\bibnamefont {Lawrence}},\ and\
  \bibinfo {author} {\bibfnamefont {Y.}~\bibnamefont {Yamauchi}} (\bibinfo
  {collaboration} {NuQS}),\ }\bibfield  {title} {\bibinfo {title} {{Quantum
  algorithms for transport coefficients in gauge theories}},\ }\href
  {https://doi.org/10.1103/PhysRevD.104.094514} {\bibfield  {journal} {\bibinfo
   {journal} {Phys. Rev. D}\ }\textbf {\bibinfo {volume} {104}},\ \bibinfo
  {pages} {094514} (\bibinfo {year} {2021})},\ \Eprint
  {https://arxiv.org/abs/2104.02024} {arXiv:2104.02024 [hep-lat]} \BibitemShut
  {NoStop}%
\bibitem [{\citenamefont {Lee}\ \emph {et~al.}(2025)\citenamefont {Lee},
  \citenamefont {Turro},\ and\ \citenamefont {Yao}}]{Lee:2024jnt}%
  \BibitemOpen
  \bibfield  {author} {\bibinfo {author} {\bibfnamefont {K.}~\bibnamefont
  {Lee}}, \bibinfo {author} {\bibfnamefont {F.}~\bibnamefont {Turro}},\ and\
  \bibinfo {author} {\bibfnamefont {X.}~\bibnamefont {Yao}},\ }\bibfield
  {title} {\bibinfo {title} {Quantum computing for energy correlators},\ }\href
  {https://doi.org/10.1103/PhysRevD.111.054514} {\bibfield  {journal} {\bibinfo
   {journal} {Phys. Rev. D}\ }\textbf {\bibinfo {volume} {111}},\ \bibinfo
  {pages} {054514} (\bibinfo {year} {2025})}\BibitemShut {NoStop}%
\bibitem [{\citenamefont {Turro}\ \emph {et~al.}(2024)\citenamefont {Turro},
  \citenamefont {Ciavarella},\ and\ \citenamefont {Yao}}]{Turro:2024pxu}%
  \BibitemOpen
  \bibfield  {author} {\bibinfo {author} {\bibfnamefont {F.}~\bibnamefont
  {Turro}}, \bibinfo {author} {\bibfnamefont {A.}~\bibnamefont {Ciavarella}},\
  and\ \bibinfo {author} {\bibfnamefont {X.}~\bibnamefont {Yao}},\ }\bibfield
  {title} {\bibinfo {title} {{Classical and quantum computing of shear
  viscosity for (2+1)D SU(2) gauge theory}},\ }\href
  {https://doi.org/10.1103/PhysRevD.109.114511} {\bibfield  {journal} {\bibinfo
   {journal} {Phys. Rev. D}\ }\textbf {\bibinfo {volume} {109}},\ \bibinfo
  {pages} {114511} (\bibinfo {year} {2024})},\ \Eprint
  {https://arxiv.org/abs/2402.04221} {arXiv:2402.04221 [hep-lat]} \BibitemShut
  {NoStop}%
\bibitem [{\citenamefont
  {Bauer}(2025)}]{bauer2025efficientusequantumcomputers}%
  \BibitemOpen
  \bibfield  {author} {\bibinfo {author} {\bibfnamefont {C.~W.}\ \bibnamefont
  {Bauer}},\ }\bibfield  {title} {\bibinfo {title} {Efficient use of quantum
  computers for collider physics},\ }\href {https://arxiv.org/abs/2503.16602}
  {\  (\bibinfo {year} {2025})},\ \Eprint {https://arxiv.org/abs/2503.16602}
  {arXiv:2503.16602 [hep-ph]} \BibitemShut {NoStop}%
\bibitem [{\citenamefont {Wegner}(1971)}]{Wegner1971}%
  \BibitemOpen
  \bibfield  {author} {\bibinfo {author} {\bibfnamefont {F.~J.}\ \bibnamefont
  {Wegner}},\ }\bibfield  {title} {\bibinfo {title} {Duality in generalized
  ising models and phase transitions without local order parameters},\ }\href
  {https://doi.org/10.1063/1.1665530} {\bibfield  {journal} {\bibinfo
  {journal} {Journal of Mathematical Physics}\ }\textbf {\bibinfo {volume}
  {12}},\ \bibinfo {pages} {2259} (\bibinfo {year} {1971})},\ \Eprint
  {https://arxiv.org/abs/https://pubs.aip.org/aip/jmp/article-pdf/12/10/2259/19106483/2259\_1\_online.pdf}
  {https://pubs.aip.org/aip/jmp/article-pdf/12/10/2259/19106483/2259\_1\_online.pdf}
  \BibitemShut {NoStop}%
\bibitem [{\citenamefont {Kogut}(1979)}]{Kogut_review}%
  \BibitemOpen
  \bibfield  {author} {\bibinfo {author} {\bibfnamefont {J.~B.}\ \bibnamefont
  {Kogut}},\ }\bibfield  {title} {\bibinfo {title} {An introduction to lattice
  gauge theory and spin systems},\ }\href
  {https://doi.org/10.1103/RevModPhys.51.659} {\bibfield  {journal} {\bibinfo
  {journal} {Rev. Mod. Phys.}\ }\textbf {\bibinfo {volume} {51}},\ \bibinfo
  {pages} {659} (\bibinfo {year} {1979})}\BibitemShut {NoStop}%
\bibitem [{\citenamefont {Wen}(2004)}]{wen2004quantum}%
  \BibitemOpen
  \bibfield  {author} {\bibinfo {author} {\bibfnamefont {X.}~\bibnamefont
  {Wen}},\ }\href {https://books.google.de/books?id=llnlrfdR4YgC} {\emph
  {\bibinfo {title} {Quantum Field Theory of Many-Body Systems:From the Origin
  of Sound to an Origin of Light and Electrons: From the Origin of Sound to an
  Origin of Light and Electrons}}},\ Oxford Graduate Texts\ (\bibinfo
  {publisher} {OUP Oxford},\ \bibinfo {year} {2004})\BibitemShut {NoStop}%
\bibitem [{\citenamefont {Savary}\ and\ \citenamefont
  {Balents}(2016)}]{Savary2016}%
  \BibitemOpen
  \bibfield  {author} {\bibinfo {author} {\bibfnamefont {L.}~\bibnamefont
  {Savary}}\ and\ \bibinfo {author} {\bibfnamefont {L.}~\bibnamefont
  {Balents}},\ }\bibfield  {title} {\bibinfo {title} {Quantum spin liquids: a
  review},\ }\href {https://doi.org/10.1088/0034-4885/80/1/016502} {\bibfield
  {journal} {\bibinfo  {journal} {Rep. Prog. Phys.}\ }\textbf {\bibinfo
  {volume} {80}},\ \bibinfo {pages} {016502} (\bibinfo {year}
  {2016})}\BibitemShut {NoStop}%
\bibitem [{\citenamefont {Calzetta}\ and\ \citenamefont
  {Hu}(2008)}]{Calzetta_book}%
  \BibitemOpen
  \bibfield  {author} {\bibinfo {author} {\bibfnamefont {E.}~\bibnamefont
  {Calzetta}}\ and\ \bibinfo {author} {\bibfnamefont {B.}~\bibnamefont {Hu}},\
  }\href {https://books.google.de/books?id=BRJ7ryt2l1IC} {\emph {\bibinfo
  {title} {Nonequilibrium Quantum Field Theory}}},\ Cambridge Monographs on
  Mathematical Physics\ (\bibinfo  {publisher} {Cambridge University Press},\
  \bibinfo {year} {2008})\BibitemShut {NoStop}%
\bibitem [{\citenamefont {Smith}\ \emph
  {et~al.}(2017{\natexlab{a}})\citenamefont {Smith}, \citenamefont {Knolle},
  \citenamefont {Kovrizhin},\ and\ \citenamefont {Moessner}}]{Smith2017}%
  \BibitemOpen
  \bibfield  {author} {\bibinfo {author} {\bibfnamefont {A.}~\bibnamefont
  {Smith}}, \bibinfo {author} {\bibfnamefont {J.}~\bibnamefont {Knolle}},
  \bibinfo {author} {\bibfnamefont {D.~L.}\ \bibnamefont {Kovrizhin}},\ and\
  \bibinfo {author} {\bibfnamefont {R.}~\bibnamefont {Moessner}},\ }\bibfield
  {title} {\bibinfo {title} {Disorder-free localization},\ }\href
  {https://doi.org/10.1103/PhysRevLett.118.266601} {\bibfield  {journal}
  {\bibinfo  {journal} {Phys. Rev. Lett.}\ }\textbf {\bibinfo {volume} {118}},\
  \bibinfo {pages} {266601} (\bibinfo {year} {2017}{\natexlab{a}})}\BibitemShut
  {NoStop}%
\bibitem [{\citenamefont {Brenes}\ \emph {et~al.}(2018)\citenamefont {Brenes},
  \citenamefont {Dalmonte}, \citenamefont {Heyl},\ and\ \citenamefont
  {Scardicchio}}]{Brenes2018}%
  \BibitemOpen
  \bibfield  {author} {\bibinfo {author} {\bibfnamefont {M.}~\bibnamefont
  {Brenes}}, \bibinfo {author} {\bibfnamefont {M.}~\bibnamefont {Dalmonte}},
  \bibinfo {author} {\bibfnamefont {M.}~\bibnamefont {Heyl}},\ and\ \bibinfo
  {author} {\bibfnamefont {A.}~\bibnamefont {Scardicchio}},\ }\bibfield
  {title} {\bibinfo {title} {Many-body localization dynamics from gauge
  invariance},\ }\href {https://doi.org/10.1103/PhysRevLett.120.030601}
  {\bibfield  {journal} {\bibinfo  {journal} {Phys. Rev. Lett.}\ }\textbf
  {\bibinfo {volume} {120}},\ \bibinfo {pages} {030601} (\bibinfo {year}
  {2018})}\BibitemShut {NoStop}%
\bibitem [{\citenamefont {Smith}\ \emph
  {et~al.}(2017{\natexlab{b}})\citenamefont {Smith}, \citenamefont {Knolle},
  \citenamefont {Moessner},\ and\ \citenamefont
  {Kovrizhin}}]{smith2017absence}%
  \BibitemOpen
  \bibfield  {author} {\bibinfo {author} {\bibfnamefont {A.}~\bibnamefont
  {Smith}}, \bibinfo {author} {\bibfnamefont {J.}~\bibnamefont {Knolle}},
  \bibinfo {author} {\bibfnamefont {R.}~\bibnamefont {Moessner}},\ and\
  \bibinfo {author} {\bibfnamefont {D.~L.}\ \bibnamefont {Kovrizhin}},\
  }\bibfield  {title} {\bibinfo {title} {Absence of ergodicity without quenched
  disorder: From quantum disentangled liquids to many-body localization},\
  }\href {https://doi.org/10.1103/PhysRevLett.119.176601} {\bibfield  {journal}
  {\bibinfo  {journal} {Phys. Rev. Lett.}\ }\textbf {\bibinfo {volume} {119}},\
  \bibinfo {pages} {176601} (\bibinfo {year} {2017}{\natexlab{b}})}\BibitemShut
  {NoStop}%
\bibitem [{\citenamefont {Karpov}\ \emph {et~al.}(2021)\citenamefont {Karpov},
  \citenamefont {Verdel}, \citenamefont {Huang}, \citenamefont {Schmitt},\ and\
  \citenamefont {Heyl}}]{karpov2021disorder}%
  \BibitemOpen
  \bibfield  {author} {\bibinfo {author} {\bibfnamefont {P.}~\bibnamefont
  {Karpov}}, \bibinfo {author} {\bibfnamefont {R.}~\bibnamefont {Verdel}},
  \bibinfo {author} {\bibfnamefont {Y.-P.}\ \bibnamefont {Huang}}, \bibinfo
  {author} {\bibfnamefont {M.}~\bibnamefont {Schmitt}},\ and\ \bibinfo {author}
  {\bibfnamefont {M.}~\bibnamefont {Heyl}},\ }\bibfield  {title} {\bibinfo
  {title} {Disorder-free localization in an interacting {2D} lattice gauge
  theory},\ }\href {https://doi.org/10.1103/PhysRevLett.126.130401} {\bibfield
  {journal} {\bibinfo  {journal} {Phys. Rev. Lett.}\ }\textbf {\bibinfo
  {volume} {126}},\ \bibinfo {pages} {130401} (\bibinfo {year}
  {2021})}\BibitemShut {NoStop}%
\bibitem [{\citenamefont {Sous}\ \emph {et~al.}(2021)\citenamefont {Sous},
  \citenamefont {Kloss}, \citenamefont {Kennes}, \citenamefont {Reichman},\
  and\ \citenamefont {Millis}}]{Sous2021}%
  \BibitemOpen
  \bibfield  {author} {\bibinfo {author} {\bibfnamefont {J.}~\bibnamefont
  {Sous}}, \bibinfo {author} {\bibfnamefont {B.}~\bibnamefont {Kloss}},
  \bibinfo {author} {\bibfnamefont {D.~M.}\ \bibnamefont {Kennes}}, \bibinfo
  {author} {\bibfnamefont {D.~R.}\ \bibnamefont {Reichman}},\ and\ \bibinfo
  {author} {\bibfnamefont {A.~J.}\ \bibnamefont {Millis}},\ }\bibfield  {title}
  {\bibinfo {title} {Phonon-induced disorder in dynamics of optically pumped
  metals from nonlinear electron-phonon coupling},\ }\href
  {https://doi.org/10.1038/s41467-021-26030-3} {\bibfield  {journal} {\bibinfo
  {journal} {Nat. Commun.}\ }\textbf {\bibinfo {volume} {12}},\ \bibinfo
  {pages} {5803} (\bibinfo {year} {2021})}\BibitemShut {NoStop}%
\bibitem [{\citenamefont {{Chakraborty}}\ \emph {et~al.}(2022)\citenamefont
  {{Chakraborty}}, \citenamefont {{Heyl}}, \citenamefont {{Karpov}},\ and\
  \citenamefont {{Moessner}}}]{Chakraborty2022}%
  \BibitemOpen
  \bibfield  {author} {\bibinfo {author} {\bibfnamefont {N.}~\bibnamefont
  {{Chakraborty}}}, \bibinfo {author} {\bibfnamefont {M.}~\bibnamefont
  {{Heyl}}}, \bibinfo {author} {\bibfnamefont {P.}~\bibnamefont {{Karpov}}},\
  and\ \bibinfo {author} {\bibfnamefont {R.}~\bibnamefont {{Moessner}}},\
  }\bibfield  {title} {\bibinfo {title} {{Disorder-free localization transition
  in a two-dimensional lattice gauge theory}},\ }\href@noop {} {\bibfield
  {journal} {\bibinfo  {journal} {arXiv preprint}\ } (\bibinfo {year}
  {2022})},\ \Eprint {https://arxiv.org/abs/2203.06198} {arXiv:2203.06198
  [cond-mat.dis-nn]} \BibitemShut {NoStop}%
\bibitem [{\citenamefont {Halimeh}\ \emph {et~al.}(2021)\citenamefont
  {Halimeh}, \citenamefont {Homeier}, \citenamefont {Zhao}, \citenamefont
  {Bohrdt}, \citenamefont {Grusdt}, \citenamefont {Hauke},\ and\ \citenamefont
  {Knolle}}]{Halimeh2021enhancing}%
  \BibitemOpen
  \bibfield  {author} {\bibinfo {author} {\bibfnamefont {J.~C.}\ \bibnamefont
  {Halimeh}}, \bibinfo {author} {\bibfnamefont {L.}~\bibnamefont {Homeier}},
  \bibinfo {author} {\bibfnamefont {H.}~\bibnamefont {Zhao}}, \bibinfo {author}
  {\bibfnamefont {A.}~\bibnamefont {Bohrdt}}, \bibinfo {author} {\bibfnamefont
  {F.}~\bibnamefont {Grusdt}}, \bibinfo {author} {\bibfnamefont
  {P.}~\bibnamefont {Hauke}},\ and\ \bibinfo {author} {\bibfnamefont
  {J.}~\bibnamefont {Knolle}},\ }\bibfield  {title} {\bibinfo {title}
  {Enhancing disorder-free localization through dynamically emergent local
  symmetries},\ }\href@noop {} {\  (\bibinfo {year} {2021})},\ \Eprint
  {https://arxiv.org/abs/2111.08715} {arXiv:2111.08715 [cond-mat.quant-gas]}
  \BibitemShut {NoStop}%
\bibitem [{\citenamefont {Surace}\ \emph {et~al.}(2020)\citenamefont {Surace},
  \citenamefont {Mazza}, \citenamefont {Giudici}, \citenamefont {Lerose},
  \citenamefont {Gambassi},\ and\ \citenamefont {Dalmonte}}]{Surace2020}%
  \BibitemOpen
  \bibfield  {author} {\bibinfo {author} {\bibfnamefont {F.~M.}\ \bibnamefont
  {Surace}}, \bibinfo {author} {\bibfnamefont {P.~P.}\ \bibnamefont {Mazza}},
  \bibinfo {author} {\bibfnamefont {G.}~\bibnamefont {Giudici}}, \bibinfo
  {author} {\bibfnamefont {A.}~\bibnamefont {Lerose}}, \bibinfo {author}
  {\bibfnamefont {A.}~\bibnamefont {Gambassi}},\ and\ \bibinfo {author}
  {\bibfnamefont {M.}~\bibnamefont {Dalmonte}},\ }\bibfield  {title} {\bibinfo
  {title} {Lattice gauge theories and string dynamics in {Rydberg} atom quantum
  simulators},\ }\href {https://doi.org/10.1103/PhysRevX.10.021041} {\bibfield
  {journal} {\bibinfo  {journal} {Phys. Rev. X}\ }\textbf {\bibinfo {volume}
  {10}},\ \bibinfo {pages} {021041} (\bibinfo {year} {2020})}\BibitemShut
  {NoStop}%
\bibitem [{\citenamefont {Desaules}\ \emph
  {et~al.}(2023{\natexlab{a}})\citenamefont {Desaules}, \citenamefont
  {Banerjee}, \citenamefont {Hudomal}, \citenamefont
  {Papi\ifmmode~\acute{c}\else \'{c}\fi{}}, \citenamefont {Sen},\ and\
  \citenamefont {Halimeh}}]{Desaules2022weak}%
  \BibitemOpen
  \bibfield  {author} {\bibinfo {author} {\bibfnamefont {J.-Y.}\ \bibnamefont
  {Desaules}}, \bibinfo {author} {\bibfnamefont {D.}~\bibnamefont {Banerjee}},
  \bibinfo {author} {\bibfnamefont {A.}~\bibnamefont {Hudomal}}, \bibinfo
  {author} {\bibfnamefont {Z.}~\bibnamefont {Papi\ifmmode~\acute{c}\else
  \'{c}\fi{}}}, \bibinfo {author} {\bibfnamefont {A.}~\bibnamefont {Sen}},\
  and\ \bibinfo {author} {\bibfnamefont {J.~C.}\ \bibnamefont {Halimeh}},\
  }\bibfield  {title} {\bibinfo {title} {Weak ergodicity breaking in the
  schwinger model},\ }\href {https://doi.org/10.1103/PhysRevB.107.L201105}
  {\bibfield  {journal} {\bibinfo  {journal} {Phys. Rev. B}\ }\textbf {\bibinfo
  {volume} {107}},\ \bibinfo {pages} {L201105} (\bibinfo {year}
  {2023}{\natexlab{a}})}\BibitemShut {NoStop}%
\bibitem [{\citenamefont {Desaules}\ \emph
  {et~al.}(2023{\natexlab{b}})\citenamefont {Desaules}, \citenamefont
  {Hudomal}, \citenamefont {Banerjee}, \citenamefont {Sen}, \citenamefont
  {Papi\ifmmode~\acute{c}\else \'{c}\fi{}},\ and\ \citenamefont
  {Halimeh}}]{Desaules2022prominent}%
  \BibitemOpen
  \bibfield  {author} {\bibinfo {author} {\bibfnamefont {J.-Y.}\ \bibnamefont
  {Desaules}}, \bibinfo {author} {\bibfnamefont {A.}~\bibnamefont {Hudomal}},
  \bibinfo {author} {\bibfnamefont {D.}~\bibnamefont {Banerjee}}, \bibinfo
  {author} {\bibfnamefont {A.}~\bibnamefont {Sen}}, \bibinfo {author}
  {\bibfnamefont {Z.}~\bibnamefont {Papi\ifmmode~\acute{c}\else \'{c}\fi{}}},\
  and\ \bibinfo {author} {\bibfnamefont {J.~C.}\ \bibnamefont {Halimeh}},\
  }\bibfield  {title} {\bibinfo {title} {Prominent quantum many-body scars in a
  truncated schwinger model},\ }\href
  {https://doi.org/10.1103/PhysRevB.107.205112} {\bibfield  {journal} {\bibinfo
   {journal} {Phys. Rev. B}\ }\textbf {\bibinfo {volume} {107}},\ \bibinfo
  {pages} {205112} (\bibinfo {year} {2023}{\natexlab{b}})}\BibitemShut
  {NoStop}%
\bibitem [{\citenamefont {Aramthottil}\ \emph {et~al.}(2022)\citenamefont
  {Aramthottil}, \citenamefont {Bhattacharya}, \citenamefont
  {Gonz\'alez-Cuadra}, \citenamefont {Lewenstein}, \citenamefont {Barbiero},\
  and\ \citenamefont {Zakrzewski}}]{aramthottil2022scar}%
  \BibitemOpen
  \bibfield  {author} {\bibinfo {author} {\bibfnamefont {A.~S.}\ \bibnamefont
  {Aramthottil}}, \bibinfo {author} {\bibfnamefont {U.}~\bibnamefont
  {Bhattacharya}}, \bibinfo {author} {\bibfnamefont {D.}~\bibnamefont
  {Gonz\'alez-Cuadra}}, \bibinfo {author} {\bibfnamefont {M.}~\bibnamefont
  {Lewenstein}}, \bibinfo {author} {\bibfnamefont {L.}~\bibnamefont
  {Barbiero}},\ and\ \bibinfo {author} {\bibfnamefont {J.}~\bibnamefont
  {Zakrzewski}},\ }\bibfield  {title} {\bibinfo {title} {Scar states in
  deconfined ${\mathbb{z}}_{2}$ lattice gauge theories},\ }\href
  {https://doi.org/10.1103/PhysRevB.106.L041101} {\bibfield  {journal}
  {\bibinfo  {journal} {Phys. Rev. B}\ }\textbf {\bibinfo {volume} {106}},\
  \bibinfo {pages} {L041101} (\bibinfo {year} {2022})}\BibitemShut {NoStop}%
\bibitem [{\citenamefont {Desaules}\ \emph {et~al.}(2024)\citenamefont
  {Desaules}, \citenamefont {Su}, \citenamefont {McCulloch}, \citenamefont
  {Yang}, \citenamefont {Papi{\'{c}}},\ and\ \citenamefont
  {Halimeh}}]{Desaules2024ergodicitybreaking}%
  \BibitemOpen
  \bibfield  {author} {\bibinfo {author} {\bibfnamefont {J.-Y.}\ \bibnamefont
  {Desaules}}, \bibinfo {author} {\bibfnamefont {G.-X.}\ \bibnamefont {Su}},
  \bibinfo {author} {\bibfnamefont {I.~P.}\ \bibnamefont {McCulloch}}, \bibinfo
  {author} {\bibfnamefont {B.}~\bibnamefont {Yang}}, \bibinfo {author}
  {\bibfnamefont {Z.}~\bibnamefont {Papi{\'{c}}}},\ and\ \bibinfo {author}
  {\bibfnamefont {J.~C.}\ \bibnamefont {Halimeh}},\ }\bibfield  {title}
  {\bibinfo {title} {Ergodicity {B}reaking {U}nder {C}onfinement in
  {C}old-{A}tom {Q}uantum {S}imulators},\ }\href
  {https://doi.org/10.22331/q-2024-02-29-1274} {\bibfield  {journal} {\bibinfo
  {journal} {{Quantum}}\ }\textbf {\bibinfo {volume} {8}},\ \bibinfo {pages}
  {1274} (\bibinfo {year} {2024})}\BibitemShut {NoStop}%
\bibitem [{\citenamefont {Desaules}\ \emph {et~al.}(2025)\citenamefont
  {Desaules}, \citenamefont {Iadecola},\ and\ \citenamefont
  {Halimeh}}]{desaules2024massassistedlocaldeconfinementconfined}%
  \BibitemOpen
  \bibfield  {author} {\bibinfo {author} {\bibfnamefont {J.-Y.}\ \bibnamefont
  {Desaules}}, \bibinfo {author} {\bibfnamefont {T.}~\bibnamefont {Iadecola}},\
  and\ \bibinfo {author} {\bibfnamefont {J.~C.}\ \bibnamefont {Halimeh}},\
  }\bibfield  {title} {\bibinfo {title} {Mass-assisted local deconfinement in a
  confined ${\mathbb{z}}_{2}$ lattice gauge theory},\ }\href
  {https://doi.org/10.1103/mfg2-t6gb} {\bibfield  {journal} {\bibinfo
  {journal} {Phys. Rev. B}\ }\textbf {\bibinfo {volume} {112}},\ \bibinfo
  {pages} {014301} (\bibinfo {year} {2025})}\BibitemShut {NoStop}%
\bibitem [{\citenamefont {Jeyaretnam}\ \emph {et~al.}(2025)\citenamefont
  {Jeyaretnam}, \citenamefont {Bhore}, \citenamefont {Osborne}, \citenamefont
  {Halimeh},\ and\ \citenamefont
  {Papić}}]{jeyaretnam2025hilbertspacefragmentationorigin}%
  \BibitemOpen
  \bibfield  {author} {\bibinfo {author} {\bibfnamefont {J.}~\bibnamefont
  {Jeyaretnam}}, \bibinfo {author} {\bibfnamefont {T.}~\bibnamefont {Bhore}},
  \bibinfo {author} {\bibfnamefont {J.~J.}\ \bibnamefont {Osborne}}, \bibinfo
  {author} {\bibfnamefont {J.~C.}\ \bibnamefont {Halimeh}},\ and\ \bibinfo
  {author} {\bibfnamefont {Z.}~\bibnamefont {Papić}},\ }\bibfield  {title}
  {\bibinfo {title} {Hilbert space fragmentation at the origin of disorder-free
  localization in the lattice schwinger model},\ }\href
  {https://arxiv.org/abs/2409.08320} {\  (\bibinfo {year} {2025})},\ \Eprint
  {https://arxiv.org/abs/2409.08320} {arXiv:2409.08320 [quant-ph]} \BibitemShut
  {NoStop}%
\bibitem [{\citenamefont {Ciavarella}\ \emph {et~al.}(2025)\citenamefont
  {Ciavarella}, \citenamefont {Bauer},\ and\ \citenamefont
  {Halimeh}}]{ciavarella2025generichilbertspacefragmentation}%
  \BibitemOpen
  \bibfield  {author} {\bibinfo {author} {\bibfnamefont {A.~N.}\ \bibnamefont
  {Ciavarella}}, \bibinfo {author} {\bibfnamefont {C.~W.}\ \bibnamefont
  {Bauer}},\ and\ \bibinfo {author} {\bibfnamefont {J.~C.}\ \bibnamefont
  {Halimeh}},\ }\bibfield  {title} {\bibinfo {title} {Generic hilbert space
  fragmentation in kogut--susskind lattice gauge theories},\ }\href
  {https://arxiv.org/abs/2502.03533} {\  (\bibinfo {year} {2025})},\ \Eprint
  {https://arxiv.org/abs/2502.03533} {arXiv:2502.03533 [quant-ph]} \BibitemShut
  {NoStop}%
\bibitem [{\citenamefont {Tarabunga}\ \emph {et~al.}(2023)\citenamefont
  {Tarabunga}, \citenamefont {Tirrito}, \citenamefont {Chanda},\ and\
  \citenamefont {Dalmonte}}]{Tarabunga2023many}%
  \BibitemOpen
  \bibfield  {author} {\bibinfo {author} {\bibfnamefont {P.~S.}\ \bibnamefont
  {Tarabunga}}, \bibinfo {author} {\bibfnamefont {E.}~\bibnamefont {Tirrito}},
  \bibinfo {author} {\bibfnamefont {T.}~\bibnamefont {Chanda}},\ and\ \bibinfo
  {author} {\bibfnamefont {M.}~\bibnamefont {Dalmonte}},\ }\bibfield  {title}
  {\bibinfo {title} {Many-body magic via pauli-markov chains---from criticality
  to gauge theories},\ }\href {https://doi.org/10.1103/PRXQuantum.4.040317}
  {\bibfield  {journal} {\bibinfo  {journal} {PRX Quantum}\ }\textbf {\bibinfo
  {volume} {4}},\ \bibinfo {pages} {040317} (\bibinfo {year}
  {2023})}\BibitemShut {NoStop}%
\bibitem [{\citenamefont {Hartse}\ \emph {et~al.}(2024)\citenamefont {Hartse},
  \citenamefont {Fidkowski},\ and\ \citenamefont
  {Mueller}}]{hartse2024stabilizerscars}%
  \BibitemOpen
  \bibfield  {author} {\bibinfo {author} {\bibfnamefont {J.}~\bibnamefont
  {Hartse}}, \bibinfo {author} {\bibfnamefont {L.}~\bibnamefont {Fidkowski}},\
  and\ \bibinfo {author} {\bibfnamefont {N.}~\bibnamefont {Mueller}},\
  }\bibfield  {title} {\bibinfo {title} {Stabilizer scars},\ }\href
  {https://arxiv.org/abs/2411.12797} {\  (\bibinfo {year} {2024})},\ \Eprint
  {https://arxiv.org/abs/2411.12797} {arXiv:2411.12797 [quant-ph]} \BibitemShut
  {NoStop}%
\bibitem [{\citenamefont {Smith}\ \emph {et~al.}(2025)\citenamefont {Smith},
  \citenamefont {Papi\ifmmode~\acute{c}\else \'{c}\fi{}},\ and\ \citenamefont
  {Hallam}}]{Smith2025nonstabilizerness}%
  \BibitemOpen
  \bibfield  {author} {\bibinfo {author} {\bibfnamefont {R.}~\bibnamefont
  {Smith}}, \bibinfo {author} {\bibfnamefont {Z.}~\bibnamefont
  {Papi\ifmmode~\acute{c}\else \'{c}\fi{}}},\ and\ \bibinfo {author}
  {\bibfnamefont {A.}~\bibnamefont {Hallam}},\ }\bibfield  {title} {\bibinfo
  {title} {Nonstabilizerness in kinetically constrained rydberg atom arrays},\
  }\href {https://doi.org/10.1103/jz4d-vdhj} {\bibfield  {journal} {\bibinfo
  {journal} {Phys. Rev. B}\ }\textbf {\bibinfo {volume} {111}},\ \bibinfo
  {pages} {245148} (\bibinfo {year} {2025})}\BibitemShut {NoStop}%
\bibitem [{\citenamefont {Falc\~ao}\ \emph {et~al.}(2025)\citenamefont
  {Falc\~ao}, \citenamefont {Tarabunga}, \citenamefont {Frau}, \citenamefont
  {Tirrito}, \citenamefont {Zakrzewski},\ and\ \citenamefont
  {Dalmonte}}]{Falcao2025Nonstabilizerness}%
  \BibitemOpen
  \bibfield  {author} {\bibinfo {author} {\bibfnamefont {P.~R.~N.}\
  \bibnamefont {Falc\~ao}}, \bibinfo {author} {\bibfnamefont {P.~S.}\
  \bibnamefont {Tarabunga}}, \bibinfo {author} {\bibfnamefont {M.}~\bibnamefont
  {Frau}}, \bibinfo {author} {\bibfnamefont {E.}~\bibnamefont {Tirrito}},
  \bibinfo {author} {\bibfnamefont {J.}~\bibnamefont {Zakrzewski}},\ and\
  \bibinfo {author} {\bibfnamefont {M.}~\bibnamefont {Dalmonte}},\ }\bibfield
  {title} {\bibinfo {title} {Nonstabilizerness in u(1) lattice gauge theory},\
  }\href {https://doi.org/10.1103/PhysRevB.111.L081102} {\bibfield  {journal}
  {\bibinfo  {journal} {Phys. Rev. B}\ }\textbf {\bibinfo {volume} {111}},\
  \bibinfo {pages} {L081102} (\bibinfo {year} {2025})}\BibitemShut {NoStop}%
\bibitem [{\citenamefont {Esposito}\ \emph {et~al.}(2025)\citenamefont
  {Esposito}, \citenamefont {Cepollaro}, \citenamefont {Cappiello},\ and\
  \citenamefont {Hamma}}]{Esposito2025magic}%
  \BibitemOpen
  \bibfield  {author} {\bibinfo {author} {\bibfnamefont {G.}~\bibnamefont
  {Esposito}}, \bibinfo {author} {\bibfnamefont {S.}~\bibnamefont {Cepollaro}},
  \bibinfo {author} {\bibfnamefont {L.}~\bibnamefont {Cappiello}},\ and\
  \bibinfo {author} {\bibfnamefont {A.}~\bibnamefont {Hamma}},\ }\bibfield
  {title} {\bibinfo {title} {Magic of discrete lattice gauge theories},\ }\href
  {https://doi.org/10.1142/S0219887825500033} {\bibfield  {journal} {\bibinfo
  {journal} {International Journal of Geometric Methods in Modern Physics}\
  }\textbf {\bibinfo {volume} {22}},\ \bibinfo {pages} {2550003} (\bibinfo
  {year} {2025})},\ \Eprint
  {https://arxiv.org/abs/https://doi.org/10.1142/S0219887825500033}
  {https://doi.org/10.1142/S0219887825500033} \BibitemShut {NoStop}%
\bibitem [{\citenamefont {Martinez}\ \emph {et~al.}(2016)\citenamefont
  {Martinez}, \citenamefont {Muschik}, \citenamefont {Schindler}, \citenamefont
  {Nigg}, \citenamefont {Erhard}, \citenamefont {Heyl}, \citenamefont {Hauke},
  \citenamefont {Dalmonte}, \citenamefont {Monz}, \citenamefont {Zoller},\ and\
  \citenamefont {Blatt}}]{Martinez2016}%
  \BibitemOpen
  \bibfield  {author} {\bibinfo {author} {\bibfnamefont {E.~A.}\ \bibnamefont
  {Martinez}}, \bibinfo {author} {\bibfnamefont {C.~A.}\ \bibnamefont
  {Muschik}}, \bibinfo {author} {\bibfnamefont {P.}~\bibnamefont {Schindler}},
  \bibinfo {author} {\bibfnamefont {D.}~\bibnamefont {Nigg}}, \bibinfo {author}
  {\bibfnamefont {A.}~\bibnamefont {Erhard}}, \bibinfo {author} {\bibfnamefont
  {M.}~\bibnamefont {Heyl}}, \bibinfo {author} {\bibfnamefont {P.}~\bibnamefont
  {Hauke}}, \bibinfo {author} {\bibfnamefont {M.}~\bibnamefont {Dalmonte}},
  \bibinfo {author} {\bibfnamefont {T.}~\bibnamefont {Monz}}, \bibinfo {author}
  {\bibfnamefont {P.}~\bibnamefont {Zoller}},\ and\ \bibinfo {author}
  {\bibfnamefont {R.}~\bibnamefont {Blatt}},\ }\bibfield  {title} {\bibinfo
  {title} {Real-time dynamics of lattice gauge theories with a few-qubit
  quantum computer},\ }\href {https://doi.org/10.1038/nature18318} {\bibfield
  {journal} {\bibinfo  {journal} {Nature}\ }\textbf {\bibinfo {volume} {534}},\
  \bibinfo {pages} {516} (\bibinfo {year} {2016})}\BibitemShut {NoStop}%
\bibitem [{\citenamefont {Klco}\ \emph {et~al.}(2018)\citenamefont {Klco},
  \citenamefont {Dumitrescu}, \citenamefont {McCaskey}, \citenamefont {Morris},
  \citenamefont {Pooser}, \citenamefont {Sanz}, \citenamefont {Solano},
  \citenamefont {Lougovski},\ and\ \citenamefont {Savage}}]{Klco2018}%
  \BibitemOpen
  \bibfield  {author} {\bibinfo {author} {\bibfnamefont {N.}~\bibnamefont
  {Klco}}, \bibinfo {author} {\bibfnamefont {E.~F.}\ \bibnamefont
  {Dumitrescu}}, \bibinfo {author} {\bibfnamefont {A.~J.}\ \bibnamefont
  {McCaskey}}, \bibinfo {author} {\bibfnamefont {T.~D.}\ \bibnamefont
  {Morris}}, \bibinfo {author} {\bibfnamefont {R.~C.}\ \bibnamefont {Pooser}},
  \bibinfo {author} {\bibfnamefont {M.}~\bibnamefont {Sanz}}, \bibinfo {author}
  {\bibfnamefont {E.}~\bibnamefont {Solano}}, \bibinfo {author} {\bibfnamefont
  {P.}~\bibnamefont {Lougovski}},\ and\ \bibinfo {author} {\bibfnamefont
  {M.~J.}\ \bibnamefont {Savage}},\ }\bibfield  {title} {\bibinfo {title}
  {Quantum-classical computation of {Schwinger} model dynamics using quantum
  computers},\ }\href {https://doi.org/10.1103/PhysRevA.98.032331} {\bibfield
  {journal} {\bibinfo  {journal} {Phys. Rev. A}\ }\textbf {\bibinfo {volume}
  {98}},\ \bibinfo {pages} {032331} (\bibinfo {year} {2018})}\BibitemShut
  {NoStop}%
\bibitem [{\citenamefont {G{\"o}rg}\ \emph {et~al.}(2019)\citenamefont
  {G{\"o}rg}, \citenamefont {Sandholzer}, \citenamefont {Minguzzi},
  \citenamefont {Desbuquois}, \citenamefont {Messer},\ and\ \citenamefont
  {Esslinger}}]{Goerg2019}%
  \BibitemOpen
  \bibfield  {author} {\bibinfo {author} {\bibfnamefont {F.}~\bibnamefont
  {G{\"o}rg}}, \bibinfo {author} {\bibfnamefont {K.}~\bibnamefont
  {Sandholzer}}, \bibinfo {author} {\bibfnamefont {J.}~\bibnamefont
  {Minguzzi}}, \bibinfo {author} {\bibfnamefont {R.}~\bibnamefont
  {Desbuquois}}, \bibinfo {author} {\bibfnamefont {M.}~\bibnamefont {Messer}},\
  and\ \bibinfo {author} {\bibfnamefont {T.}~\bibnamefont {Esslinger}},\
  }\bibfield  {title} {\bibinfo {title} {Realization of density-dependent
  {Peierls} phases to engineer quantized gauge fields coupled to ultracold
  matter},\ }\href {https://doi.org/10.1038/s41567-019-0615-4} {\bibfield
  {journal} {\bibinfo  {journal} {Nat. Phys.}\ }\textbf {\bibinfo {volume}
  {15}},\ \bibinfo {pages} {1161} (\bibinfo {year} {2019})}\BibitemShut
  {NoStop}%
\bibitem [{\citenamefont {Schweizer}\ \emph {et~al.}(2019)\citenamefont
  {Schweizer}, \citenamefont {Grusdt}, \citenamefont {Berngruber},
  \citenamefont {Barbiero}, \citenamefont {Demler}, \citenamefont {Goldman},
  \citenamefont {Bloch},\ and\ \citenamefont {Aidelsburger}}]{Schweizer2019}%
  \BibitemOpen
  \bibfield  {author} {\bibinfo {author} {\bibfnamefont {C.}~\bibnamefont
  {Schweizer}}, \bibinfo {author} {\bibfnamefont {F.}~\bibnamefont {Grusdt}},
  \bibinfo {author} {\bibfnamefont {M.}~\bibnamefont {Berngruber}}, \bibinfo
  {author} {\bibfnamefont {L.}~\bibnamefont {Barbiero}}, \bibinfo {author}
  {\bibfnamefont {E.}~\bibnamefont {Demler}}, \bibinfo {author} {\bibfnamefont
  {N.}~\bibnamefont {Goldman}}, \bibinfo {author} {\bibfnamefont
  {I.}~\bibnamefont {Bloch}},\ and\ \bibinfo {author} {\bibfnamefont
  {M.}~\bibnamefont {Aidelsburger}},\ }\bibfield  {title} {\bibinfo {title}
  {Floquet approach to $\mathbb{Z}$2 lattice gauge theories with ultracold
  atoms in optical lattices},\ }\href
  {https://doi.org/10.1038/s41567-019-0649-7} {\bibfield  {journal} {\bibinfo
  {journal} {Nat. Phys.}\ }\textbf {\bibinfo {volume} {15}},\ \bibinfo {pages}
  {1168} (\bibinfo {year} {2019})}\BibitemShut {NoStop}%
\bibitem [{\citenamefont {Mil}\ \emph {et~al.}(2020)\citenamefont {Mil},
  \citenamefont {Zache}, \citenamefont {Hegde}, \citenamefont {Xia},
  \citenamefont {Bhatt}, \citenamefont {Oberthaler}, \citenamefont {Hauke},
  \citenamefont {Berges},\ and\ \citenamefont {Jendrzejewski}}]{Mil2020}%
  \BibitemOpen
  \bibfield  {author} {\bibinfo {author} {\bibfnamefont {A.}~\bibnamefont
  {Mil}}, \bibinfo {author} {\bibfnamefont {T.~V.}\ \bibnamefont {Zache}},
  \bibinfo {author} {\bibfnamefont {A.}~\bibnamefont {Hegde}}, \bibinfo
  {author} {\bibfnamefont {A.}~\bibnamefont {Xia}}, \bibinfo {author}
  {\bibfnamefont {R.~P.}\ \bibnamefont {Bhatt}}, \bibinfo {author}
  {\bibfnamefont {M.~K.}\ \bibnamefont {Oberthaler}}, \bibinfo {author}
  {\bibfnamefont {P.}~\bibnamefont {Hauke}}, \bibinfo {author} {\bibfnamefont
  {J.}~\bibnamefont {Berges}},\ and\ \bibinfo {author} {\bibfnamefont
  {F.}~\bibnamefont {Jendrzejewski}},\ }\bibfield  {title} {\bibinfo {title} {A
  scalable realization of local {U(1)} gauge invariance in cold atomic
  mixtures},\ }\href {https://doi.org/10.1126/science.aaz5312} {\bibfield
  {journal} {\bibinfo  {journal} {Science}\ }\textbf {\bibinfo {volume}
  {367}},\ \bibinfo {pages} {1128} (\bibinfo {year} {2020})}\BibitemShut
  {NoStop}%
\bibitem [{\citenamefont {Yang}\ \emph {et~al.}(2020)\citenamefont {Yang},
  \citenamefont {Sun}, \citenamefont {Ott}, \citenamefont {Wang}, \citenamefont
  {Zache}, \citenamefont {Halimeh}, \citenamefont {Yuan}, \citenamefont
  {Hauke},\ and\ \citenamefont {Pan}}]{Yang2020}%
  \BibitemOpen
  \bibfield  {author} {\bibinfo {author} {\bibfnamefont {B.}~\bibnamefont
  {Yang}}, \bibinfo {author} {\bibfnamefont {H.}~\bibnamefont {Sun}}, \bibinfo
  {author} {\bibfnamefont {R.}~\bibnamefont {Ott}}, \bibinfo {author}
  {\bibfnamefont {H.-Y.}\ \bibnamefont {Wang}}, \bibinfo {author}
  {\bibfnamefont {T.~V.}\ \bibnamefont {Zache}}, \bibinfo {author}
  {\bibfnamefont {J.~C.}\ \bibnamefont {Halimeh}}, \bibinfo {author}
  {\bibfnamefont {Z.-S.}\ \bibnamefont {Yuan}}, \bibinfo {author}
  {\bibfnamefont {P.}~\bibnamefont {Hauke}},\ and\ \bibinfo {author}
  {\bibfnamefont {J.-W.}\ \bibnamefont {Pan}},\ }\bibfield  {title} {\bibinfo
  {title} {Observation of gauge invariance in a 71-site {Bose--Hubbard} quantum
  simulator},\ }\href {https://doi.org/10.1038/s41586-020-2910-8} {\bibfield
  {journal} {\bibinfo  {journal} {Nature}\ }\textbf {\bibinfo {volume} {587}},\
  \bibinfo {pages} {392} (\bibinfo {year} {2020})}\BibitemShut {NoStop}%
\bibitem [{\citenamefont {Wang}\ \emph {et~al.}(2022)\citenamefont {Wang},
  \citenamefont {Ge}, \citenamefont {Xiang}, \citenamefont {Song},
  \citenamefont {Huang}, \citenamefont {Song}, \citenamefont {Guo},
  \citenamefont {Su}, \citenamefont {Xu}, \citenamefont {Zheng},\ and\
  \citenamefont {Fan}}]{Wang2021}%
  \BibitemOpen
  \bibfield  {author} {\bibinfo {author} {\bibfnamefont {Z.}~\bibnamefont
  {Wang}}, \bibinfo {author} {\bibfnamefont {Z.-Y.}\ \bibnamefont {Ge}},
  \bibinfo {author} {\bibfnamefont {Z.}~\bibnamefont {Xiang}}, \bibinfo
  {author} {\bibfnamefont {X.}~\bibnamefont {Song}}, \bibinfo {author}
  {\bibfnamefont {R.-Z.}\ \bibnamefont {Huang}}, \bibinfo {author}
  {\bibfnamefont {P.}~\bibnamefont {Song}}, \bibinfo {author} {\bibfnamefont
  {X.-Y.}\ \bibnamefont {Guo}}, \bibinfo {author} {\bibfnamefont
  {L.}~\bibnamefont {Su}}, \bibinfo {author} {\bibfnamefont {K.}~\bibnamefont
  {Xu}}, \bibinfo {author} {\bibfnamefont {D.}~\bibnamefont {Zheng}},\ and\
  \bibinfo {author} {\bibfnamefont {H.}~\bibnamefont {Fan}},\ }\bibfield
  {title} {\bibinfo {title} {Observation of emergent ${\mathbb{z}}_{2}$ gauge
  invariance in a superconducting circuit},\ }\href
  {https://doi.org/10.1103/PhysRevResearch.4.L022060} {\bibfield  {journal}
  {\bibinfo  {journal} {Phys. Rev. Research}\ }\textbf {\bibinfo {volume}
  {4}},\ \bibinfo {pages} {L022060} (\bibinfo {year} {2022})}\BibitemShut
  {NoStop}%
\bibitem [{\citenamefont {Su}\ \emph {et~al.}(2023)\citenamefont {Su},
  \citenamefont {Sun}, \citenamefont {Hudomal}, \citenamefont {Desaules},
  \citenamefont {Zhou}, \citenamefont {Yang}, \citenamefont {Halimeh},
  \citenamefont {Yuan}, \citenamefont {Papi\ifmmode~\acute{c}\else
  \'{c}\fi{}},\ and\ \citenamefont {Pan}}]{Su2022}%
  \BibitemOpen
  \bibfield  {author} {\bibinfo {author} {\bibfnamefont {G.-X.}\ \bibnamefont
  {Su}}, \bibinfo {author} {\bibfnamefont {H.}~\bibnamefont {Sun}}, \bibinfo
  {author} {\bibfnamefont {A.}~\bibnamefont {Hudomal}}, \bibinfo {author}
  {\bibfnamefont {J.-Y.}\ \bibnamefont {Desaules}}, \bibinfo {author}
  {\bibfnamefont {Z.-Y.}\ \bibnamefont {Zhou}}, \bibinfo {author}
  {\bibfnamefont {B.}~\bibnamefont {Yang}}, \bibinfo {author} {\bibfnamefont
  {J.~C.}\ \bibnamefont {Halimeh}}, \bibinfo {author} {\bibfnamefont {Z.-S.}\
  \bibnamefont {Yuan}}, \bibinfo {author} {\bibfnamefont {Z.}~\bibnamefont
  {Papi\ifmmode~\acute{c}\else \'{c}\fi{}}},\ and\ \bibinfo {author}
  {\bibfnamefont {J.-W.}\ \bibnamefont {Pan}},\ }\bibfield  {title} {\bibinfo
  {title} {Observation of many-body scarring in a bose-hubbard quantum
  simulator},\ }\href {https://doi.org/10.1103/PhysRevResearch.5.023010}
  {\bibfield  {journal} {\bibinfo  {journal} {Phys. Rev. Res.}\ }\textbf
  {\bibinfo {volume} {5}},\ \bibinfo {pages} {023010} (\bibinfo {year}
  {2023})}\BibitemShut {NoStop}%
\bibitem [{\citenamefont {Zhou}\ \emph {et~al.}(2022)\citenamefont {Zhou},
  \citenamefont {Su}, \citenamefont {Halimeh}, \citenamefont {Ott},
  \citenamefont {Sun}, \citenamefont {Hauke}, \citenamefont {Yang},
  \citenamefont {Yuan}, \citenamefont {Berges},\ and\ \citenamefont
  {Pan}}]{Zhou2022}%
  \BibitemOpen
  \bibfield  {author} {\bibinfo {author} {\bibfnamefont {Z.-Y.}\ \bibnamefont
  {Zhou}}, \bibinfo {author} {\bibfnamefont {G.-X.}\ \bibnamefont {Su}},
  \bibinfo {author} {\bibfnamefont {J.~C.}\ \bibnamefont {Halimeh}}, \bibinfo
  {author} {\bibfnamefont {R.}~\bibnamefont {Ott}}, \bibinfo {author}
  {\bibfnamefont {H.}~\bibnamefont {Sun}}, \bibinfo {author} {\bibfnamefont
  {P.}~\bibnamefont {Hauke}}, \bibinfo {author} {\bibfnamefont
  {B.}~\bibnamefont {Yang}}, \bibinfo {author} {\bibfnamefont {Z.-S.}\
  \bibnamefont {Yuan}}, \bibinfo {author} {\bibfnamefont {J.}~\bibnamefont
  {Berges}},\ and\ \bibinfo {author} {\bibfnamefont {J.-W.}\ \bibnamefont
  {Pan}},\ }\bibfield  {title} {\bibinfo {title} {Thermalization dynamics of a
  gauge theory on a quantum simulator},\ }\href
  {https://doi.org/10.1126/science.abl6277} {\bibfield  {journal} {\bibinfo
  {journal} {Science}\ }\textbf {\bibinfo {volume} {377}},\ \bibinfo {pages}
  {311} (\bibinfo {year} {2022})}\BibitemShut {NoStop}%
\bibitem [{\citenamefont {Wang}\ \emph {et~al.}(2023)\citenamefont {Wang},
  \citenamefont {Zhang}, \citenamefont {Yao}, \citenamefont {Liu},
  \citenamefont {Zhu}, \citenamefont {Zheng}, \citenamefont {Wang},
  \citenamefont {Zhai}, \citenamefont {Yuan},\ and\ \citenamefont
  {Pan}}]{Wang2023}%
  \BibitemOpen
  \bibfield  {author} {\bibinfo {author} {\bibfnamefont {H.-Y.}\ \bibnamefont
  {Wang}}, \bibinfo {author} {\bibfnamefont {W.-Y.}\ \bibnamefont {Zhang}},
  \bibinfo {author} {\bibfnamefont {Z.}~\bibnamefont {Yao}}, \bibinfo {author}
  {\bibfnamefont {Y.}~\bibnamefont {Liu}}, \bibinfo {author} {\bibfnamefont
  {Z.-H.}\ \bibnamefont {Zhu}}, \bibinfo {author} {\bibfnamefont {Y.-G.}\
  \bibnamefont {Zheng}}, \bibinfo {author} {\bibfnamefont {X.-K.}\ \bibnamefont
  {Wang}}, \bibinfo {author} {\bibfnamefont {H.}~\bibnamefont {Zhai}}, \bibinfo
  {author} {\bibfnamefont {Z.-S.}\ \bibnamefont {Yuan}},\ and\ \bibinfo
  {author} {\bibfnamefont {J.-W.}\ \bibnamefont {Pan}},\ }\bibfield  {title}
  {\bibinfo {title} {Interrelated thermalization and quantum criticality in a
  lattice gauge simulator},\ }\href
  {https://doi.org/10.1103/PhysRevLett.131.050401} {\bibfield  {journal}
  {\bibinfo  {journal} {Phys. Rev. Lett.}\ }\textbf {\bibinfo {volume} {131}},\
  \bibinfo {pages} {050401} (\bibinfo {year} {2023})}\BibitemShut {NoStop}%
\bibitem [{\citenamefont {Zhang}\ \emph {et~al.}(2024)\citenamefont {Zhang},
  \citenamefont {Liu}, \citenamefont {Cheng}, \citenamefont {He}, \citenamefont
  {Wang}, \citenamefont {Wang}, \citenamefont {Zhu}, \citenamefont {Su},
  \citenamefont {Zhou}, \citenamefont {Zheng}, \citenamefont {Sun},
  \citenamefont {Yang}, \citenamefont {Hauke}, \citenamefont {Zheng},
  \citenamefont {Halimeh}, \citenamefont {Yuan},\ and\ \citenamefont
  {Pan}}]{Zhang2023}%
  \BibitemOpen
  \bibfield  {author} {\bibinfo {author} {\bibfnamefont {W.-Y.}\ \bibnamefont
  {Zhang}}, \bibinfo {author} {\bibfnamefont {Y.}~\bibnamefont {Liu}}, \bibinfo
  {author} {\bibfnamefont {Y.}~\bibnamefont {Cheng}}, \bibinfo {author}
  {\bibfnamefont {M.-G.}\ \bibnamefont {He}}, \bibinfo {author} {\bibfnamefont
  {H.-Y.}\ \bibnamefont {Wang}}, \bibinfo {author} {\bibfnamefont {T.-Y.}\
  \bibnamefont {Wang}}, \bibinfo {author} {\bibfnamefont {Z.-H.}\ \bibnamefont
  {Zhu}}, \bibinfo {author} {\bibfnamefont {G.-X.}\ \bibnamefont {Su}},
  \bibinfo {author} {\bibfnamefont {Z.-Y.}\ \bibnamefont {Zhou}}, \bibinfo
  {author} {\bibfnamefont {Y.-G.}\ \bibnamefont {Zheng}}, \bibinfo {author}
  {\bibfnamefont {H.}~\bibnamefont {Sun}}, \bibinfo {author} {\bibfnamefont
  {B.}~\bibnamefont {Yang}}, \bibinfo {author} {\bibfnamefont {P.}~\bibnamefont
  {Hauke}}, \bibinfo {author} {\bibfnamefont {W.}~\bibnamefont {Zheng}},
  \bibinfo {author} {\bibfnamefont {J.~C.}\ \bibnamefont {Halimeh}}, \bibinfo
  {author} {\bibfnamefont {Z.-S.}\ \bibnamefont {Yuan}},\ and\ \bibinfo
  {author} {\bibfnamefont {J.-W.}\ \bibnamefont {Pan}},\ }\bibfield  {title}
  {\bibinfo {title} {Observation of microscopic confinement dynamics by a
  tunable topological theta-angle},\ }\bibfield  {journal} {\bibinfo  {journal}
  {Nature Physics}\ }\href {https://doi.org/10.1038/s41567-024-02702-x}
  {10.1038/s41567-024-02702-x} (\bibinfo {year} {2024})\BibitemShut {NoStop}%
\bibitem [{\citenamefont {Ciavarella}\ and\ \citenamefont
  {Bauer}(2024)}]{Ciavarella2024quantum}%
  \BibitemOpen
  \bibfield  {author} {\bibinfo {author} {\bibfnamefont {A.~N.}\ \bibnamefont
  {Ciavarella}}\ and\ \bibinfo {author} {\bibfnamefont {C.~W.}\ \bibnamefont
  {Bauer}},\ }\bibfield  {title} {\bibinfo {title} {Quantum simulation of su(3)
  lattice yang-mills theory at leading order in large-${N}_{c}$ expansion},\
  }\href {https://doi.org/10.1103/PhysRevLett.133.111901} {\bibfield  {journal}
  {\bibinfo  {journal} {Phys. Rev. Lett.}\ }\textbf {\bibinfo {volume} {133}},\
  \bibinfo {pages} {111901} (\bibinfo {year} {2024})}\BibitemShut {NoStop}%
\bibitem [{\citenamefont {Ciavarella}(2024)}]{Ciavarella:2024lsp}%
  \BibitemOpen
  \bibfield  {author} {\bibinfo {author} {\bibfnamefont {A.~N.}\ \bibnamefont
  {Ciavarella}},\ }\bibfield  {title} {\bibinfo {title} {{String Breaking in
  the Heavy Quark Limit with Scalable Circuits}},\ }\href@noop {} {\  (\bibinfo
  {year} {2024})},\ \Eprint {https://arxiv.org/abs/2411.05915}
  {arXiv:2411.05915 [quant-ph]} \BibitemShut {NoStop}%
\bibitem [{\citenamefont {Farrell}\ \emph
  {et~al.}(2024{\natexlab{a}})\citenamefont {Farrell}, \citenamefont {Illa},
  \citenamefont {Ciavarella},\ and\ \citenamefont {Savage}}]{Farrell:2023fgd}%
  \BibitemOpen
  \bibfield  {author} {\bibinfo {author} {\bibfnamefont {R.~C.}\ \bibnamefont
  {Farrell}}, \bibinfo {author} {\bibfnamefont {M.}~\bibnamefont {Illa}},
  \bibinfo {author} {\bibfnamefont {A.~N.}\ \bibnamefont {Ciavarella}},\ and\
  \bibinfo {author} {\bibfnamefont {M.~J.}\ \bibnamefont {Savage}},\ }\bibfield
   {title} {\bibinfo {title} {{Scalable Circuits for Preparing Ground States on
  Digital Quantum Computers: The Schwinger Model Vacuum on 100 Qubits}},\
  }\href {https://doi.org/10.1103/PRXQuantum.5.020315} {\bibfield  {journal}
  {\bibinfo  {journal} {PRX Quantum}\ }\textbf {\bibinfo {volume} {5}},\
  \bibinfo {pages} {020315} (\bibinfo {year} {2024}{\natexlab{a}})},\ \Eprint
  {https://arxiv.org/abs/2308.04481} {arXiv:2308.04481 [quant-ph]} \BibitemShut
  {NoStop}%
\bibitem [{\citenamefont {Farrell}\ \emph
  {et~al.}(2024{\natexlab{b}})\citenamefont {Farrell}, \citenamefont {Illa},
  \citenamefont {Ciavarella},\ and\ \citenamefont {Savage}}]{Farrell:2024fit}%
  \BibitemOpen
  \bibfield  {author} {\bibinfo {author} {\bibfnamefont {R.~C.}\ \bibnamefont
  {Farrell}}, \bibinfo {author} {\bibfnamefont {M.}~\bibnamefont {Illa}},
  \bibinfo {author} {\bibfnamefont {A.~N.}\ \bibnamefont {Ciavarella}},\ and\
  \bibinfo {author} {\bibfnamefont {M.~J.}\ \bibnamefont {Savage}},\ }\bibfield
   {title} {\bibinfo {title} {{Quantum simulations of hadron dynamics in the
  Schwinger model using 112 qubits}},\ }\href
  {https://doi.org/10.1103/PhysRevD.109.114510} {\bibfield  {journal} {\bibinfo
   {journal} {Phys. Rev. D}\ }\textbf {\bibinfo {volume} {109}},\ \bibinfo
  {pages} {114510} (\bibinfo {year} {2024}{\natexlab{b}})},\ \Eprint
  {https://arxiv.org/abs/2401.08044} {arXiv:2401.08044 [quant-ph]} \BibitemShut
  {NoStop}%
\bibitem [{\citenamefont {Zhu}\ \emph {et~al.}(2024)\citenamefont {Zhu},
  \citenamefont {Liu}, \citenamefont {Lagnese}, \citenamefont {Surace},
  \citenamefont {Zhang}, \citenamefont {He}, \citenamefont {Halimeh},
  \citenamefont {Dalmonte}, \citenamefont {Morampudi}, \citenamefont {Wilczek},
  \citenamefont {Yuan},\ and\ \citenamefont
  {Pan}}]{zhu2024probingfalsevacuumdecay}%
  \BibitemOpen
  \bibfield  {author} {\bibinfo {author} {\bibfnamefont {Z.-H.}\ \bibnamefont
  {Zhu}}, \bibinfo {author} {\bibfnamefont {Y.}~\bibnamefont {Liu}}, \bibinfo
  {author} {\bibfnamefont {G.}~\bibnamefont {Lagnese}}, \bibinfo {author}
  {\bibfnamefont {F.~M.}\ \bibnamefont {Surace}}, \bibinfo {author}
  {\bibfnamefont {W.-Y.}\ \bibnamefont {Zhang}}, \bibinfo {author}
  {\bibfnamefont {M.-G.}\ \bibnamefont {He}}, \bibinfo {author} {\bibfnamefont
  {J.~C.}\ \bibnamefont {Halimeh}}, \bibinfo {author} {\bibfnamefont
  {M.}~\bibnamefont {Dalmonte}}, \bibinfo {author} {\bibfnamefont {S.~C.}\
  \bibnamefont {Morampudi}}, \bibinfo {author} {\bibfnamefont {F.}~\bibnamefont
  {Wilczek}}, \bibinfo {author} {\bibfnamefont {Z.-S.}\ \bibnamefont {Yuan}},\
  and\ \bibinfo {author} {\bibfnamefont {J.-W.}\ \bibnamefont {Pan}},\
  }\bibfield  {title} {\bibinfo {title} {Probing false vacuum decay on a
  cold-atom gauge-theory quantum simulator},\ }\href
  {https://arxiv.org/abs/2411.12565} {\  (\bibinfo {year} {2024})},\ \Eprint
  {https://arxiv.org/abs/2411.12565} {arXiv:2411.12565 [cond-mat.quant-gas]}
  \BibitemShut {NoStop}%
\bibitem [{\citenamefont {Ciavarella}\ \emph {et~al.}(2021)\citenamefont
  {Ciavarella}, \citenamefont {Klco},\ and\ \citenamefont
  {Savage}}]{Ciavarella:2021nmj}%
  \BibitemOpen
  \bibfield  {author} {\bibinfo {author} {\bibfnamefont {A.}~\bibnamefont
  {Ciavarella}}, \bibinfo {author} {\bibfnamefont {N.}~\bibnamefont {Klco}},\
  and\ \bibinfo {author} {\bibfnamefont {M.~J.}\ \bibnamefont {Savage}},\
  }\bibfield  {title} {\bibinfo {title} {{Trailhead for quantum simulation of
  SU(3) Yang-Mills lattice gauge theory in the local multiplet basis}},\ }\href
  {https://doi.org/10.1103/PhysRevD.103.094501} {\bibfield  {journal} {\bibinfo
   {journal} {Phys. Rev. D}\ }\textbf {\bibinfo {volume} {103}},\ \bibinfo
  {pages} {094501} (\bibinfo {year} {2021})},\ \Eprint
  {https://arxiv.org/abs/2101.10227} {arXiv:2101.10227 [quant-ph]} \BibitemShut
  {NoStop}%
\bibitem [{\citenamefont {Ciavarella}(2023)}]{Ciavarella:2023mfc}%
  \BibitemOpen
  \bibfield  {author} {\bibinfo {author} {\bibfnamefont {A.~N.}\ \bibnamefont
  {Ciavarella}},\ }\bibfield  {title} {\bibinfo {title} {{Quantum simulation of
  lattice QCD with improved Hamiltonians}},\ }\href
  {https://doi.org/10.1103/PhysRevD.108.094513} {\bibfield  {journal} {\bibinfo
   {journal} {Phys. Rev. D}\ }\textbf {\bibinfo {volume} {108}},\ \bibinfo
  {pages} {094513} (\bibinfo {year} {2023})},\ \Eprint
  {https://arxiv.org/abs/2307.05593} {arXiv:2307.05593 [hep-lat]} \BibitemShut
  {NoStop}%
\bibitem [{\citenamefont {Ciavarella}\ and\ \citenamefont
  {Chernyshev}(2022)}]{Ciavarella:2021lel}%
  \BibitemOpen
  \bibfield  {author} {\bibinfo {author} {\bibfnamefont {A.~N.}\ \bibnamefont
  {Ciavarella}}\ and\ \bibinfo {author} {\bibfnamefont {I.~A.}\ \bibnamefont
  {Chernyshev}},\ }\bibfield  {title} {\bibinfo {title} {{Preparation of the
  SU(3) lattice Yang-Mills vacuum with variational quantum methods}},\ }\href
  {https://doi.org/10.1103/PhysRevD.105.074504} {\bibfield  {journal} {\bibinfo
   {journal} {Phys. Rev. D}\ }\textbf {\bibinfo {volume} {105}},\ \bibinfo
  {pages} {074504} (\bibinfo {year} {2022})},\ \Eprint
  {https://arxiv.org/abs/2112.09083} {arXiv:2112.09083 [quant-ph]} \BibitemShut
  {NoStop}%
\bibitem [{\citenamefont {Gustafson}\ \emph
  {et~al.}(2024{\natexlab{a}})\citenamefont {Gustafson}, \citenamefont {Lamm},\
  and\ \citenamefont {Lovelace}}]{Gustafson:2023kvd}%
  \BibitemOpen
  \bibfield  {author} {\bibinfo {author} {\bibfnamefont {E.~J.}\ \bibnamefont
  {Gustafson}}, \bibinfo {author} {\bibfnamefont {H.}~\bibnamefont {Lamm}},\
  and\ \bibinfo {author} {\bibfnamefont {F.}~\bibnamefont {Lovelace}},\
  }\bibfield  {title} {\bibinfo {title} {{Primitive quantum gates for an SU(2)
  discrete subgroup: Binary octahedral}},\ }\href
  {https://doi.org/10.1103/PhysRevD.109.054503} {\bibfield  {journal} {\bibinfo
   {journal} {Phys. Rev. D}\ }\textbf {\bibinfo {volume} {109}},\ \bibinfo
  {pages} {054503} (\bibinfo {year} {2024}{\natexlab{a}})},\ \Eprint
  {https://arxiv.org/abs/2312.10285} {arXiv:2312.10285 [hep-lat]} \BibitemShut
  {NoStop}%
\bibitem [{\citenamefont {Gustafson}\ \emph
  {et~al.}(2024{\natexlab{b}})\citenamefont {Gustafson}, \citenamefont {Ji},
  \citenamefont {Lamm}, \citenamefont {Murairi}, \citenamefont {Perez},\ and\
  \citenamefont {Zhu}}]{Gustafson:2024kym}%
  \BibitemOpen
  \bibfield  {author} {\bibinfo {author} {\bibfnamefont {E.~J.}\ \bibnamefont
  {Gustafson}}, \bibinfo {author} {\bibfnamefont {Y.}~\bibnamefont {Ji}},
  \bibinfo {author} {\bibfnamefont {H.}~\bibnamefont {Lamm}}, \bibinfo {author}
  {\bibfnamefont {E.~M.}\ \bibnamefont {Murairi}}, \bibinfo {author}
  {\bibfnamefont {S.~O.}\ \bibnamefont {Perez}},\ and\ \bibinfo {author}
  {\bibfnamefont {S.}~\bibnamefont {Zhu}},\ }\bibfield  {title} {\bibinfo
  {title} {{Primitive quantum gates for an SU(3) discrete subgroup:
  \ensuremath{\Sigma}(36\texttimes{}3)}},\ }\href
  {https://doi.org/10.1103/PhysRevD.110.034515} {\bibfield  {journal} {\bibinfo
   {journal} {Phys. Rev. D}\ }\textbf {\bibinfo {volume} {110}},\ \bibinfo
  {pages} {034515} (\bibinfo {year} {2024}{\natexlab{b}})},\ \Eprint
  {https://arxiv.org/abs/2405.05973} {arXiv:2405.05973 [hep-lat]} \BibitemShut
  {NoStop}%
\bibitem [{\citenamefont {Lamm}\ \emph {et~al.}(2024)\citenamefont {Lamm},
  \citenamefont {Li}, \citenamefont {Shu}, \citenamefont {Wang},\ and\
  \citenamefont {Xu}}]{Lamm:2024jnl}%
  \BibitemOpen
  \bibfield  {author} {\bibinfo {author} {\bibfnamefont {H.}~\bibnamefont
  {Lamm}}, \bibinfo {author} {\bibfnamefont {Y.-Y.}\ \bibnamefont {Li}},
  \bibinfo {author} {\bibfnamefont {J.}~\bibnamefont {Shu}}, \bibinfo {author}
  {\bibfnamefont {Y.-L.}\ \bibnamefont {Wang}},\ and\ \bibinfo {author}
  {\bibfnamefont {B.}~\bibnamefont {Xu}},\ }\bibfield  {title} {\bibinfo
  {title} {{Block encodings of discrete subgroups on a quantum computer}},\
  }\href {https://doi.org/10.1103/PhysRevD.110.054505} {\bibfield  {journal}
  {\bibinfo  {journal} {Phys. Rev. D}\ }\textbf {\bibinfo {volume} {110}},\
  \bibinfo {pages} {054505} (\bibinfo {year} {2024})},\ \Eprint
  {https://arxiv.org/abs/2405.12890} {arXiv:2405.12890 [hep-lat]} \BibitemShut
  {NoStop}%
\bibitem [{\citenamefont {Farrell}\ \emph
  {et~al.}(2023{\natexlab{a}})\citenamefont {Farrell}, \citenamefont
  {Chernyshev}, \citenamefont {Powell}, \citenamefont {Zemlevskiy},
  \citenamefont {Illa},\ and\ \citenamefont {Savage}}]{Farrell:2022wyt}%
  \BibitemOpen
  \bibfield  {author} {\bibinfo {author} {\bibfnamefont {R.~C.}\ \bibnamefont
  {Farrell}}, \bibinfo {author} {\bibfnamefont {I.~A.}\ \bibnamefont
  {Chernyshev}}, \bibinfo {author} {\bibfnamefont {S.~J.~M.}\ \bibnamefont
  {Powell}}, \bibinfo {author} {\bibfnamefont {N.~A.}\ \bibnamefont
  {Zemlevskiy}}, \bibinfo {author} {\bibfnamefont {M.}~\bibnamefont {Illa}},\
  and\ \bibinfo {author} {\bibfnamefont {M.~J.}\ \bibnamefont {Savage}},\
  }\bibfield  {title} {\bibinfo {title} {{Preparations for quantum simulations
  of quantum chromodynamics in 1+1 dimensions. I. Axial gauge}},\ }\href
  {https://doi.org/10.1103/PhysRevD.107.054512} {\bibfield  {journal} {\bibinfo
   {journal} {Phys. Rev. D}\ }\textbf {\bibinfo {volume} {107}},\ \bibinfo
  {pages} {054512} (\bibinfo {year} {2023}{\natexlab{a}})},\ \Eprint
  {https://arxiv.org/abs/2207.01731} {arXiv:2207.01731 [quant-ph]} \BibitemShut
  {NoStop}%
\bibitem [{\citenamefont {Farrell}\ \emph
  {et~al.}(2023{\natexlab{b}})\citenamefont {Farrell}, \citenamefont
  {Chernyshev}, \citenamefont {Powell}, \citenamefont {Zemlevskiy},
  \citenamefont {Illa},\ and\ \citenamefont {Savage}}]{Farrell:2022vyh}%
  \BibitemOpen
  \bibfield  {author} {\bibinfo {author} {\bibfnamefont {R.~C.}\ \bibnamefont
  {Farrell}}, \bibinfo {author} {\bibfnamefont {I.~A.}\ \bibnamefont
  {Chernyshev}}, \bibinfo {author} {\bibfnamefont {S.~J.~M.}\ \bibnamefont
  {Powell}}, \bibinfo {author} {\bibfnamefont {N.~A.}\ \bibnamefont
  {Zemlevskiy}}, \bibinfo {author} {\bibfnamefont {M.}~\bibnamefont {Illa}},\
  and\ \bibinfo {author} {\bibfnamefont {M.~J.}\ \bibnamefont {Savage}},\
  }\bibfield  {title} {\bibinfo {title} {{Preparations for quantum simulations
  of quantum chromodynamics in 1+1 dimensions. II. Single-baryon
  \ensuremath{\beta}-decay in real time}},\ }\href
  {https://doi.org/10.1103/PhysRevD.107.054513} {\bibfield  {journal} {\bibinfo
   {journal} {Phys. Rev. D}\ }\textbf {\bibinfo {volume} {107}},\ \bibinfo
  {pages} {054513} (\bibinfo {year} {2023}{\natexlab{b}})},\ \Eprint
  {https://arxiv.org/abs/2209.10781} {arXiv:2209.10781 [quant-ph]} \BibitemShut
  {NoStop}%
\bibitem [{\citenamefont {Li}\ \emph {et~al.}(2024)\citenamefont {Li},
  \citenamefont {Grabowska},\ and\ \citenamefont {Savage}}]{Li:2024lrl}%
  \BibitemOpen
  \bibfield  {author} {\bibinfo {author} {\bibfnamefont {Z.}~\bibnamefont
  {Li}}, \bibinfo {author} {\bibfnamefont {D.~M.}\ \bibnamefont {Grabowska}},\
  and\ \bibinfo {author} {\bibfnamefont {M.~J.}\ \bibnamefont {Savage}},\
  }\bibfield  {title} {\bibinfo {title} {{Sequency Hierarchy Truncation (SeqHT)
  for Adiabatic State Preparation and Time Evolution in Quantum Simulations}},\
  }\href@noop {} {\  (\bibinfo {year} {2024})},\ \Eprint
  {https://arxiv.org/abs/2407.13835} {arXiv:2407.13835 [quant-ph]} \BibitemShut
  {NoStop}%
\bibitem [{\citenamefont {Zemlevskiy}(2024)}]{Zemlevskiy:2024vxt}%
  \BibitemOpen
  \bibfield  {author} {\bibinfo {author} {\bibfnamefont {N.~A.}\ \bibnamefont
  {Zemlevskiy}},\ }\bibfield  {title} {\bibinfo {title} {{Scalable Quantum
  Simulations of Scattering in Scalar Field Theory on 120 Qubits}},\
  }\href@noop {} {\  (\bibinfo {year} {2024})},\ \Eprint
  {https://arxiv.org/abs/2411.02486} {arXiv:2411.02486 [quant-ph]} \BibitemShut
  {NoStop}%
\bibitem [{\citenamefont {Lewis}\ and\ \citenamefont
  {Woloshyn}(2019)}]{Lewis:2019wfx}%
  \BibitemOpen
  \bibfield  {author} {\bibinfo {author} {\bibfnamefont {R.}~\bibnamefont
  {Lewis}}\ and\ \bibinfo {author} {\bibfnamefont {R.~M.}\ \bibnamefont
  {Woloshyn}},\ }\bibfield  {title} {\bibinfo {title} {{A qubit model for U(1)
  lattice gauge theory}}\ }(\bibinfo {year} {2019})\ \Eprint
  {https://arxiv.org/abs/1905.09789} {arXiv:1905.09789 [hep-lat]} \BibitemShut
  {NoStop}%
\bibitem [{\citenamefont {Atas}\ \emph {et~al.}(2021)\citenamefont {Atas},
  \citenamefont {Zhang}, \citenamefont {Lewis}, \citenamefont {Jahanpour},
  \citenamefont {Haase},\ and\ \citenamefont {Muschik}}]{Atas:2021ext}%
  \BibitemOpen
  \bibfield  {author} {\bibinfo {author} {\bibfnamefont {Y.~Y.}\ \bibnamefont
  {Atas}}, \bibinfo {author} {\bibfnamefont {J.}~\bibnamefont {Zhang}},
  \bibinfo {author} {\bibfnamefont {R.}~\bibnamefont {Lewis}}, \bibinfo
  {author} {\bibfnamefont {A.}~\bibnamefont {Jahanpour}}, \bibinfo {author}
  {\bibfnamefont {J.~F.}\ \bibnamefont {Haase}},\ and\ \bibinfo {author}
  {\bibfnamefont {C.~A.}\ \bibnamefont {Muschik}},\ }\bibfield  {title}
  {\bibinfo {title} {{SU(2) hadrons on a quantum computer via a variational
  approach}},\ }\href {https://doi.org/10.1038/s41467-021-26825-4} {\bibfield
  {journal} {\bibinfo  {journal} {Nature Commun.}\ }\textbf {\bibinfo {volume}
  {12}},\ \bibinfo {pages} {6499} (\bibinfo {year} {2021})},\ \Eprint
  {https://arxiv.org/abs/2102.08920} {arXiv:2102.08920 [quant-ph]} \BibitemShut
  {NoStop}%
\bibitem [{\citenamefont {A~Rahman}\ \emph {et~al.}(2022)\citenamefont
  {A~Rahman}, \citenamefont {Lewis}, \citenamefont {Mendicelli},\ and\
  \citenamefont {Powell}}]{ARahman:2022tkr}%
  \BibitemOpen
  \bibfield  {author} {\bibinfo {author} {\bibfnamefont {S.}~\bibnamefont
  {A~Rahman}}, \bibinfo {author} {\bibfnamefont {R.}~\bibnamefont {Lewis}},
  \bibinfo {author} {\bibfnamefont {E.}~\bibnamefont {Mendicelli}},\ and\
  \bibinfo {author} {\bibfnamefont {S.}~\bibnamefont {Powell}},\ }\bibfield
  {title} {\bibinfo {title} {{Self-mitigating Trotter circuits for SU(2)
  lattice gauge theory on a quantum computer}},\ }\href
  {https://doi.org/10.1103/PhysRevD.106.074502} {\bibfield  {journal} {\bibinfo
   {journal} {Phys. Rev. D}\ }\textbf {\bibinfo {volume} {106}},\ \bibinfo
  {pages} {074502} (\bibinfo {year} {2022})},\ \Eprint
  {https://arxiv.org/abs/2205.09247} {arXiv:2205.09247 [hep-lat]} \BibitemShut
  {NoStop}%
\bibitem [{\citenamefont {Atas}\ \emph {et~al.}(2023)\citenamefont {Atas},
  \citenamefont {Haase}, \citenamefont {Zhang}, \citenamefont {Wei},
  \citenamefont {Pfaendler}, \citenamefont {Lewis},\ and\ \citenamefont
  {Muschik}}]{Atas:2022dqm}%
  \BibitemOpen
  \bibfield  {author} {\bibinfo {author} {\bibfnamefont {Y.~Y.}\ \bibnamefont
  {Atas}}, \bibinfo {author} {\bibfnamefont {J.~F.}\ \bibnamefont {Haase}},
  \bibinfo {author} {\bibfnamefont {J.}~\bibnamefont {Zhang}}, \bibinfo
  {author} {\bibfnamefont {V.}~\bibnamefont {Wei}}, \bibinfo {author}
  {\bibfnamefont {S.~M.~L.}\ \bibnamefont {Pfaendler}}, \bibinfo {author}
  {\bibfnamefont {R.}~\bibnamefont {Lewis}},\ and\ \bibinfo {author}
  {\bibfnamefont {C.~A.}\ \bibnamefont {Muschik}},\ }\bibfield  {title}
  {\bibinfo {title} {{Simulating one-dimensional quantum chromodynamics on a
  quantum computer: Real-time evolutions of tetra- and pentaquarks}},\ }\href
  {https://doi.org/10.1103/PhysRevResearch.5.033184} {\bibfield  {journal}
  {\bibinfo  {journal} {Phys. Rev. Res.}\ }\textbf {\bibinfo {volume} {5}},\
  \bibinfo {pages} {033184} (\bibinfo {year} {2023})},\ \Eprint
  {https://arxiv.org/abs/2207.03473} {arXiv:2207.03473 [quant-ph]} \BibitemShut
  {NoStop}%
\bibitem [{\citenamefont {Mendicelli}\ \emph {et~al.}(2023)\citenamefont
  {Mendicelli}, \citenamefont {Lewis}, \citenamefont {Rahman},\ and\
  \citenamefont {Powell}}]{Mendicelli:2022ntz}%
  \BibitemOpen
  \bibfield  {author} {\bibinfo {author} {\bibfnamefont {E.}~\bibnamefont
  {Mendicelli}}, \bibinfo {author} {\bibfnamefont {R.}~\bibnamefont {Lewis}},
  \bibinfo {author} {\bibfnamefont {S.~A.}\ \bibnamefont {Rahman}},\ and\
  \bibinfo {author} {\bibfnamefont {S.}~\bibnamefont {Powell}},\ }\bibfield
  {title} {\bibinfo {title} {{Real time evolution and a traveling excitation in
  SU(2) pure gauge theory on a quantum computer.}},\ }\href
  {https://doi.org/10.22323/1.430.0025} {\bibfield  {journal} {\bibinfo
  {journal} {PoS}\ }\textbf {\bibinfo {volume} {LATTICE2022}},\ \bibinfo
  {pages} {025} (\bibinfo {year} {2023})},\ \Eprint
  {https://arxiv.org/abs/2210.11606} {arXiv:2210.11606 [hep-lat]} \BibitemShut
  {NoStop}%
\bibitem [{\citenamefont {Kavaki}\ and\ \citenamefont
  {Lewis}(2024)}]{Kavaki:2024ijd}%
  \BibitemOpen
  \bibfield  {author} {\bibinfo {author} {\bibfnamefont {A.~H.~Z.}\
  \bibnamefont {Kavaki}}\ and\ \bibinfo {author} {\bibfnamefont
  {R.}~\bibnamefont {Lewis}},\ }\bibfield  {title} {\bibinfo {title} {{From
  square plaquettes to triamond lattices for SU(2) gauge theory}},\ }\href
  {https://doi.org/10.1038/s42005-024-01697-4} {\bibfield  {journal} {\bibinfo
  {journal} {Commun. Phys.}\ }\textbf {\bibinfo {volume} {7}},\ \bibinfo
  {pages} {208} (\bibinfo {year} {2024})},\ \Eprint
  {https://arxiv.org/abs/2401.14570} {arXiv:2401.14570 [hep-lat]} \BibitemShut
  {NoStop}%
\bibitem [{\citenamefont {Than}\ \emph {et~al.}(2024)\citenamefont {Than},
  \citenamefont {Atas}, \citenamefont {Chakraborty}, \citenamefont {Zhang},
  \citenamefont {Diaz}, \citenamefont {Wen}, \citenamefont {Liu}, \citenamefont
  {Lewis}, \citenamefont {Green}, \citenamefont {Muschik},\ and\ \citenamefont
  {Linke}}]{Than:2024zaj}%
  \BibitemOpen
  \bibfield  {author} {\bibinfo {author} {\bibfnamefont {A.~T.}\ \bibnamefont
  {Than}}, \bibinfo {author} {\bibfnamefont {Y.~Y.}\ \bibnamefont {Atas}},
  \bibinfo {author} {\bibfnamefont {A.}~\bibnamefont {Chakraborty}}, \bibinfo
  {author} {\bibfnamefont {J.}~\bibnamefont {Zhang}}, \bibinfo {author}
  {\bibfnamefont {M.~T.}\ \bibnamefont {Diaz}}, \bibinfo {author}
  {\bibfnamefont {K.}~\bibnamefont {Wen}}, \bibinfo {author} {\bibfnamefont
  {X.}~\bibnamefont {Liu}}, \bibinfo {author} {\bibfnamefont {R.}~\bibnamefont
  {Lewis}}, \bibinfo {author} {\bibfnamefont {A.~M.}\ \bibnamefont {Green}},
  \bibinfo {author} {\bibfnamefont {C.~A.}\ \bibnamefont {Muschik}},\ and\
  \bibinfo {author} {\bibfnamefont {N.~M.}\ \bibnamefont {Linke}},\ }\bibfield
  {title} {\bibinfo {title} {The phase diagram of quantum chromodynamics in one
  dimension on a quantum computer},\ }\href {https://arxiv.org/abs/2501.00579}
  {\  (\bibinfo {year} {2024})},\ \Eprint {https://arxiv.org/abs/2501.00579}
  {arXiv:2501.00579 [quant-ph]} \BibitemShut {NoStop}%
\bibitem [{\citenamefont {Angelides}\ \emph {et~al.}(2025)\citenamefont
  {Angelides}, \citenamefont {Naredi}, \citenamefont {Crippa}, \citenamefont
  {Jansen}, \citenamefont {K{\"u}hn}, \citenamefont {Tavernelli},\ and\
  \citenamefont {Wang}}]{Angelides2025first}%
  \BibitemOpen
  \bibfield  {author} {\bibinfo {author} {\bibfnamefont {T.}~\bibnamefont
  {Angelides}}, \bibinfo {author} {\bibfnamefont {P.}~\bibnamefont {Naredi}},
  \bibinfo {author} {\bibfnamefont {A.}~\bibnamefont {Crippa}}, \bibinfo
  {author} {\bibfnamefont {K.}~\bibnamefont {Jansen}}, \bibinfo {author}
  {\bibfnamefont {S.}~\bibnamefont {K{\"u}hn}}, \bibinfo {author}
  {\bibfnamefont {I.}~\bibnamefont {Tavernelli}},\ and\ \bibinfo {author}
  {\bibfnamefont {D.~S.}\ \bibnamefont {Wang}},\ }\bibfield  {title} {\bibinfo
  {title} {First-order phase transition of the schwinger model with a quantum
  computer},\ }\href {https://doi.org/10.1038/s41534-024-00950-6} {\bibfield
  {journal} {\bibinfo  {journal} {npj Quantum Information}\ }\textbf {\bibinfo
  {volume} {11}},\ \bibinfo {pages} {6} (\bibinfo {year} {2025})}\BibitemShut
  {NoStop}%
\bibitem [{\citenamefont {Gyawali}\ \emph {et~al.}(2024)\citenamefont
  {Gyawali}, \citenamefont {Cochran}, \citenamefont {Lensky}, \citenamefont
  {Rosenberg}, \citenamefont {Karamlou}, \citenamefont {Kechedzhi},
  \citenamefont {Berndtsson}, \citenamefont {Westerhout}, \citenamefont
  {Asfaw}, \citenamefont {Abanin}, \citenamefont {Acharya}, \citenamefont
  {Beni}, \citenamefont {Andersen}, \citenamefont {Ansmann}, \citenamefont
  {Arute}, \citenamefont {Arya}, \citenamefont {Astrakhantsev}, \citenamefont
  {Atalaya}, \citenamefont {Babbush}, \citenamefont {Ballard}, \citenamefont
  {Bardin}, \citenamefont {Bengtsson}, \citenamefont {Bilmes}, \citenamefont
  {Bortoli}, \citenamefont {Bourassa}, \citenamefont {Bovaird}, \citenamefont
  {Brill}, \citenamefont {Broughton}, \citenamefont {Browne}, \citenamefont
  {Buchea}, \citenamefont {Buckley}, \citenamefont {Buell}, \citenamefont
  {Burger}, \citenamefont {Burkett}, \citenamefont {Bushnell}, \citenamefont
  {Cabrera}, \citenamefont {Campero}, \citenamefont {Chang}, \citenamefont
  {Chen}, \citenamefont {Chiaro}, \citenamefont {Claes}, \citenamefont
  {Cleland}, \citenamefont {Cogan}, \citenamefont {Collins}, \citenamefont
  {Conner}, \citenamefont {Courtney}, \citenamefont {Crook}, \citenamefont
  {Das}, \citenamefont {Debroy}, \citenamefont {Lorenzo}, \citenamefont
  {Barba}, \citenamefont {Demura}, \citenamefont {Paolo}, \citenamefont
  {Donohoe}, \citenamefont {Drozdov}, \citenamefont {Dunsworth}, \citenamefont
  {Earle}, \citenamefont {Eickbusch}, \citenamefont {Elbag}, \citenamefont
  {Elzouka}, \citenamefont {Erickson}, \citenamefont {Faoro}, \citenamefont
  {Fatemi}, \citenamefont {Ferreira}, \citenamefont {Burgos}, \citenamefont
  {Forati}, \citenamefont {Fowler}, \citenamefont {Foxen}, \citenamefont
  {Ganjam}, \citenamefont {Gasca}, \citenamefont {Giang}, \citenamefont
  {Gidney}, \citenamefont {Gilboa}, \citenamefont {Gosula}, \citenamefont
  {Dau}, \citenamefont {Graumann}, \citenamefont {Greene}, \citenamefont
  {Gross}, \citenamefont {Habegger}, \citenamefont {Hamilton}, \citenamefont
  {Hansen}, \citenamefont {Harrigan}, \citenamefont {Harrington}, \citenamefont
  {Heslin}, \citenamefont {Heu}, \citenamefont {Hill}, \citenamefont {Hilton},
  \citenamefont {Hoffmann}, \citenamefont {Huang}, \citenamefont {Huff},
  \citenamefont {Huggins}, \citenamefont {Ioffe}, \citenamefont {Isakov},
  \citenamefont {Jeffrey}, \citenamefont {Jiang}, \citenamefont {Jones},
  \citenamefont {Jordan}, \citenamefont {Joshi}, \citenamefont {Juhas},
  \citenamefont {Kafri}, \citenamefont {Kang}, \citenamefont {Khaire},
  \citenamefont {Khattar}, \citenamefont {Khezri}, \citenamefont {Kieferová},
  \citenamefont {Kim}, \citenamefont {Klimov}, \citenamefont {Klots},
  \citenamefont {Kobrin}, \citenamefont {Korotkov}, \citenamefont {Kostritsa},
  \citenamefont {Kreikebaum}, \citenamefont {Kurilovich}, \citenamefont
  {Landhuis}, \citenamefont {Lange-Dei}, \citenamefont {Langley}, \citenamefont
  {Laptev}, \citenamefont {Lau}, \citenamefont {Guevel}, \citenamefont
  {Ledford}, \citenamefont {Lee}, \citenamefont {Lee}, \citenamefont {Lester},
  \citenamefont {Li}, \citenamefont {Lill}, \citenamefont {Liu}, \citenamefont
  {Livingston}, \citenamefont {Locharla}, \citenamefont {Lundahl},
  \citenamefont {Lunt}, \citenamefont {Madhuk}, \citenamefont {Maloney},
  \citenamefont {Mandrà}, \citenamefont {Martin}, \citenamefont {Martin},
  \citenamefont {Martin}, \citenamefont {Maxfield}, \citenamefont {McClean},
  \citenamefont {McEwen}, \citenamefont {Meeks}, \citenamefont {Megrant},
  \citenamefont {Mi}, \citenamefont {Miao}, \citenamefont {Mieszala},
  \citenamefont {Molina}, \citenamefont {Montazeri}, \citenamefont {Morvan},
  \citenamefont {Movassagh}, \citenamefont {Neill}, \citenamefont {Nersisyan},
  \citenamefont {Newman}, \citenamefont {Nguyen}, \citenamefont {Nguyen},
  \citenamefont {Ni}, \citenamefont {Niu}, \citenamefont {Oliver},
  \citenamefont {Ottosson}, \citenamefont {Pizzuto}, \citenamefont {Potter},
  \citenamefont {Pritchard}, \citenamefont {Pryadko}, \citenamefont {Quintana},
  \citenamefont {Reagor}, \citenamefont {Rhodes}, \citenamefont {Roberts},
  \citenamefont {Rocque}, \citenamefont {Rubin}, \citenamefont {Saei},
  \citenamefont {Sankaragomathi}, \citenamefont {Satzinger}, \citenamefont
  {Schurkus}, \citenamefont {Schuster}, \citenamefont {Shearn}, \citenamefont
  {Shorter}, \citenamefont {Shutty}, \citenamefont {Shvarts}, \citenamefont
  {Sivak}, \citenamefont {Skruzny}, \citenamefont {Small}, \citenamefont
  {Smith}, \citenamefont {Springer}, \citenamefont {Sterling}, \citenamefont
  {Suchard}, \citenamefont {Szalay}, \citenamefont {Szasz}, \citenamefont
  {Sztein}, \citenamefont {Thor}, \citenamefont {Torunbalci}, \citenamefont
  {Vaishnav}, \citenamefont {Vdovichev}, \citenamefont {Vidal}, \citenamefont
  {Heidweiller}, \citenamefont {Waltman}, \citenamefont {Wang}, \citenamefont
  {White}, \citenamefont {Wong}, \citenamefont {Woo}, \citenamefont {Xing},
  \citenamefont {Yao}, \citenamefont {Yeh}, \citenamefont {Ying}, \citenamefont
  {Yoo}, \citenamefont {Yosri}, \citenamefont {Young}, \citenamefont {Zalcman},
  \citenamefont {Zhang}, \citenamefont {Zhu}, \citenamefont {Zobrist},
  \citenamefont {Boixo}, \citenamefont {Kelly}, \citenamefont {Lucero},
  \citenamefont {Chen}, \citenamefont {Smelyanskiy}, \citenamefont {Neven},
  \citenamefont {Kovrizhin}, \citenamefont {Knolle}, \citenamefont {Halimeh},
  \citenamefont {Aleiner}, \citenamefont {Moessner},\ and\ \citenamefont
  {Roushan}}]{gyawali2024observationdisorderfreelocalizationefficient}%
  \BibitemOpen
  \bibfield  {author} {\bibinfo {author} {\bibfnamefont {G.}~\bibnamefont
  {Gyawali}}, \bibinfo {author} {\bibfnamefont {T.}~\bibnamefont {Cochran}},
  \bibinfo {author} {\bibfnamefont {Y.}~\bibnamefont {Lensky}}, \bibinfo
  {author} {\bibfnamefont {E.}~\bibnamefont {Rosenberg}}, \bibinfo {author}
  {\bibfnamefont {A.~H.}\ \bibnamefont {Karamlou}}, \bibinfo {author}
  {\bibfnamefont {K.}~\bibnamefont {Kechedzhi}}, \bibinfo {author}
  {\bibfnamefont {J.}~\bibnamefont {Berndtsson}}, \bibinfo {author}
  {\bibfnamefont {T.}~\bibnamefont {Westerhout}}, \bibinfo {author}
  {\bibfnamefont {A.}~\bibnamefont {Asfaw}}, \bibinfo {author} {\bibfnamefont
  {D.}~\bibnamefont {Abanin}}, \bibinfo {author} {\bibfnamefont
  {R.}~\bibnamefont {Acharya}}, \bibinfo {author} {\bibfnamefont {L.~A.}\
  \bibnamefont {Beni}}, \bibinfo {author} {\bibfnamefont {T.~I.}\ \bibnamefont
  {Andersen}}, \bibinfo {author} {\bibfnamefont {M.}~\bibnamefont {Ansmann}},
  \bibinfo {author} {\bibfnamefont {F.}~\bibnamefont {Arute}}, \bibinfo
  {author} {\bibfnamefont {K.}~\bibnamefont {Arya}}, \bibinfo {author}
  {\bibfnamefont {N.}~\bibnamefont {Astrakhantsev}}, \bibinfo {author}
  {\bibfnamefont {J.}~\bibnamefont {Atalaya}}, \bibinfo {author} {\bibfnamefont
  {R.}~\bibnamefont {Babbush}}, \bibinfo {author} {\bibfnamefont
  {B.}~\bibnamefont {Ballard}}, \bibinfo {author} {\bibfnamefont {J.~C.}\
  \bibnamefont {Bardin}}, \bibinfo {author} {\bibfnamefont {A.}~\bibnamefont
  {Bengtsson}}, \bibinfo {author} {\bibfnamefont {A.}~\bibnamefont {Bilmes}},
  \bibinfo {author} {\bibfnamefont {G.}~\bibnamefont {Bortoli}}, \bibinfo
  {author} {\bibfnamefont {A.}~\bibnamefont {Bourassa}}, \bibinfo {author}
  {\bibfnamefont {J.}~\bibnamefont {Bovaird}}, \bibinfo {author} {\bibfnamefont
  {L.}~\bibnamefont {Brill}}, \bibinfo {author} {\bibfnamefont
  {M.}~\bibnamefont {Broughton}}, \bibinfo {author} {\bibfnamefont {D.~A.}\
  \bibnamefont {Browne}}, \bibinfo {author} {\bibfnamefont {B.}~\bibnamefont
  {Buchea}}, \bibinfo {author} {\bibfnamefont {B.~B.}\ \bibnamefont {Buckley}},
  \bibinfo {author} {\bibfnamefont {D.~A.}\ \bibnamefont {Buell}}, \bibinfo
  {author} {\bibfnamefont {T.}~\bibnamefont {Burger}}, \bibinfo {author}
  {\bibfnamefont {B.}~\bibnamefont {Burkett}}, \bibinfo {author} {\bibfnamefont
  {N.}~\bibnamefont {Bushnell}}, \bibinfo {author} {\bibfnamefont
  {A.}~\bibnamefont {Cabrera}}, \bibinfo {author} {\bibfnamefont
  {J.}~\bibnamefont {Campero}}, \bibinfo {author} {\bibfnamefont {H.-S.}\
  \bibnamefont {Chang}}, \bibinfo {author} {\bibfnamefont {Z.}~\bibnamefont
  {Chen}}, \bibinfo {author} {\bibfnamefont {B.}~\bibnamefont {Chiaro}},
  \bibinfo {author} {\bibfnamefont {J.}~\bibnamefont {Claes}}, \bibinfo
  {author} {\bibfnamefont {A.~Y.}\ \bibnamefont {Cleland}}, \bibinfo {author}
  {\bibfnamefont {J.}~\bibnamefont {Cogan}}, \bibinfo {author} {\bibfnamefont
  {R.}~\bibnamefont {Collins}}, \bibinfo {author} {\bibfnamefont
  {P.}~\bibnamefont {Conner}}, \bibinfo {author} {\bibfnamefont
  {W.}~\bibnamefont {Courtney}}, \bibinfo {author} {\bibfnamefont {A.~L.}\
  \bibnamefont {Crook}}, \bibinfo {author} {\bibfnamefont {S.}~\bibnamefont
  {Das}}, \bibinfo {author} {\bibfnamefont {D.~M.}\ \bibnamefont {Debroy}},
  \bibinfo {author} {\bibfnamefont {L.~D.}\ \bibnamefont {Lorenzo}}, \bibinfo
  {author} {\bibfnamefont {A.~D.~T.}\ \bibnamefont {Barba}}, \bibinfo {author}
  {\bibfnamefont {S.}~\bibnamefont {Demura}}, \bibinfo {author} {\bibfnamefont
  {A.~D.}\ \bibnamefont {Paolo}}, \bibinfo {author} {\bibfnamefont
  {P.}~\bibnamefont {Donohoe}}, \bibinfo {author} {\bibfnamefont
  {I.}~\bibnamefont {Drozdov}}, \bibinfo {author} {\bibfnamefont
  {A.}~\bibnamefont {Dunsworth}}, \bibinfo {author} {\bibfnamefont
  {C.}~\bibnamefont {Earle}}, \bibinfo {author} {\bibfnamefont
  {A.}~\bibnamefont {Eickbusch}}, \bibinfo {author} {\bibfnamefont {A.~M.}\
  \bibnamefont {Elbag}}, \bibinfo {author} {\bibfnamefont {M.}~\bibnamefont
  {Elzouka}}, \bibinfo {author} {\bibfnamefont {C.}~\bibnamefont {Erickson}},
  \bibinfo {author} {\bibfnamefont {L.}~\bibnamefont {Faoro}}, \bibinfo
  {author} {\bibfnamefont {R.}~\bibnamefont {Fatemi}}, \bibinfo {author}
  {\bibfnamefont {V.~S.}\ \bibnamefont {Ferreira}}, \bibinfo {author}
  {\bibfnamefont {L.~F.}\ \bibnamefont {Burgos}}, \bibinfo {author}
  {\bibfnamefont {E.}~\bibnamefont {Forati}}, \bibinfo {author} {\bibfnamefont
  {A.~G.}\ \bibnamefont {Fowler}}, \bibinfo {author} {\bibfnamefont
  {B.}~\bibnamefont {Foxen}}, \bibinfo {author} {\bibfnamefont
  {S.}~\bibnamefont {Ganjam}}, \bibinfo {author} {\bibfnamefont
  {R.}~\bibnamefont {Gasca}}, \bibinfo {author} {\bibfnamefont
  {W.}~\bibnamefont {Giang}}, \bibinfo {author} {\bibfnamefont
  {C.}~\bibnamefont {Gidney}}, \bibinfo {author} {\bibfnamefont
  {D.}~\bibnamefont {Gilboa}}, \bibinfo {author} {\bibfnamefont
  {R.}~\bibnamefont {Gosula}}, \bibinfo {author} {\bibfnamefont {A.~G.}\
  \bibnamefont {Dau}}, \bibinfo {author} {\bibfnamefont {D.}~\bibnamefont
  {Graumann}}, \bibinfo {author} {\bibfnamefont {A.}~\bibnamefont {Greene}},
  \bibinfo {author} {\bibfnamefont {J.~A.}\ \bibnamefont {Gross}}, \bibinfo
  {author} {\bibfnamefont {S.}~\bibnamefont {Habegger}}, \bibinfo {author}
  {\bibfnamefont {M.~C.}\ \bibnamefont {Hamilton}}, \bibinfo {author}
  {\bibfnamefont {M.}~\bibnamefont {Hansen}}, \bibinfo {author} {\bibfnamefont
  {M.~P.}\ \bibnamefont {Harrigan}}, \bibinfo {author} {\bibfnamefont {S.~D.}\
  \bibnamefont {Harrington}}, \bibinfo {author} {\bibfnamefont
  {S.}~\bibnamefont {Heslin}}, \bibinfo {author} {\bibfnamefont
  {P.}~\bibnamefont {Heu}}, \bibinfo {author} {\bibfnamefont {G.}~\bibnamefont
  {Hill}}, \bibinfo {author} {\bibfnamefont {J.}~\bibnamefont {Hilton}},
  \bibinfo {author} {\bibfnamefont {M.~R.}\ \bibnamefont {Hoffmann}}, \bibinfo
  {author} {\bibfnamefont {H.-Y.}\ \bibnamefont {Huang}}, \bibinfo {author}
  {\bibfnamefont {A.}~\bibnamefont {Huff}}, \bibinfo {author} {\bibfnamefont
  {W.~J.}\ \bibnamefont {Huggins}}, \bibinfo {author} {\bibfnamefont {L.~B.}\
  \bibnamefont {Ioffe}}, \bibinfo {author} {\bibfnamefont {S.~V.}\ \bibnamefont
  {Isakov}}, \bibinfo {author} {\bibfnamefont {E.}~\bibnamefont {Jeffrey}},
  \bibinfo {author} {\bibfnamefont {Z.}~\bibnamefont {Jiang}}, \bibinfo
  {author} {\bibfnamefont {C.}~\bibnamefont {Jones}}, \bibinfo {author}
  {\bibfnamefont {S.}~\bibnamefont {Jordan}}, \bibinfo {author} {\bibfnamefont
  {C.}~\bibnamefont {Joshi}}, \bibinfo {author} {\bibfnamefont
  {P.}~\bibnamefont {Juhas}}, \bibinfo {author} {\bibfnamefont
  {D.}~\bibnamefont {Kafri}}, \bibinfo {author} {\bibfnamefont
  {H.}~\bibnamefont {Kang}}, \bibinfo {author} {\bibfnamefont {T.}~\bibnamefont
  {Khaire}}, \bibinfo {author} {\bibfnamefont {T.}~\bibnamefont {Khattar}},
  \bibinfo {author} {\bibfnamefont {M.}~\bibnamefont {Khezri}}, \bibinfo
  {author} {\bibfnamefont {M.}~\bibnamefont {Kieferová}}, \bibinfo {author}
  {\bibfnamefont {S.}~\bibnamefont {Kim}}, \bibinfo {author} {\bibfnamefont
  {P.~V.}\ \bibnamefont {Klimov}}, \bibinfo {author} {\bibfnamefont {A.~R.}\
  \bibnamefont {Klots}}, \bibinfo {author} {\bibfnamefont {B.}~\bibnamefont
  {Kobrin}}, \bibinfo {author} {\bibfnamefont {A.~N.}\ \bibnamefont
  {Korotkov}}, \bibinfo {author} {\bibfnamefont {F.}~\bibnamefont {Kostritsa}},
  \bibinfo {author} {\bibfnamefont {J.~M.}\ \bibnamefont {Kreikebaum}},
  \bibinfo {author} {\bibfnamefont {V.~D.}\ \bibnamefont {Kurilovich}},
  \bibinfo {author} {\bibfnamefont {D.}~\bibnamefont {Landhuis}}, \bibinfo
  {author} {\bibfnamefont {T.}~\bibnamefont {Lange-Dei}}, \bibinfo {author}
  {\bibfnamefont {B.~W.}\ \bibnamefont {Langley}}, \bibinfo {author}
  {\bibfnamefont {P.}~\bibnamefont {Laptev}}, \bibinfo {author} {\bibfnamefont
  {K.-M.}\ \bibnamefont {Lau}}, \bibinfo {author} {\bibfnamefont {L.~L.}\
  \bibnamefont {Guevel}}, \bibinfo {author} {\bibfnamefont {J.}~\bibnamefont
  {Ledford}}, \bibinfo {author} {\bibfnamefont {J.}~\bibnamefont {Lee}},
  \bibinfo {author} {\bibfnamefont {K.}~\bibnamefont {Lee}}, \bibinfo {author}
  {\bibfnamefont {B.~J.}\ \bibnamefont {Lester}}, \bibinfo {author}
  {\bibfnamefont {W.~Y.}\ \bibnamefont {Li}}, \bibinfo {author} {\bibfnamefont
  {A.~T.}\ \bibnamefont {Lill}}, \bibinfo {author} {\bibfnamefont
  {W.}~\bibnamefont {Liu}}, \bibinfo {author} {\bibfnamefont {W.~P.}\
  \bibnamefont {Livingston}}, \bibinfo {author} {\bibfnamefont
  {A.}~\bibnamefont {Locharla}}, \bibinfo {author} {\bibfnamefont
  {D.}~\bibnamefont {Lundahl}}, \bibinfo {author} {\bibfnamefont
  {A.}~\bibnamefont {Lunt}}, \bibinfo {author} {\bibfnamefont {S.}~\bibnamefont
  {Madhuk}}, \bibinfo {author} {\bibfnamefont {A.}~\bibnamefont {Maloney}},
  \bibinfo {author} {\bibfnamefont {S.}~\bibnamefont {Mandrà}}, \bibinfo
  {author} {\bibfnamefont {L.~S.}\ \bibnamefont {Martin}}, \bibinfo {author}
  {\bibfnamefont {S.}~\bibnamefont {Martin}}, \bibinfo {author} {\bibfnamefont
  {O.}~\bibnamefont {Martin}}, \bibinfo {author} {\bibfnamefont
  {C.}~\bibnamefont {Maxfield}}, \bibinfo {author} {\bibfnamefont {J.~R.}\
  \bibnamefont {McClean}}, \bibinfo {author} {\bibfnamefont {M.}~\bibnamefont
  {McEwen}}, \bibinfo {author} {\bibfnamefont {S.}~\bibnamefont {Meeks}},
  \bibinfo {author} {\bibfnamefont {A.}~\bibnamefont {Megrant}}, \bibinfo
  {author} {\bibfnamefont {X.}~\bibnamefont {Mi}}, \bibinfo {author}
  {\bibfnamefont {K.~C.}\ \bibnamefont {Miao}}, \bibinfo {author}
  {\bibfnamefont {A.}~\bibnamefont {Mieszala}}, \bibinfo {author}
  {\bibfnamefont {S.}~\bibnamefont {Molina}}, \bibinfo {author} {\bibfnamefont
  {S.}~\bibnamefont {Montazeri}}, \bibinfo {author} {\bibfnamefont
  {A.}~\bibnamefont {Morvan}}, \bibinfo {author} {\bibfnamefont
  {R.}~\bibnamefont {Movassagh}}, \bibinfo {author} {\bibfnamefont
  {C.}~\bibnamefont {Neill}}, \bibinfo {author} {\bibfnamefont
  {A.}~\bibnamefont {Nersisyan}}, \bibinfo {author} {\bibfnamefont
  {M.}~\bibnamefont {Newman}}, \bibinfo {author} {\bibfnamefont
  {A.}~\bibnamefont {Nguyen}}, \bibinfo {author} {\bibfnamefont
  {M.}~\bibnamefont {Nguyen}}, \bibinfo {author} {\bibfnamefont {C.-H.}\
  \bibnamefont {Ni}}, \bibinfo {author} {\bibfnamefont {M.~Y.}\ \bibnamefont
  {Niu}}, \bibinfo {author} {\bibfnamefont {W.~D.}\ \bibnamefont {Oliver}},
  \bibinfo {author} {\bibfnamefont {K.}~\bibnamefont {Ottosson}}, \bibinfo
  {author} {\bibfnamefont {A.}~\bibnamefont {Pizzuto}}, \bibinfo {author}
  {\bibfnamefont {R.}~\bibnamefont {Potter}}, \bibinfo {author} {\bibfnamefont
  {O.}~\bibnamefont {Pritchard}}, \bibinfo {author} {\bibfnamefont {L.~P.}\
  \bibnamefont {Pryadko}}, \bibinfo {author} {\bibfnamefont {C.}~\bibnamefont
  {Quintana}}, \bibinfo {author} {\bibfnamefont {M.~J.}\ \bibnamefont
  {Reagor}}, \bibinfo {author} {\bibfnamefont {D.~M.}\ \bibnamefont {Rhodes}},
  \bibinfo {author} {\bibfnamefont {G.}~\bibnamefont {Roberts}}, \bibinfo
  {author} {\bibfnamefont {C.}~\bibnamefont {Rocque}}, \bibinfo {author}
  {\bibfnamefont {N.~C.}\ \bibnamefont {Rubin}}, \bibinfo {author}
  {\bibfnamefont {N.}~\bibnamefont {Saei}}, \bibinfo {author} {\bibfnamefont
  {K.}~\bibnamefont {Sankaragomathi}}, \bibinfo {author} {\bibfnamefont
  {K.~J.}\ \bibnamefont {Satzinger}}, \bibinfo {author} {\bibfnamefont {H.~F.}\
  \bibnamefont {Schurkus}}, \bibinfo {author} {\bibfnamefont {C.}~\bibnamefont
  {Schuster}}, \bibinfo {author} {\bibfnamefont {M.~J.}\ \bibnamefont
  {Shearn}}, \bibinfo {author} {\bibfnamefont {A.}~\bibnamefont {Shorter}},
  \bibinfo {author} {\bibfnamefont {N.}~\bibnamefont {Shutty}}, \bibinfo
  {author} {\bibfnamefont {V.}~\bibnamefont {Shvarts}}, \bibinfo {author}
  {\bibfnamefont {V.}~\bibnamefont {Sivak}}, \bibinfo {author} {\bibfnamefont
  {J.}~\bibnamefont {Skruzny}}, \bibinfo {author} {\bibfnamefont
  {S.}~\bibnamefont {Small}}, \bibinfo {author} {\bibfnamefont {W.~C.}\
  \bibnamefont {Smith}}, \bibinfo {author} {\bibfnamefont {S.}~\bibnamefont
  {Springer}}, \bibinfo {author} {\bibfnamefont {G.}~\bibnamefont {Sterling}},
  \bibinfo {author} {\bibfnamefont {J.}~\bibnamefont {Suchard}}, \bibinfo
  {author} {\bibfnamefont {M.}~\bibnamefont {Szalay}}, \bibinfo {author}
  {\bibfnamefont {A.}~\bibnamefont {Szasz}}, \bibinfo {author} {\bibfnamefont
  {A.}~\bibnamefont {Sztein}}, \bibinfo {author} {\bibfnamefont
  {D.}~\bibnamefont {Thor}}, \bibinfo {author} {\bibfnamefont {M.~M.}\
  \bibnamefont {Torunbalci}}, \bibinfo {author} {\bibfnamefont
  {A.}~\bibnamefont {Vaishnav}}, \bibinfo {author} {\bibfnamefont
  {S.}~\bibnamefont {Vdovichev}}, \bibinfo {author} {\bibfnamefont
  {G.}~\bibnamefont {Vidal}}, \bibinfo {author} {\bibfnamefont {C.~V.}\
  \bibnamefont {Heidweiller}}, \bibinfo {author} {\bibfnamefont
  {S.}~\bibnamefont {Waltman}}, \bibinfo {author} {\bibfnamefont {S.~X.}\
  \bibnamefont {Wang}}, \bibinfo {author} {\bibfnamefont {T.}~\bibnamefont
  {White}}, \bibinfo {author} {\bibfnamefont {K.}~\bibnamefont {Wong}},
  \bibinfo {author} {\bibfnamefont {B.~W.~K.}\ \bibnamefont {Woo}}, \bibinfo
  {author} {\bibfnamefont {C.}~\bibnamefont {Xing}}, \bibinfo {author}
  {\bibfnamefont {Z.~J.}\ \bibnamefont {Yao}}, \bibinfo {author} {\bibfnamefont
  {P.}~\bibnamefont {Yeh}}, \bibinfo {author} {\bibfnamefont {B.}~\bibnamefont
  {Ying}}, \bibinfo {author} {\bibfnamefont {J.}~\bibnamefont {Yoo}}, \bibinfo
  {author} {\bibfnamefont {N.}~\bibnamefont {Yosri}}, \bibinfo {author}
  {\bibfnamefont {G.}~\bibnamefont {Young}}, \bibinfo {author} {\bibfnamefont
  {A.}~\bibnamefont {Zalcman}}, \bibinfo {author} {\bibfnamefont
  {Y.}~\bibnamefont {Zhang}}, \bibinfo {author} {\bibfnamefont
  {N.}~\bibnamefont {Zhu}}, \bibinfo {author} {\bibfnamefont {N.}~\bibnamefont
  {Zobrist}}, \bibinfo {author} {\bibfnamefont {S.}~\bibnamefont {Boixo}},
  \bibinfo {author} {\bibfnamefont {J.}~\bibnamefont {Kelly}}, \bibinfo
  {author} {\bibfnamefont {E.}~\bibnamefont {Lucero}}, \bibinfo {author}
  {\bibfnamefont {Y.}~\bibnamefont {Chen}}, \bibinfo {author} {\bibfnamefont
  {V.}~\bibnamefont {Smelyanskiy}}, \bibinfo {author} {\bibfnamefont
  {H.}~\bibnamefont {Neven}}, \bibinfo {author} {\bibfnamefont
  {D.}~\bibnamefont {Kovrizhin}}, \bibinfo {author} {\bibfnamefont
  {J.}~\bibnamefont {Knolle}}, \bibinfo {author} {\bibfnamefont {J.~C.}\
  \bibnamefont {Halimeh}}, \bibinfo {author} {\bibfnamefont {I.}~\bibnamefont
  {Aleiner}}, \bibinfo {author} {\bibfnamefont {R.}~\bibnamefont {Moessner}},\
  and\ \bibinfo {author} {\bibfnamefont {P.}~\bibnamefont {Roushan}},\
  }\bibfield  {title} {\bibinfo {title} {Observation of disorder-free
  localization and efficient disorder averaging on a quantum processor},\
  }\href {https://arxiv.org/abs/2410.06557} {\  (\bibinfo {year} {2024})},\
  \Eprint {https://arxiv.org/abs/2410.06557} {arXiv:2410.06557 [quant-ph]}
  \BibitemShut {NoStop}%
\bibitem [{\citenamefont {Schuhmacher}\ \emph {et~al.}(2025)\citenamefont
  {Schuhmacher}, \citenamefont {Su}, \citenamefont {Osborne}, \citenamefont
  {Gandon}, \citenamefont {Halimeh},\ and\ \citenamefont
  {Tavernelli}}]{schuhmacher2025observationhadronscatteringlattice}%
  \BibitemOpen
  \bibfield  {author} {\bibinfo {author} {\bibfnamefont {J.}~\bibnamefont
  {Schuhmacher}}, \bibinfo {author} {\bibfnamefont {G.-X.}\ \bibnamefont {Su}},
  \bibinfo {author} {\bibfnamefont {J.~J.}\ \bibnamefont {Osborne}}, \bibinfo
  {author} {\bibfnamefont {A.}~\bibnamefont {Gandon}}, \bibinfo {author}
  {\bibfnamefont {J.~C.}\ \bibnamefont {Halimeh}},\ and\ \bibinfo {author}
  {\bibfnamefont {I.}~\bibnamefont {Tavernelli}},\ }\bibfield  {title}
  {\bibinfo {title} {Observation of hadron scattering in a lattice gauge theory
  on a quantum computer},\ }\href {https://arxiv.org/abs/2505.20387} {\
  (\bibinfo {year} {2025})},\ \Eprint {https://arxiv.org/abs/2505.20387}
  {arXiv:2505.20387 [quant-ph]} \BibitemShut {NoStop}%
\bibitem [{\citenamefont {Davoudi}\ \emph {et~al.}(2025)\citenamefont
  {Davoudi}, \citenamefont {Hsieh},\ and\ \citenamefont
  {Kadam}}]{davoudi2025quantumcomputationhadronscattering}%
  \BibitemOpen
  \bibfield  {author} {\bibinfo {author} {\bibfnamefont {Z.}~\bibnamefont
  {Davoudi}}, \bibinfo {author} {\bibfnamefont {C.-C.}\ \bibnamefont {Hsieh}},\
  and\ \bibinfo {author} {\bibfnamefont {S.~V.}\ \bibnamefont {Kadam}},\
  }\bibfield  {title} {\bibinfo {title} {Quantum computation of hadron
  scattering in a lattice gauge theory},\ }\href
  {https://arxiv.org/abs/2505.20408} {\  (\bibinfo {year} {2025})},\ \Eprint
  {https://arxiv.org/abs/2505.20408} {arXiv:2505.20408 [quant-ph]} \BibitemShut
  {NoStop}%
\bibitem [{\citenamefont {Pichler}\ \emph {et~al.}(2016)\citenamefont
  {Pichler}, \citenamefont {Dalmonte}, \citenamefont {Rico}, \citenamefont
  {Zoller},\ and\ \citenamefont {Montangero}}]{Pichler2016}%
  \BibitemOpen
  \bibfield  {author} {\bibinfo {author} {\bibfnamefont {T.}~\bibnamefont
  {Pichler}}, \bibinfo {author} {\bibfnamefont {M.}~\bibnamefont {Dalmonte}},
  \bibinfo {author} {\bibfnamefont {E.}~\bibnamefont {Rico}}, \bibinfo {author}
  {\bibfnamefont {P.}~\bibnamefont {Zoller}},\ and\ \bibinfo {author}
  {\bibfnamefont {S.}~\bibnamefont {Montangero}},\ }\bibfield  {title}
  {\bibinfo {title} {Real-time dynamics in u(1) lattice gauge theories with
  tensor networks},\ }\href {https://doi.org/10.1103/PhysRevX.6.011023}
  {\bibfield  {journal} {\bibinfo  {journal} {Phys. Rev. X}\ }\textbf {\bibinfo
  {volume} {6}},\ \bibinfo {pages} {011023} (\bibinfo {year}
  {2016})}\BibitemShut {NoStop}%
\bibitem [{\citenamefont {Chanda}\ \emph {et~al.}(2020)\citenamefont {Chanda},
  \citenamefont {Zakrzewski}, \citenamefont {Lewenstein},\ and\ \citenamefont
  {Tagliacozzo}}]{Chanda2020confinement}%
  \BibitemOpen
  \bibfield  {author} {\bibinfo {author} {\bibfnamefont {T.}~\bibnamefont
  {Chanda}}, \bibinfo {author} {\bibfnamefont {J.}~\bibnamefont {Zakrzewski}},
  \bibinfo {author} {\bibfnamefont {M.}~\bibnamefont {Lewenstein}},\ and\
  \bibinfo {author} {\bibfnamefont {L.}~\bibnamefont {Tagliacozzo}},\
  }\bibfield  {title} {\bibinfo {title} {Confinement and lack of thermalization
  after quenches in the bosonic schwinger model},\ }\href
  {https://doi.org/10.1103/PhysRevLett.124.180602} {\bibfield  {journal}
  {\bibinfo  {journal} {Phys. Rev. Lett.}\ }\textbf {\bibinfo {volume} {124}},\
  \bibinfo {pages} {180602} (\bibinfo {year} {2020})}\BibitemShut {NoStop}%
\bibitem [{\citenamefont {Notarnicola}\ \emph {et~al.}(2020)\citenamefont
  {Notarnicola}, \citenamefont {Collura},\ and\ \citenamefont
  {Montangero}}]{Notarnicola2020real}%
  \BibitemOpen
  \bibfield  {author} {\bibinfo {author} {\bibfnamefont {S.}~\bibnamefont
  {Notarnicola}}, \bibinfo {author} {\bibfnamefont {M.}~\bibnamefont
  {Collura}},\ and\ \bibinfo {author} {\bibfnamefont {S.}~\bibnamefont
  {Montangero}},\ }\bibfield  {title} {\bibinfo {title} {Real-time-dynamics
  quantum simulation of $(1+1)\text{-dimensional}$ lattice qed with rydberg
  atoms},\ }\href {https://doi.org/10.1103/PhysRevResearch.2.013288} {\bibfield
   {journal} {\bibinfo  {journal} {Phys. Rev. Res.}\ }\textbf {\bibinfo
  {volume} {2}},\ \bibinfo {pages} {013288} (\bibinfo {year}
  {2020})}\BibitemShut {NoStop}%
\bibitem [{\citenamefont {Rigobello}\ \emph {et~al.}(2021)\citenamefont
  {Rigobello}, \citenamefont {Notarnicola}, \citenamefont {Magnifico},\ and\
  \citenamefont {Montangero}}]{Rigobello2021entanglement}%
  \BibitemOpen
  \bibfield  {author} {\bibinfo {author} {\bibfnamefont {M.}~\bibnamefont
  {Rigobello}}, \bibinfo {author} {\bibfnamefont {S.}~\bibnamefont
  {Notarnicola}}, \bibinfo {author} {\bibfnamefont {G.}~\bibnamefont
  {Magnifico}},\ and\ \bibinfo {author} {\bibfnamefont {S.}~\bibnamefont
  {Montangero}},\ }\bibfield  {title} {\bibinfo {title} {Entanglement
  generation in $(1+1)\mathrm{D}$ qed scattering processes},\ }\href
  {https://doi.org/10.1103/PhysRevD.104.114501} {\bibfield  {journal} {\bibinfo
   {journal} {Phys. Rev. D}\ }\textbf {\bibinfo {volume} {104}},\ \bibinfo
  {pages} {114501} (\bibinfo {year} {2021})}\BibitemShut {NoStop}%
\bibitem [{\citenamefont {Van~Damme}\ \emph {et~al.}(2022)\citenamefont
  {Van~Damme}, \citenamefont {Zache}, \citenamefont {Banerjee}, \citenamefont
  {Hauke},\ and\ \citenamefont {Halimeh}}]{vandamme2022dqpt}%
  \BibitemOpen
  \bibfield  {author} {\bibinfo {author} {\bibfnamefont {M.}~\bibnamefont
  {Van~Damme}}, \bibinfo {author} {\bibfnamefont {T.~V.}\ \bibnamefont
  {Zache}}, \bibinfo {author} {\bibfnamefont {D.}~\bibnamefont {Banerjee}},
  \bibinfo {author} {\bibfnamefont {P.}~\bibnamefont {Hauke}},\ and\ \bibinfo
  {author} {\bibfnamefont {J.~C.}\ \bibnamefont {Halimeh}},\ }\bibfield
  {title} {\bibinfo {title} {Dynamical quantum phase transitions in spin-$s
  u(1)$ quantum link models},\ }\href
  {https://doi.org/10.1103/PhysRevB.106.245110} {\bibfield  {journal} {\bibinfo
   {journal} {Phys. Rev. B}\ }\textbf {\bibinfo {volume} {106}},\ \bibinfo
  {pages} {245110} (\bibinfo {year} {2022})}\BibitemShut {NoStop}%
\bibitem [{\citenamefont {Su}\ \emph {et~al.}(2024)\citenamefont {Su},
  \citenamefont {Osborne},\ and\ \citenamefont
  {Halimeh}}]{su2024particlecollider}%
  \BibitemOpen
  \bibfield  {author} {\bibinfo {author} {\bibfnamefont {G.-X.}\ \bibnamefont
  {Su}}, \bibinfo {author} {\bibfnamefont {J.~J.}\ \bibnamefont {Osborne}},\
  and\ \bibinfo {author} {\bibfnamefont {J.~C.}\ \bibnamefont {Halimeh}},\
  }\bibfield  {title} {\bibinfo {title} {Cold-atom particle collider},\ }\href
  {https://doi.org/10.1103/PRXQuantum.5.040310} {\bibfield  {journal} {\bibinfo
   {journal} {PRX Quantum}\ }\textbf {\bibinfo {volume} {5}},\ \bibinfo {pages}
  {040310} (\bibinfo {year} {2024})}\BibitemShut {NoStop}%
\bibitem [{\citenamefont {Belyansky}\ \emph {et~al.}(2024)\citenamefont
  {Belyansky}, \citenamefont {Whitsitt}, \citenamefont {Mueller}, \citenamefont
  {Fahimniya}, \citenamefont {Bennewitz}, \citenamefont {Davoudi},\ and\
  \citenamefont {Gorshkov}}]{Belyansky2024high}%
  \BibitemOpen
  \bibfield  {author} {\bibinfo {author} {\bibfnamefont {R.}~\bibnamefont
  {Belyansky}}, \bibinfo {author} {\bibfnamefont {S.}~\bibnamefont {Whitsitt}},
  \bibinfo {author} {\bibfnamefont {N.}~\bibnamefont {Mueller}}, \bibinfo
  {author} {\bibfnamefont {A.}~\bibnamefont {Fahimniya}}, \bibinfo {author}
  {\bibfnamefont {E.~R.}\ \bibnamefont {Bennewitz}}, \bibinfo {author}
  {\bibfnamefont {Z.}~\bibnamefont {Davoudi}},\ and\ \bibinfo {author}
  {\bibfnamefont {A.~V.}\ \bibnamefont {Gorshkov}},\ }\bibfield  {title}
  {\bibinfo {title} {High-energy collision of quarks and mesons in the
  schwinger model: From tensor networks to circuit qed},\ }\href
  {https://doi.org/10.1103/PhysRevLett.132.091903} {\bibfield  {journal}
  {\bibinfo  {journal} {Phys. Rev. Lett.}\ }\textbf {\bibinfo {volume} {132}},\
  \bibinfo {pages} {091903} (\bibinfo {year} {2024})}\BibitemShut {NoStop}%
\bibitem [{\citenamefont {Calaj\'o}\ \emph {et~al.}(2024)\citenamefont
  {Calaj\'o}, \citenamefont {Magnifico}, \citenamefont {Edmunds}, \citenamefont
  {Ringbauer}, \citenamefont {Montangero},\ and\ \citenamefont
  {Silvi}}]{Calajo2024digital}%
  \BibitemOpen
  \bibfield  {author} {\bibinfo {author} {\bibfnamefont {G.}~\bibnamefont
  {Calaj\'o}}, \bibinfo {author} {\bibfnamefont {G.}~\bibnamefont {Magnifico}},
  \bibinfo {author} {\bibfnamefont {C.}~\bibnamefont {Edmunds}}, \bibinfo
  {author} {\bibfnamefont {M.}~\bibnamefont {Ringbauer}}, \bibinfo {author}
  {\bibfnamefont {S.}~\bibnamefont {Montangero}},\ and\ \bibinfo {author}
  {\bibfnamefont {P.}~\bibnamefont {Silvi}},\ }\bibfield  {title} {\bibinfo
  {title} {Digital quantum simulation of a (1+1)d su(2) lattice gauge theory
  with ion qudits},\ }\href {https://doi.org/10.1103/PRXQuantum.5.040309}
  {\bibfield  {journal} {\bibinfo  {journal} {PRX Quantum}\ }\textbf {\bibinfo
  {volume} {5}},\ \bibinfo {pages} {040309} (\bibinfo {year}
  {2024})}\BibitemShut {NoStop}%
\bibitem [{\citenamefont {Calaj\'o}\ \emph {et~al.}(2025)\citenamefont
  {Calaj\'o}, \citenamefont {Cataldi}, \citenamefont {Rigobello}, \citenamefont
  {Wanisch}, \citenamefont {Magnifico}, \citenamefont {Silvi}, \citenamefont
  {Montangero},\ and\ \citenamefont {Halimeh}}]{Calajo2025QMBS}%
  \BibitemOpen
  \bibfield  {author} {\bibinfo {author} {\bibfnamefont {G.}~\bibnamefont
  {Calaj\'o}}, \bibinfo {author} {\bibfnamefont {G.}~\bibnamefont {Cataldi}},
  \bibinfo {author} {\bibfnamefont {M.}~\bibnamefont {Rigobello}}, \bibinfo
  {author} {\bibfnamefont {D.}~\bibnamefont {Wanisch}}, \bibinfo {author}
  {\bibfnamefont {G.}~\bibnamefont {Magnifico}}, \bibinfo {author}
  {\bibfnamefont {P.}~\bibnamefont {Silvi}}, \bibinfo {author} {\bibfnamefont
  {S.}~\bibnamefont {Montangero}},\ and\ \bibinfo {author} {\bibfnamefont
  {J.~C.}\ \bibnamefont {Halimeh}},\ }\bibfield  {title} {\bibinfo {title}
  {Quantum many-body scarring in a non-abelian lattice gauge theory},\ }\href
  {https://doi.org/10.1103/PhysRevResearch.7.013322} {\bibfield  {journal}
  {\bibinfo  {journal} {Phys. Rev. Res.}\ }\textbf {\bibinfo {volume} {7}},\
  \bibinfo {pages} {013322} (\bibinfo {year} {2025})}\BibitemShut {NoStop}%
\bibitem [{\citenamefont {Cataldi}\ \emph {et~al.}(2025)\citenamefont
  {Cataldi}, \citenamefont {Calajó}, \citenamefont {Silvi}, \citenamefont
  {Montangero},\ and\ \citenamefont
  {Halimeh}}]{cataldi2025disorderfreelocalizationfragmentationnonabelian}%
  \BibitemOpen
  \bibfield  {author} {\bibinfo {author} {\bibfnamefont {G.}~\bibnamefont
  {Cataldi}}, \bibinfo {author} {\bibfnamefont {G.}~\bibnamefont {Calajó}},
  \bibinfo {author} {\bibfnamefont {P.}~\bibnamefont {Silvi}}, \bibinfo
  {author} {\bibfnamefont {S.}~\bibnamefont {Montangero}},\ and\ \bibinfo
  {author} {\bibfnamefont {J.~C.}\ \bibnamefont {Halimeh}},\ }\bibfield
  {title} {\bibinfo {title} {Disorder-free localization and fragmentation in a
  non-abelian lattice gauge theory},\ }\href {https://arxiv.org/abs/2505.04704}
  {\  (\bibinfo {year} {2025})},\ \Eprint {https://arxiv.org/abs/2505.04704}
  {arXiv:2505.04704 [cond-mat.quant-gas]} \BibitemShut {NoStop}%
\bibitem [{\citenamefont {Osborne}\ \emph {et~al.}(2024)\citenamefont
  {Osborne}, \citenamefont {McCulloch},\ and\ \citenamefont
  {Halimeh}}]{osborne2024quantummanybodyscarring21d}%
  \BibitemOpen
  \bibfield  {author} {\bibinfo {author} {\bibfnamefont {J.}~\bibnamefont
  {Osborne}}, \bibinfo {author} {\bibfnamefont {I.~P.}\ \bibnamefont
  {McCulloch}},\ and\ \bibinfo {author} {\bibfnamefont {J.~C.}\ \bibnamefont
  {Halimeh}},\ }\bibfield  {title} {\bibinfo {title} {Quantum many-body
  scarring in $2+1$d gauge theories with dynamical matter},\ }\href
  {https://arxiv.org/abs/2403.08858} {\  (\bibinfo {year} {2024})},\ \Eprint
  {https://arxiv.org/abs/2403.08858} {arXiv:2403.08858 [cond-mat.quant-gas]}
  \BibitemShut {NoStop}%
\bibitem [{\citenamefont {Budde}\ \emph {et~al.}(2024)\citenamefont {Budde},
  \citenamefont {Krstic~Marinkovic},\ and\ \citenamefont
  {Pinto~Barros}}]{Budde2024qmbs}%
  \BibitemOpen
  \bibfield  {author} {\bibinfo {author} {\bibfnamefont {T.}~\bibnamefont
  {Budde}}, \bibinfo {author} {\bibfnamefont {M.}~\bibnamefont
  {Krstic~Marinkovic}},\ and\ \bibinfo {author} {\bibfnamefont {J.~C.}\
  \bibnamefont {Pinto~Barros}},\ }\bibfield  {title} {\bibinfo {title} {Quantum
  many-body scars for arbitrary integer spin in $2+1\mathrm{D}$ abelian gauge
  theories},\ }\href {https://doi.org/10.1103/PhysRevD.110.094506} {\bibfield
  {journal} {\bibinfo  {journal} {Phys. Rev. D}\ }\textbf {\bibinfo {volume}
  {110}},\ \bibinfo {pages} {094506} (\bibinfo {year} {2024})}\BibitemShut
  {NoStop}%
\bibitem [{\citenamefont {Osborne}\ \emph {et~al.}(2023)\citenamefont
  {Osborne}, \citenamefont {McCulloch},\ and\ \citenamefont
  {Halimeh}}]{osborne2023disorderfreelocalization21dlattice}%
  \BibitemOpen
  \bibfield  {author} {\bibinfo {author} {\bibfnamefont {J.}~\bibnamefont
  {Osborne}}, \bibinfo {author} {\bibfnamefont {I.~P.}\ \bibnamefont
  {McCulloch}},\ and\ \bibinfo {author} {\bibfnamefont {J.~C.}\ \bibnamefont
  {Halimeh}},\ }\bibfield  {title} {\bibinfo {title} {Disorder-free
  localization in $2+1$d lattice gauge theories with dynamical matter},\ }\href
  {https://arxiv.org/abs/2301.07720} {\  (\bibinfo {year} {2023})},\ \Eprint
  {https://arxiv.org/abs/2301.07720} {arXiv:2301.07720 [cond-mat.quant-gas]}
  \BibitemShut {NoStop}%
\bibitem [{\citenamefont {Carmen~Bañuls}\ and\ \citenamefont
  {Cichy}(2020)}]{Banuls_review}%
  \BibitemOpen
  \bibfield  {author} {\bibinfo {author} {\bibfnamefont {M.}~\bibnamefont
  {Carmen~Bañuls}}\ and\ \bibinfo {author} {\bibfnamefont {K.}~\bibnamefont
  {Cichy}},\ }\bibfield  {title} {\bibinfo {title} {Review on novel methods for
  lattice gauge theories},\ }\href {https://doi.org/10.1088/1361-6633/ab6311}
  {\bibfield  {journal} {\bibinfo  {journal} {Reports on Progress in Physics}\
  }\textbf {\bibinfo {volume} {83}},\ \bibinfo {pages} {024401} (\bibinfo
  {year} {2020})}\BibitemShut {NoStop}%
\bibitem [{\citenamefont {Magnifico}\ \emph {et~al.}(2024)\citenamefont
  {Magnifico}, \citenamefont {Cataldi}, \citenamefont {Rigobello},
  \citenamefont {Majcen}, \citenamefont {Jaschke}, \citenamefont {Silvi},\ and\
  \citenamefont {Montangero}}]{magnifico2024tensornetworkslatticegauge}%
  \BibitemOpen
  \bibfield  {author} {\bibinfo {author} {\bibfnamefont {G.}~\bibnamefont
  {Magnifico}}, \bibinfo {author} {\bibfnamefont {G.}~\bibnamefont {Cataldi}},
  \bibinfo {author} {\bibfnamefont {M.}~\bibnamefont {Rigobello}}, \bibinfo
  {author} {\bibfnamefont {P.}~\bibnamefont {Majcen}}, \bibinfo {author}
  {\bibfnamefont {D.}~\bibnamefont {Jaschke}}, \bibinfo {author} {\bibfnamefont
  {P.}~\bibnamefont {Silvi}},\ and\ \bibinfo {author} {\bibfnamefont
  {S.}~\bibnamefont {Montangero}},\ }\bibfield  {title} {\bibinfo {title}
  {Tensor networks for lattice gauge theories beyond one dimension: a
  roadmap},\ }\href {https://arxiv.org/abs/2407.03058} {\  (\bibinfo {year}
  {2024})},\ \Eprint {https://arxiv.org/abs/2407.03058} {arXiv:2407.03058
  [hep-lat]} \BibitemShut {NoStop}%
\bibitem [{\citenamefont {De}\ \emph {et~al.}(2024)\citenamefont {De},
  \citenamefont {Lerose}, \citenamefont {Luo}, \citenamefont {Surace},
  \citenamefont {Schuckert}, \citenamefont {Bennewitz}, \citenamefont {Ware},
  \citenamefont {Morong}, \citenamefont {Collins}, \citenamefont {Davoudi},
  \citenamefont {Gorshkov}, \citenamefont {Katz},\ and\ \citenamefont
  {Monroe}}]{de2024observationstringbreakingdynamicsquantum}%
  \BibitemOpen
  \bibfield  {author} {\bibinfo {author} {\bibfnamefont {A.}~\bibnamefont
  {De}}, \bibinfo {author} {\bibfnamefont {A.}~\bibnamefont {Lerose}}, \bibinfo
  {author} {\bibfnamefont {D.}~\bibnamefont {Luo}}, \bibinfo {author}
  {\bibfnamefont {F.~M.}\ \bibnamefont {Surace}}, \bibinfo {author}
  {\bibfnamefont {A.}~\bibnamefont {Schuckert}}, \bibinfo {author}
  {\bibfnamefont {E.~R.}\ \bibnamefont {Bennewitz}}, \bibinfo {author}
  {\bibfnamefont {B.}~\bibnamefont {Ware}}, \bibinfo {author} {\bibfnamefont
  {W.}~\bibnamefont {Morong}}, \bibinfo {author} {\bibfnamefont {K.~S.}\
  \bibnamefont {Collins}}, \bibinfo {author} {\bibfnamefont {Z.}~\bibnamefont
  {Davoudi}}, \bibinfo {author} {\bibfnamefont {A.~V.}\ \bibnamefont
  {Gorshkov}}, \bibinfo {author} {\bibfnamefont {O.}~\bibnamefont {Katz}},\
  and\ \bibinfo {author} {\bibfnamefont {C.}~\bibnamefont {Monroe}},\
  }\bibfield  {title} {\bibinfo {title} {Observation of string-breaking
  dynamics in a quantum simulator},\ }\href {https://arxiv.org/abs/2410.13815}
  {\  (\bibinfo {year} {2024})},\ \Eprint {https://arxiv.org/abs/2410.13815}
  {arXiv:2410.13815 [quant-ph]} \BibitemShut {NoStop}%
\bibitem [{\citenamefont {Liu}\ \emph {et~al.}(2024)\citenamefont {Liu},
  \citenamefont {Zhang}, \citenamefont {Zhu}, \citenamefont {He}, \citenamefont
  {Yuan},\ and\ \citenamefont {Pan}}]{liu2024stringbreakingmechanismlattice}%
  \BibitemOpen
  \bibfield  {author} {\bibinfo {author} {\bibfnamefont {Y.}~\bibnamefont
  {Liu}}, \bibinfo {author} {\bibfnamefont {W.-Y.}\ \bibnamefont {Zhang}},
  \bibinfo {author} {\bibfnamefont {Z.-H.}\ \bibnamefont {Zhu}}, \bibinfo
  {author} {\bibfnamefont {M.-G.}\ \bibnamefont {He}}, \bibinfo {author}
  {\bibfnamefont {Z.-S.}\ \bibnamefont {Yuan}},\ and\ \bibinfo {author}
  {\bibfnamefont {J.-W.}\ \bibnamefont {Pan}},\ }\bibfield  {title} {\bibinfo
  {title} {String breaking mechanism in a lattice schwinger model simulator},\
  }\href {https://arxiv.org/abs/2411.15443} {\  (\bibinfo {year} {2024})},\
  \Eprint {https://arxiv.org/abs/2411.15443} {arXiv:2411.15443
  [cond-mat.quant-gas]} \BibitemShut {NoStop}%
\bibitem [{\citenamefont {Alexandrou}\ \emph {et~al.}(2025)\citenamefont
  {Alexandrou}, \citenamefont {Athenodorou}, \citenamefont {Blekos},
  \citenamefont {Polykratis},\ and\ \citenamefont
  {Kühn}}]{alexandrou2025realizingstringbreakingdynamics}%
  \BibitemOpen
  \bibfield  {author} {\bibinfo {author} {\bibfnamefont {C.}~\bibnamefont
  {Alexandrou}}, \bibinfo {author} {\bibfnamefont {A.}~\bibnamefont
  {Athenodorou}}, \bibinfo {author} {\bibfnamefont {K.}~\bibnamefont {Blekos}},
  \bibinfo {author} {\bibfnamefont {G.}~\bibnamefont {Polykratis}},\ and\
  \bibinfo {author} {\bibfnamefont {S.}~\bibnamefont {Kühn}},\ }\bibfield
  {title} {\bibinfo {title} {Realizing string breaking dynamics in a $z_2$
  lattice gauge theory on quantum hardware},\ }\href
  {https://arxiv.org/abs/2504.13760} {\  (\bibinfo {year} {2025})},\ \Eprint
  {https://arxiv.org/abs/2504.13760} {arXiv:2504.13760 [hep-lat]} \BibitemShut
  {NoStop}%
\bibitem [{\citenamefont {Cochran}\ \emph {et~al.}(2025)\citenamefont
  {Cochran}, \citenamefont {Jobst}, \citenamefont {Rosenberg}, \citenamefont
  {Lensky}, \citenamefont {Gyawali}, \citenamefont {Eassa}, \citenamefont
  {Will}, \citenamefont {Szasz}, \citenamefont {Abanin}, \citenamefont
  {Acharya}, \citenamefont {Aghababaie~Beni}, \citenamefont {Andersen},
  \citenamefont {Ansmann}, \citenamefont {Arute}, \citenamefont {Arya},
  \citenamefont {Asfaw}, \citenamefont {Atalaya}, \citenamefont {Babbush},
  \citenamefont {Ballard}, \citenamefont {Bardin}, \citenamefont {Bengtsson},
  \citenamefont {Bilmes}, \citenamefont {Bourassa}, \citenamefont {Bovaird},
  \citenamefont {Broughton}, \citenamefont {Browne}, \citenamefont {Buchea},
  \citenamefont {Buckley}, \citenamefont {Burger}, \citenamefont {Burkett},
  \citenamefont {Bushnell}, \citenamefont {Cabrera}, \citenamefont {Campero},
  \citenamefont {Chang}, \citenamefont {Chen}, \citenamefont {Chiaro},
  \citenamefont {Claes}, \citenamefont {Cleland}, \citenamefont {Cogan},
  \citenamefont {Collins}, \citenamefont {Conner}, \citenamefont {Courtney},
  \citenamefont {Crook}, \citenamefont {Curtin}, \citenamefont {Das},
  \citenamefont {Demura}, \citenamefont {De~Lorenzo}, \citenamefont {Di~Paolo},
  \citenamefont {Donohoe}, \citenamefont {Drozdov}, \citenamefont {Dunsworth},
  \citenamefont {Eickbusch}, \citenamefont {Elbag}, \citenamefont {Elzouka},
  \citenamefont {Erickson}, \citenamefont {Ferreira}, \citenamefont {Burgos},
  \citenamefont {Forati}, \citenamefont {Fowler}, \citenamefont {Foxen},
  \citenamefont {Ganjam}, \citenamefont {Gasca}, \citenamefont {Genois},
  \citenamefont {Giang}, \citenamefont {Gilboa}, \citenamefont {Gosula},
  \citenamefont {Grajales~Dau}, \citenamefont {Graumann}, \citenamefont
  {Greene}, \citenamefont {Gross}, \citenamefont {Habegger}, \citenamefont
  {Hansen}, \citenamefont {Harrigan}, \citenamefont {Harrington}, \citenamefont
  {Heu}, \citenamefont {Higgott}, \citenamefont {Hilton}, \citenamefont
  {Huang}, \citenamefont {Huff}, \citenamefont {Huggins}, \citenamefont
  {Jeffrey}, \citenamefont {Jiang}, \citenamefont {Jones}, \citenamefont
  {Joshi}, \citenamefont {Juhas}, \citenamefont {Kafri}, \citenamefont {Kang},
  \citenamefont {Karamlou}, \citenamefont {Kechedzhi}, \citenamefont {Khaire},
  \citenamefont {Khattar}, \citenamefont {Khezri}, \citenamefont {Kim},
  \citenamefont {Klimov}, \citenamefont {Kobrin}, \citenamefont {Korotkov},
  \citenamefont {Kostritsa}, \citenamefont {Kreikebaum}, \citenamefont
  {Kurilovich}, \citenamefont {Landhuis}, \citenamefont {Lange-Dei},
  \citenamefont {Langley}, \citenamefont {Lau}, \citenamefont {Ledford},
  \citenamefont {Lee}, \citenamefont {Lester}, \citenamefont {Le~Guevel},
  \citenamefont {Li}, \citenamefont {Lill}, \citenamefont {Livingston},
  \citenamefont {Locharla}, \citenamefont {Lundahl}, \citenamefont {Lunt},
  \citenamefont {Madhuk}, \citenamefont {Maloney}, \citenamefont {Mandr{\`a}},
  \citenamefont {Martin}, \citenamefont {Martin}, \citenamefont {Maxfield},
  \citenamefont {McClean}, \citenamefont {McEwen}, \citenamefont {Meeks},
  \citenamefont {Megrant}, \citenamefont {Miao}, \citenamefont {Molavi},
  \citenamefont {Molina}, \citenamefont {Montazeri}, \citenamefont {Movassagh},
  \citenamefont {Neill}, \citenamefont {Newman}, \citenamefont {Nguyen},
  \citenamefont {Nguyen}, \citenamefont {Ni}, \citenamefont {Ottosson},
  \citenamefont {Pizzuto}, \citenamefont {Potter}, \citenamefont {Pritchard},
  \citenamefont {Quintana}, \citenamefont {Ramachandran}, \citenamefont
  {Reagor}, \citenamefont {Rhodes}, \citenamefont {Roberts}, \citenamefont
  {Sankaragomathi}, \citenamefont {Satzinger}, \citenamefont {Schurkus},
  \citenamefont {Shearn}, \citenamefont {Shorter}, \citenamefont {Shutty},
  \citenamefont {Shvarts}, \citenamefont {Sivak}, \citenamefont {Small},
  \citenamefont {Smith}, \citenamefont {Springer}, \citenamefont {Sterling},
  \citenamefont {Suchard}, \citenamefont {Sztein}, \citenamefont {Thor},
  \citenamefont {Torunbalci}, \citenamefont {Vaishnav}, \citenamefont {Vargas},
  \citenamefont {Vdovichev}, \citenamefont {Vidal}, \citenamefont
  {Vollgraff~Heidweiller}, \citenamefont {Waltman}, \citenamefont {Wang},
  \citenamefont {Ware}, \citenamefont {White}, \citenamefont {Wong},
  \citenamefont {Woo}, \citenamefont {Xing}, \citenamefont {Yao}, \citenamefont
  {Yeh}, \citenamefont {Ying}, \citenamefont {Yoo}, \citenamefont {Yosri},
  \citenamefont {Young}, \citenamefont {Zalcman}, \citenamefont {Zhang},
  \citenamefont {Zhu}, \citenamefont {Zobrist}, \citenamefont {Boixo},
  \citenamefont {Kelly}, \citenamefont {Lucero}, \citenamefont {Chen},
  \citenamefont {Smelyanskiy}, \citenamefont {Neven}, \citenamefont
  {Gammon-Smith}, \citenamefont {Pollmann}, \citenamefont {Knap},\ and\
  \citenamefont {Roushan}}]{cochran2024visualizingdynamicschargesstrings}%
  \BibitemOpen
  \bibfield  {author} {\bibinfo {author} {\bibfnamefont {T.~A.}\ \bibnamefont
  {Cochran}}, \bibinfo {author} {\bibfnamefont {B.}~\bibnamefont {Jobst}},
  \bibinfo {author} {\bibfnamefont {E.}~\bibnamefont {Rosenberg}}, \bibinfo
  {author} {\bibfnamefont {Y.~D.}\ \bibnamefont {Lensky}}, \bibinfo {author}
  {\bibfnamefont {G.}~\bibnamefont {Gyawali}}, \bibinfo {author} {\bibfnamefont
  {N.}~\bibnamefont {Eassa}}, \bibinfo {author} {\bibfnamefont
  {M.}~\bibnamefont {Will}}, \bibinfo {author} {\bibfnamefont {A.}~\bibnamefont
  {Szasz}}, \bibinfo {author} {\bibfnamefont {D.}~\bibnamefont {Abanin}},
  \bibinfo {author} {\bibfnamefont {R.}~\bibnamefont {Acharya}}, \bibinfo
  {author} {\bibfnamefont {L.}~\bibnamefont {Aghababaie~Beni}}, \bibinfo
  {author} {\bibfnamefont {T.~I.}\ \bibnamefont {Andersen}}, \bibinfo {author}
  {\bibfnamefont {M.}~\bibnamefont {Ansmann}}, \bibinfo {author} {\bibfnamefont
  {F.}~\bibnamefont {Arute}}, \bibinfo {author} {\bibfnamefont
  {K.}~\bibnamefont {Arya}}, \bibinfo {author} {\bibfnamefont {A.}~\bibnamefont
  {Asfaw}}, \bibinfo {author} {\bibfnamefont {J.}~\bibnamefont {Atalaya}},
  \bibinfo {author} {\bibfnamefont {R.}~\bibnamefont {Babbush}}, \bibinfo
  {author} {\bibfnamefont {B.}~\bibnamefont {Ballard}}, \bibinfo {author}
  {\bibfnamefont {J.~C.}\ \bibnamefont {Bardin}}, \bibinfo {author}
  {\bibfnamefont {A.}~\bibnamefont {Bengtsson}}, \bibinfo {author}
  {\bibfnamefont {A.}~\bibnamefont {Bilmes}}, \bibinfo {author} {\bibfnamefont
  {A.}~\bibnamefont {Bourassa}}, \bibinfo {author} {\bibfnamefont
  {J.}~\bibnamefont {Bovaird}}, \bibinfo {author} {\bibfnamefont
  {M.}~\bibnamefont {Broughton}}, \bibinfo {author} {\bibfnamefont {D.~A.}\
  \bibnamefont {Browne}}, \bibinfo {author} {\bibfnamefont {B.}~\bibnamefont
  {Buchea}}, \bibinfo {author} {\bibfnamefont {B.~B.}\ \bibnamefont {Buckley}},
  \bibinfo {author} {\bibfnamefont {T.}~\bibnamefont {Burger}}, \bibinfo
  {author} {\bibfnamefont {B.}~\bibnamefont {Burkett}}, \bibinfo {author}
  {\bibfnamefont {N.}~\bibnamefont {Bushnell}}, \bibinfo {author}
  {\bibfnamefont {A.}~\bibnamefont {Cabrera}}, \bibinfo {author} {\bibfnamefont
  {J.}~\bibnamefont {Campero}}, \bibinfo {author} {\bibfnamefont {H.~S.}\
  \bibnamefont {Chang}}, \bibinfo {author} {\bibfnamefont {Z.}~\bibnamefont
  {Chen}}, \bibinfo {author} {\bibfnamefont {B.}~\bibnamefont {Chiaro}},
  \bibinfo {author} {\bibfnamefont {J.}~\bibnamefont {Claes}}, \bibinfo
  {author} {\bibfnamefont {A.~Y.}\ \bibnamefont {Cleland}}, \bibinfo {author}
  {\bibfnamefont {J.}~\bibnamefont {Cogan}}, \bibinfo {author} {\bibfnamefont
  {R.}~\bibnamefont {Collins}}, \bibinfo {author} {\bibfnamefont
  {P.}~\bibnamefont {Conner}}, \bibinfo {author} {\bibfnamefont
  {W.}~\bibnamefont {Courtney}}, \bibinfo {author} {\bibfnamefont {A.~L.}\
  \bibnamefont {Crook}}, \bibinfo {author} {\bibfnamefont {B.}~\bibnamefont
  {Curtin}}, \bibinfo {author} {\bibfnamefont {S.}~\bibnamefont {Das}},
  \bibinfo {author} {\bibfnamefont {S.}~\bibnamefont {Demura}}, \bibinfo
  {author} {\bibfnamefont {L.}~\bibnamefont {De~Lorenzo}}, \bibinfo {author}
  {\bibfnamefont {A.}~\bibnamefont {Di~Paolo}}, \bibinfo {author}
  {\bibfnamefont {P.}~\bibnamefont {Donohoe}}, \bibinfo {author} {\bibfnamefont
  {I.}~\bibnamefont {Drozdov}}, \bibinfo {author} {\bibfnamefont
  {A.}~\bibnamefont {Dunsworth}}, \bibinfo {author} {\bibfnamefont
  {A.}~\bibnamefont {Eickbusch}}, \bibinfo {author} {\bibfnamefont {A.~M.}\
  \bibnamefont {Elbag}}, \bibinfo {author} {\bibfnamefont {M.}~\bibnamefont
  {Elzouka}}, \bibinfo {author} {\bibfnamefont {C.}~\bibnamefont {Erickson}},
  \bibinfo {author} {\bibfnamefont {V.~S.}\ \bibnamefont {Ferreira}}, \bibinfo
  {author} {\bibfnamefont {L.~F.}\ \bibnamefont {Burgos}}, \bibinfo {author}
  {\bibfnamefont {E.}~\bibnamefont {Forati}}, \bibinfo {author} {\bibfnamefont
  {A.~G.}\ \bibnamefont {Fowler}}, \bibinfo {author} {\bibfnamefont
  {B.}~\bibnamefont {Foxen}}, \bibinfo {author} {\bibfnamefont
  {S.}~\bibnamefont {Ganjam}}, \bibinfo {author} {\bibfnamefont
  {R.}~\bibnamefont {Gasca}}, \bibinfo {author} {\bibfnamefont
  {{\'E}.}~\bibnamefont {Genois}}, \bibinfo {author} {\bibfnamefont
  {W.}~\bibnamefont {Giang}}, \bibinfo {author} {\bibfnamefont
  {D.}~\bibnamefont {Gilboa}}, \bibinfo {author} {\bibfnamefont
  {R.}~\bibnamefont {Gosula}}, \bibinfo {author} {\bibfnamefont
  {A.}~\bibnamefont {Grajales~Dau}}, \bibinfo {author} {\bibfnamefont
  {D.}~\bibnamefont {Graumann}}, \bibinfo {author} {\bibfnamefont
  {A.}~\bibnamefont {Greene}}, \bibinfo {author} {\bibfnamefont {J.~A.}\
  \bibnamefont {Gross}}, \bibinfo {author} {\bibfnamefont {S.}~\bibnamefont
  {Habegger}}, \bibinfo {author} {\bibfnamefont {M.}~\bibnamefont {Hansen}},
  \bibinfo {author} {\bibfnamefont {M.~P.}\ \bibnamefont {Harrigan}}, \bibinfo
  {author} {\bibfnamefont {S.~D.}\ \bibnamefont {Harrington}}, \bibinfo
  {author} {\bibfnamefont {P.}~\bibnamefont {Heu}}, \bibinfo {author}
  {\bibfnamefont {O.}~\bibnamefont {Higgott}}, \bibinfo {author} {\bibfnamefont
  {J.}~\bibnamefont {Hilton}}, \bibinfo {author} {\bibfnamefont {H.~Y.}\
  \bibnamefont {Huang}}, \bibinfo {author} {\bibfnamefont {A.}~\bibnamefont
  {Huff}}, \bibinfo {author} {\bibfnamefont {W.}~\bibnamefont {Huggins}},
  \bibinfo {author} {\bibfnamefont {E.}~\bibnamefont {Jeffrey}}, \bibinfo
  {author} {\bibfnamefont {Z.}~\bibnamefont {Jiang}}, \bibinfo {author}
  {\bibfnamefont {C.}~\bibnamefont {Jones}}, \bibinfo {author} {\bibfnamefont
  {C.}~\bibnamefont {Joshi}}, \bibinfo {author} {\bibfnamefont
  {P.}~\bibnamefont {Juhas}}, \bibinfo {author} {\bibfnamefont
  {D.}~\bibnamefont {Kafri}}, \bibinfo {author} {\bibfnamefont
  {H.}~\bibnamefont {Kang}}, \bibinfo {author} {\bibfnamefont {A.~H.}\
  \bibnamefont {Karamlou}}, \bibinfo {author} {\bibfnamefont {K.}~\bibnamefont
  {Kechedzhi}}, \bibinfo {author} {\bibfnamefont {T.}~\bibnamefont {Khaire}},
  \bibinfo {author} {\bibfnamefont {T.}~\bibnamefont {Khattar}}, \bibinfo
  {author} {\bibfnamefont {M.}~\bibnamefont {Khezri}}, \bibinfo {author}
  {\bibfnamefont {S.}~\bibnamefont {Kim}}, \bibinfo {author} {\bibfnamefont
  {P.}~\bibnamefont {Klimov}}, \bibinfo {author} {\bibfnamefont
  {B.}~\bibnamefont {Kobrin}}, \bibinfo {author} {\bibfnamefont
  {A.}~\bibnamefont {Korotkov}}, \bibinfo {author} {\bibfnamefont
  {F.}~\bibnamefont {Kostritsa}}, \bibinfo {author} {\bibfnamefont
  {J.}~\bibnamefont {Kreikebaum}}, \bibinfo {author} {\bibfnamefont
  {V.}~\bibnamefont {Kurilovich}}, \bibinfo {author} {\bibfnamefont
  {D.}~\bibnamefont {Landhuis}}, \bibinfo {author} {\bibfnamefont
  {T.}~\bibnamefont {Lange-Dei}}, \bibinfo {author} {\bibfnamefont
  {B.}~\bibnamefont {Langley}}, \bibinfo {author} {\bibfnamefont {K.~M.}\
  \bibnamefont {Lau}}, \bibinfo {author} {\bibfnamefont {J.}~\bibnamefont
  {Ledford}}, \bibinfo {author} {\bibfnamefont {K.}~\bibnamefont {Lee}},
  \bibinfo {author} {\bibfnamefont {B.}~\bibnamefont {Lester}}, \bibinfo
  {author} {\bibfnamefont {L.}~\bibnamefont {Le~Guevel}}, \bibinfo {author}
  {\bibfnamefont {W.}~\bibnamefont {Li}}, \bibinfo {author} {\bibfnamefont
  {A.~T.}\ \bibnamefont {Lill}}, \bibinfo {author} {\bibfnamefont
  {W.}~\bibnamefont {Livingston}}, \bibinfo {author} {\bibfnamefont
  {A.}~\bibnamefont {Locharla}}, \bibinfo {author} {\bibfnamefont
  {D.}~\bibnamefont {Lundahl}}, \bibinfo {author} {\bibfnamefont
  {A.}~\bibnamefont {Lunt}}, \bibinfo {author} {\bibfnamefont {S.}~\bibnamefont
  {Madhuk}}, \bibinfo {author} {\bibfnamefont {A.}~\bibnamefont {Maloney}},
  \bibinfo {author} {\bibfnamefont {S.}~\bibnamefont {Mandr{\`a}}}, \bibinfo
  {author} {\bibfnamefont {L.}~\bibnamefont {Martin}}, \bibinfo {author}
  {\bibfnamefont {O.}~\bibnamefont {Martin}}, \bibinfo {author} {\bibfnamefont
  {C.}~\bibnamefont {Maxfield}}, \bibinfo {author} {\bibfnamefont
  {J.}~\bibnamefont {McClean}}, \bibinfo {author} {\bibfnamefont
  {M.}~\bibnamefont {McEwen}}, \bibinfo {author} {\bibfnamefont
  {S.}~\bibnamefont {Meeks}}, \bibinfo {author} {\bibfnamefont
  {A.}~\bibnamefont {Megrant}}, \bibinfo {author} {\bibfnamefont
  {K.}~\bibnamefont {Miao}}, \bibinfo {author} {\bibfnamefont {R.}~\bibnamefont
  {Molavi}}, \bibinfo {author} {\bibfnamefont {S.}~\bibnamefont {Molina}},
  \bibinfo {author} {\bibfnamefont {S.}~\bibnamefont {Montazeri}}, \bibinfo
  {author} {\bibfnamefont {R.}~\bibnamefont {Movassagh}}, \bibinfo {author}
  {\bibfnamefont {C.}~\bibnamefont {Neill}}, \bibinfo {author} {\bibfnamefont
  {M.}~\bibnamefont {Newman}}, \bibinfo {author} {\bibfnamefont
  {A.}~\bibnamefont {Nguyen}}, \bibinfo {author} {\bibfnamefont
  {M.}~\bibnamefont {Nguyen}}, \bibinfo {author} {\bibfnamefont {C.~H.}\
  \bibnamefont {Ni}}, \bibinfo {author} {\bibfnamefont {K.}~\bibnamefont
  {Ottosson}}, \bibinfo {author} {\bibfnamefont {A.}~\bibnamefont {Pizzuto}},
  \bibinfo {author} {\bibfnamefont {R.}~\bibnamefont {Potter}}, \bibinfo
  {author} {\bibfnamefont {O.}~\bibnamefont {Pritchard}}, \bibinfo {author}
  {\bibfnamefont {C.}~\bibnamefont {Quintana}}, \bibinfo {author}
  {\bibfnamefont {G.}~\bibnamefont {Ramachandran}}, \bibinfo {author}
  {\bibfnamefont {M.}~\bibnamefont {Reagor}}, \bibinfo {author} {\bibfnamefont
  {D.}~\bibnamefont {Rhodes}}, \bibinfo {author} {\bibfnamefont
  {G.}~\bibnamefont {Roberts}}, \bibinfo {author} {\bibfnamefont
  {K.}~\bibnamefont {Sankaragomathi}}, \bibinfo {author} {\bibfnamefont
  {K.}~\bibnamefont {Satzinger}}, \bibinfo {author} {\bibfnamefont
  {H.}~\bibnamefont {Schurkus}}, \bibinfo {author} {\bibfnamefont
  {M.}~\bibnamefont {Shearn}}, \bibinfo {author} {\bibfnamefont
  {A.}~\bibnamefont {Shorter}}, \bibinfo {author} {\bibfnamefont
  {N.}~\bibnamefont {Shutty}}, \bibinfo {author} {\bibfnamefont
  {V.}~\bibnamefont {Shvarts}}, \bibinfo {author} {\bibfnamefont
  {V.}~\bibnamefont {Sivak}}, \bibinfo {author} {\bibfnamefont
  {S.}~\bibnamefont {Small}}, \bibinfo {author} {\bibfnamefont {W.~C.}\
  \bibnamefont {Smith}}, \bibinfo {author} {\bibfnamefont {S.}~\bibnamefont
  {Springer}}, \bibinfo {author} {\bibfnamefont {G.}~\bibnamefont {Sterling}},
  \bibinfo {author} {\bibfnamefont {J.}~\bibnamefont {Suchard}}, \bibinfo
  {author} {\bibfnamefont {A.}~\bibnamefont {Sztein}}, \bibinfo {author}
  {\bibfnamefont {D.}~\bibnamefont {Thor}}, \bibinfo {author} {\bibfnamefont
  {M.}~\bibnamefont {Torunbalci}}, \bibinfo {author} {\bibfnamefont
  {A.}~\bibnamefont {Vaishnav}}, \bibinfo {author} {\bibfnamefont
  {J.}~\bibnamefont {Vargas}}, \bibinfo {author} {\bibfnamefont
  {S.}~\bibnamefont {Vdovichev}}, \bibinfo {author} {\bibfnamefont
  {G.}~\bibnamefont {Vidal}}, \bibinfo {author} {\bibfnamefont
  {C.}~\bibnamefont {Vollgraff~Heidweiller}}, \bibinfo {author} {\bibfnamefont
  {S.}~\bibnamefont {Waltman}}, \bibinfo {author} {\bibfnamefont {S.~X.}\
  \bibnamefont {Wang}}, \bibinfo {author} {\bibfnamefont {B.}~\bibnamefont
  {Ware}}, \bibinfo {author} {\bibfnamefont {T.}~\bibnamefont {White}},
  \bibinfo {author} {\bibfnamefont {K.}~\bibnamefont {Wong}}, \bibinfo {author}
  {\bibfnamefont {B.~W.~K.}\ \bibnamefont {Woo}}, \bibinfo {author}
  {\bibfnamefont {C.}~\bibnamefont {Xing}}, \bibinfo {author} {\bibfnamefont
  {Z.~J.}\ \bibnamefont {Yao}}, \bibinfo {author} {\bibfnamefont
  {P.}~\bibnamefont {Yeh}}, \bibinfo {author} {\bibfnamefont {B.}~\bibnamefont
  {Ying}}, \bibinfo {author} {\bibfnamefont {J.}~\bibnamefont {Yoo}}, \bibinfo
  {author} {\bibfnamefont {N.}~\bibnamefont {Yosri}}, \bibinfo {author}
  {\bibfnamefont {G.}~\bibnamefont {Young}}, \bibinfo {author} {\bibfnamefont
  {A.}~\bibnamefont {Zalcman}}, \bibinfo {author} {\bibfnamefont
  {Y.}~\bibnamefont {Zhang}}, \bibinfo {author} {\bibfnamefont
  {N.}~\bibnamefont {Zhu}}, \bibinfo {author} {\bibfnamefont {N.}~\bibnamefont
  {Zobrist}}, \bibinfo {author} {\bibfnamefont {S.}~\bibnamefont {Boixo}},
  \bibinfo {author} {\bibfnamefont {J.}~\bibnamefont {Kelly}}, \bibinfo
  {author} {\bibfnamefont {E.}~\bibnamefont {Lucero}}, \bibinfo {author}
  {\bibfnamefont {Y.}~\bibnamefont {Chen}}, \bibinfo {author} {\bibfnamefont
  {V.}~\bibnamefont {Smelyanskiy}}, \bibinfo {author} {\bibfnamefont
  {H.}~\bibnamefont {Neven}}, \bibinfo {author} {\bibfnamefont
  {A.}~\bibnamefont {Gammon-Smith}}, \bibinfo {author} {\bibfnamefont
  {F.}~\bibnamefont {Pollmann}}, \bibinfo {author} {\bibfnamefont
  {M.}~\bibnamefont {Knap}},\ and\ \bibinfo {author} {\bibfnamefont
  {P.}~\bibnamefont {Roushan}},\ }\bibfield  {title} {\bibinfo {title}
  {Visualizing dynamics of charges and strings in (2 + 1)d lattice gauge
  theories},\ }\href {https://doi.org/10.1038/s41586-025-08999-9} {\bibfield
  {journal} {\bibinfo  {journal} {Nature}\ }\textbf {\bibinfo {volume} {642}},\
  \bibinfo {pages} {315} (\bibinfo {year} {2025})}\BibitemShut {NoStop}%
\bibitem [{\citenamefont {Gonz{\'a}lez-Cuadra}\ \emph
  {et~al.}(2025)\citenamefont {Gonz{\'a}lez-Cuadra}, \citenamefont {Hamdan},
  \citenamefont {Zache}, \citenamefont {Braverman}, \citenamefont {Kornja{\v
  c}a}, \citenamefont {Lukin}, \citenamefont {Cant{\'u}}, \citenamefont {Liu},
  \citenamefont {Wang}, \citenamefont {Keesling}, \citenamefont {Lukin},
  \citenamefont {Zoller},\ and\ \citenamefont
  {Bylinskii}}]{gonzalezcuadra2024observationstringbreaking2}%
  \BibitemOpen
  \bibfield  {author} {\bibinfo {author} {\bibfnamefont {D.}~\bibnamefont
  {Gonz{\'a}lez-Cuadra}}, \bibinfo {author} {\bibfnamefont {M.}~\bibnamefont
  {Hamdan}}, \bibinfo {author} {\bibfnamefont {T.~V.}\ \bibnamefont {Zache}},
  \bibinfo {author} {\bibfnamefont {B.}~\bibnamefont {Braverman}}, \bibinfo
  {author} {\bibfnamefont {M.}~\bibnamefont {Kornja{\v c}a}}, \bibinfo {author}
  {\bibfnamefont {A.}~\bibnamefont {Lukin}}, \bibinfo {author} {\bibfnamefont
  {S.~H.}\ \bibnamefont {Cant{\'u}}}, \bibinfo {author} {\bibfnamefont
  {F.}~\bibnamefont {Liu}}, \bibinfo {author} {\bibfnamefont {S.-T.}\
  \bibnamefont {Wang}}, \bibinfo {author} {\bibfnamefont {A.}~\bibnamefont
  {Keesling}}, \bibinfo {author} {\bibfnamefont {M.~D.}\ \bibnamefont {Lukin}},
  \bibinfo {author} {\bibfnamefont {P.}~\bibnamefont {Zoller}},\ and\ \bibinfo
  {author} {\bibfnamefont {A.}~\bibnamefont {Bylinskii}},\ }\bibfield  {title}
  {\bibinfo {title} {Observation of string breaking on a (2 + 1)d rydberg
  quantum simulator},\ }\href {https://doi.org/10.1038/s41586-025-09051-6}
  {\bibfield  {journal} {\bibinfo  {journal} {Nature}\ }\textbf {\bibinfo
  {volume} {642}},\ \bibinfo {pages} {321} (\bibinfo {year}
  {2025})}\BibitemShut {NoStop}%
\bibitem [{\citenamefont {Crippa}\ \emph {et~al.}(2024)\citenamefont {Crippa},
  \citenamefont {Jansen},\ and\ \citenamefont
  {Rinaldi}}]{crippa2024analysisconfinementstring2}%
  \BibitemOpen
  \bibfield  {author} {\bibinfo {author} {\bibfnamefont {A.}~\bibnamefont
  {Crippa}}, \bibinfo {author} {\bibfnamefont {K.}~\bibnamefont {Jansen}},\
  and\ \bibinfo {author} {\bibfnamefont {E.}~\bibnamefont {Rinaldi}},\
  }\bibfield  {title} {\bibinfo {title} {Analysis of the confinement string in
  (2 + 1)-dimensional quantum electrodynamics with a trapped-ion quantum
  computer},\ }\href {https://arxiv.org/abs/2411.05628} {\  (\bibinfo {year}
  {2024})},\ \Eprint {https://arxiv.org/abs/2411.05628} {arXiv:2411.05628
  [hep-lat]} \BibitemShut {NoStop}%
\bibitem [{\citenamefont {Xu}\ \emph {et~al.}(2025)\citenamefont {Xu},
  \citenamefont {Knap},\ and\ \citenamefont
  {Pollmann}}]{xu2025tensornetworkstudyrougheningtransition}%
  \BibitemOpen
  \bibfield  {author} {\bibinfo {author} {\bibfnamefont {W.-T.}\ \bibnamefont
  {Xu}}, \bibinfo {author} {\bibfnamefont {M.}~\bibnamefont {Knap}},\ and\
  \bibinfo {author} {\bibfnamefont {F.}~\bibnamefont {Pollmann}},\ }\bibfield
  {title} {\bibinfo {title} {Tensor-network study of the roughening transition
  in (2 + 1)d lattice gauge theories},\ }\href
  {https://arxiv.org/abs/2503.19027} {\  (\bibinfo {year} {2025})},\ \Eprint
  {https://arxiv.org/abs/2503.19027} {arXiv:2503.19027 [cond-mat.str-el]}
  \BibitemShut {NoStop}%
\bibitem [{\citenamefont {Marcantonio}\ \emph {et~al.}(2025)\citenamefont
  {Marcantonio}, \citenamefont {Pradhan}, \citenamefont {Vallecorsa},
  \citenamefont {Bañuls},\ and\ \citenamefont
  {Ortega}}]{dimarcantonio2025rougheningdynamicselectricflux}%
  \BibitemOpen
  \bibfield  {author} {\bibinfo {author} {\bibfnamefont {F.~D.}\ \bibnamefont
  {Marcantonio}}, \bibinfo {author} {\bibfnamefont {S.}~\bibnamefont
  {Pradhan}}, \bibinfo {author} {\bibfnamefont {S.}~\bibnamefont {Vallecorsa}},
  \bibinfo {author} {\bibfnamefont {M.~C.}\ \bibnamefont {Bañuls}},\ and\
  \bibinfo {author} {\bibfnamefont {E.~R.}\ \bibnamefont {Ortega}},\ }\bibfield
   {title} {\bibinfo {title} {Roughening and dynamics of an electric flux
  string in a (2+1)d lattice gauge theory},\ }\href
  {https://arxiv.org/abs/2505.23853} {\  (\bibinfo {year} {2025})},\ \Eprint
  {https://arxiv.org/abs/2505.23853} {arXiv:2505.23853 [hep-lat]} \BibitemShut
  {NoStop}%
\bibitem [{\citenamefont {Borla}\ \emph {et~al.}(2025)\citenamefont {Borla},
  \citenamefont {Osborne}, \citenamefont {Moroz},\ and\ \citenamefont
  {Halimeh}}]{borla2025stringbreaking21dmathbbz2}%
  \BibitemOpen
  \bibfield  {author} {\bibinfo {author} {\bibfnamefont {U.}~\bibnamefont
  {Borla}}, \bibinfo {author} {\bibfnamefont {J.~J.}\ \bibnamefont {Osborne}},
  \bibinfo {author} {\bibfnamefont {S.}~\bibnamefont {Moroz}},\ and\ \bibinfo
  {author} {\bibfnamefont {J.~C.}\ \bibnamefont {Halimeh}},\ }\bibfield
  {title} {\bibinfo {title} {String breaking in a $2+1$d $\mathbb{Z}_2$ lattice
  gauge theory},\ }\href {https://arxiv.org/abs/2501.17929} {\  (\bibinfo
  {year} {2025})},\ \Eprint {https://arxiv.org/abs/2501.17929}
  {arXiv:2501.17929 [quant-ph]} \BibitemShut {NoStop}%
\bibitem [{\citenamefont {Schollwöck}(2011)}]{Uli_review}%
  \BibitemOpen
  \bibfield  {author} {\bibinfo {author} {\bibfnamefont {U.}~\bibnamefont
  {Schollwöck}},\ }\bibfield  {title} {\bibinfo {title} {The density-matrix
  renormalization group in the age of matrix product states},\ }\href
  {https://doi.org/https://doi.org/10.1016/j.aop.2010.09.012} {\bibfield
  {journal} {\bibinfo  {journal} {Ann. Phys.}\ }\textbf {\bibinfo {volume}
  {326}},\ \bibinfo {pages} {96} (\bibinfo {year} {2011})}\BibitemShut
  {NoStop}%
\bibitem [{\citenamefont {Paeckel}\ \emph {et~al.}(2019)\citenamefont
  {Paeckel}, \citenamefont {Köhler}, \citenamefont {Swoboda}, \citenamefont
  {Manmana}, \citenamefont {Schollwöck},\ and\ \citenamefont
  {Hubig}}]{Paeckel_review}%
  \BibitemOpen
  \bibfield  {author} {\bibinfo {author} {\bibfnamefont {S.}~\bibnamefont
  {Paeckel}}, \bibinfo {author} {\bibfnamefont {T.}~\bibnamefont {Köhler}},
  \bibinfo {author} {\bibfnamefont {A.}~\bibnamefont {Swoboda}}, \bibinfo
  {author} {\bibfnamefont {S.~R.}\ \bibnamefont {Manmana}}, \bibinfo {author}
  {\bibfnamefont {U.}~\bibnamefont {Schollwöck}},\ and\ \bibinfo {author}
  {\bibfnamefont {C.}~\bibnamefont {Hubig}},\ }\bibfield  {title} {\bibinfo
  {title} {Time-evolution methods for matrix-product states},\ }\href
  {https://doi.org/https://doi.org/10.1016/j.aop.2019.167998} {\bibfield
  {journal} {\bibinfo  {journal} {Ann. Phys.}\ }\textbf {\bibinfo {volume}
  {411}},\ \bibinfo {pages} {167998} (\bibinfo {year} {2019})}\BibitemShut
  {NoStop}%
\bibitem [{\citenamefont {Montangero}(2018)}]{Montangero_book}%
  \BibitemOpen
  \bibfield  {author} {\bibinfo {author} {\bibfnamefont {S.}~\bibnamefont
  {Montangero}},\ }\href {https://books.google.de/books?id=voF8DwAAQBAJ} {\emph
  {\bibinfo {title} {Introduction to Tensor Network Methods: Numerical
  simulations of low-dimensional many-body quantum systems}}}\ (\bibinfo
  {publisher} {Springer International Publishing},\ \bibinfo {year}
  {2018})\BibitemShut {NoStop}%
\bibitem [{\citenamefont {Hauschild}\ \emph {et~al.}(2024)\citenamefont
  {Hauschild}, \citenamefont {Unfried}, \citenamefont {Anand}, \citenamefont
  {Andrews}, \citenamefont {Bintz}, \citenamefont {Borla}, \citenamefont
  {Divic}, \citenamefont {Drescher}, \citenamefont {Geiger}, \citenamefont
  {Hefel}, \citenamefont {Hémery}, \citenamefont {Kadow}, \citenamefont
  {Kemp}, \citenamefont {Kirchner}, \citenamefont {Liu}, \citenamefont
  {Möller}, \citenamefont {Parker}, \citenamefont {Rader}, \citenamefont
  {Romen}, \citenamefont {Scalet}, \citenamefont {Schoonderwoerd},
  \citenamefont {Schulz}, \citenamefont {Soejima}, \citenamefont {Thoma},
  \citenamefont {Wu}, \citenamefont {Zechmann}, \citenamefont {Zweng},
  \citenamefont {Mong}, \citenamefont {Zaletel},\ and\ \citenamefont
  {Pollmann}}]{tenpy2024}%
  \BibitemOpen
  \bibfield  {author} {\bibinfo {author} {\bibfnamefont {J.}~\bibnamefont
  {Hauschild}}, \bibinfo {author} {\bibfnamefont {J.}~\bibnamefont {Unfried}},
  \bibinfo {author} {\bibfnamefont {S.}~\bibnamefont {Anand}}, \bibinfo
  {author} {\bibfnamefont {B.}~\bibnamefont {Andrews}}, \bibinfo {author}
  {\bibfnamefont {M.}~\bibnamefont {Bintz}}, \bibinfo {author} {\bibfnamefont
  {U.}~\bibnamefont {Borla}}, \bibinfo {author} {\bibfnamefont
  {S.}~\bibnamefont {Divic}}, \bibinfo {author} {\bibfnamefont
  {M.}~\bibnamefont {Drescher}}, \bibinfo {author} {\bibfnamefont
  {J.}~\bibnamefont {Geiger}}, \bibinfo {author} {\bibfnamefont
  {M.}~\bibnamefont {Hefel}}, \bibinfo {author} {\bibfnamefont
  {K.}~\bibnamefont {Hémery}}, \bibinfo {author} {\bibfnamefont
  {W.}~\bibnamefont {Kadow}}, \bibinfo {author} {\bibfnamefont
  {J.}~\bibnamefont {Kemp}}, \bibinfo {author} {\bibfnamefont {N.}~\bibnamefont
  {Kirchner}}, \bibinfo {author} {\bibfnamefont {V.~S.}\ \bibnamefont {Liu}},
  \bibinfo {author} {\bibfnamefont {G.}~\bibnamefont {Möller}}, \bibinfo
  {author} {\bibfnamefont {D.}~\bibnamefont {Parker}}, \bibinfo {author}
  {\bibfnamefont {M.}~\bibnamefont {Rader}}, \bibinfo {author} {\bibfnamefont
  {A.}~\bibnamefont {Romen}}, \bibinfo {author} {\bibfnamefont
  {S.}~\bibnamefont {Scalet}}, \bibinfo {author} {\bibfnamefont
  {L.}~\bibnamefont {Schoonderwoerd}}, \bibinfo {author} {\bibfnamefont
  {M.}~\bibnamefont {Schulz}}, \bibinfo {author} {\bibfnamefont
  {T.}~\bibnamefont {Soejima}}, \bibinfo {author} {\bibfnamefont
  {P.}~\bibnamefont {Thoma}}, \bibinfo {author} {\bibfnamefont
  {Y.}~\bibnamefont {Wu}}, \bibinfo {author} {\bibfnamefont {P.}~\bibnamefont
  {Zechmann}}, \bibinfo {author} {\bibfnamefont {L.}~\bibnamefont {Zweng}},
  \bibinfo {author} {\bibfnamefont {R.~S.~K.}\ \bibnamefont {Mong}}, \bibinfo
  {author} {\bibfnamefont {M.~P.}\ \bibnamefont {Zaletel}},\ and\ \bibinfo
  {author} {\bibfnamefont {F.}~\bibnamefont {Pollmann}},\ }\bibfield  {title}
  {\bibinfo {title} {{Tensor network Python (TeNPy) version 1}},\ }\href
  {https://doi.org/10.21468/SciPostPhysCodeb.41} {\bibfield  {journal}
  {\bibinfo  {journal} {SciPost Phys. Codebases}\ ,\ \bibinfo {pages} {41}}
  (\bibinfo {year} {2024})}\BibitemShut {NoStop}%
\bibitem [{\citenamefont {Haegeman}\ \emph {et~al.}(2011)\citenamefont
  {Haegeman}, \citenamefont {Cirac}, \citenamefont {Osborne}, \citenamefont
  {Pi\ifmmode~\check{z}\else \v{z}\fi{}orn}, \citenamefont {Verschelde},\ and\
  \citenamefont {Verstraete}}]{Haegeman2011}%
  \BibitemOpen
  \bibfield  {author} {\bibinfo {author} {\bibfnamefont {J.}~\bibnamefont
  {Haegeman}}, \bibinfo {author} {\bibfnamefont {J.~I.}\ \bibnamefont {Cirac}},
  \bibinfo {author} {\bibfnamefont {T.~J.}\ \bibnamefont {Osborne}}, \bibinfo
  {author} {\bibfnamefont {I.}~\bibnamefont {Pi\ifmmode~\check{z}\else
  \v{z}\fi{}orn}}, \bibinfo {author} {\bibfnamefont {H.}~\bibnamefont
  {Verschelde}},\ and\ \bibinfo {author} {\bibfnamefont {F.}~\bibnamefont
  {Verstraete}},\ }\bibfield  {title} {\bibinfo {title} {Time-dependent
  variational principle for quantum lattices},\ }\href
  {https://doi.org/10.1103/PhysRevLett.107.070601} {\bibfield  {journal}
  {\bibinfo  {journal} {Phys. Rev. Lett.}\ }\textbf {\bibinfo {volume} {107}},\
  \bibinfo {pages} {070601} (\bibinfo {year} {2011})}\BibitemShut {NoStop}%
\bibitem [{\citenamefont {Haegeman}\ \emph {et~al.}(2013)\citenamefont
  {Haegeman}, \citenamefont {Osborne},\ and\ \citenamefont
  {Verstraete}}]{Haegeman2013}%
  \BibitemOpen
  \bibfield  {author} {\bibinfo {author} {\bibfnamefont {J.}~\bibnamefont
  {Haegeman}}, \bibinfo {author} {\bibfnamefont {T.~J.}\ \bibnamefont
  {Osborne}},\ and\ \bibinfo {author} {\bibfnamefont {F.}~\bibnamefont
  {Verstraete}},\ }\bibfield  {title} {\bibinfo {title} {Post-matrix product
  state methods: To tangent space and beyond},\ }\href
  {https://doi.org/10.1103/PhysRevB.88.075133} {\bibfield  {journal} {\bibinfo
  {journal} {Phys. Rev. B}\ }\textbf {\bibinfo {volume} {88}},\ \bibinfo
  {pages} {075133} (\bibinfo {year} {2013})}\BibitemShut {NoStop}%
\bibitem [{\citenamefont {Haegeman}\ \emph {et~al.}(2016)\citenamefont
  {Haegeman}, \citenamefont {Lubich}, \citenamefont {Oseledets}, \citenamefont
  {Vandereycken},\ and\ \citenamefont {Verstraete}}]{Haegeman2016}%
  \BibitemOpen
  \bibfield  {author} {\bibinfo {author} {\bibfnamefont {J.}~\bibnamefont
  {Haegeman}}, \bibinfo {author} {\bibfnamefont {C.}~\bibnamefont {Lubich}},
  \bibinfo {author} {\bibfnamefont {I.}~\bibnamefont {Oseledets}}, \bibinfo
  {author} {\bibfnamefont {B.}~\bibnamefont {Vandereycken}},\ and\ \bibinfo
  {author} {\bibfnamefont {F.}~\bibnamefont {Verstraete}},\ }\bibfield  {title}
  {\bibinfo {title} {Unifying time evolution and optimization with matrix
  product states},\ }\href {https://doi.org/10.1103/PhysRevB.94.165116}
  {\bibfield  {journal} {\bibinfo  {journal} {Phys. Rev. B}\ }\textbf {\bibinfo
  {volume} {94}},\ \bibinfo {pages} {165116} (\bibinfo {year}
  {2016})}\BibitemShut {NoStop}%
\bibitem [{\citenamefont {Greensite}(2011)}]{greensite2011introduction}%
  \BibitemOpen
  \bibfield  {author} {\bibinfo {author} {\bibfnamefont {J.}~\bibnamefont
  {Greensite}},\ }\href {https://books.google.de/books?id=CP7_QooHo8wC} {\emph
  {\bibinfo {title} {An Introduction to the Confinement Problem}}},\ Lecture
  Notes in Physics\ (\bibinfo  {publisher} {Springer Berlin Heidelberg},\
  \bibinfo {year} {2011})\BibitemShut {NoStop}%
\bibitem [{\citenamefont {Fradkin}\ and\ \citenamefont
  {Shenker}(1979)}]{Fradkin1979}%
  \BibitemOpen
  \bibfield  {author} {\bibinfo {author} {\bibfnamefont {E.}~\bibnamefont
  {Fradkin}}\ and\ \bibinfo {author} {\bibfnamefont {S.~H.}\ \bibnamefont
  {Shenker}},\ }\bibfield  {title} {\bibinfo {title} {Phase diagrams of lattice
  gauge theories with {H}iggs fields},\ }\href
  {https://doi.org/10.1103/PhysRevD.19.3682} {\bibfield  {journal} {\bibinfo
  {journal} {Phys. Rev. D}\ }\textbf {\bibinfo {volume} {19}},\ \bibinfo
  {pages} {3682} (\bibinfo {year} {1979})}\BibitemShut {NoStop}%
\bibitem [{\citenamefont {Trebst}\ \emph {et~al.}(2007)\citenamefont {Trebst},
  \citenamefont {Werner}, \citenamefont {Troyer}, \citenamefont {Shtengel},\
  and\ \citenamefont {Nayak}}]{Trebst2007breakdown}%
  \BibitemOpen
  \bibfield  {author} {\bibinfo {author} {\bibfnamefont {S.}~\bibnamefont
  {Trebst}}, \bibinfo {author} {\bibfnamefont {P.}~\bibnamefont {Werner}},
  \bibinfo {author} {\bibfnamefont {M.}~\bibnamefont {Troyer}}, \bibinfo
  {author} {\bibfnamefont {K.}~\bibnamefont {Shtengel}},\ and\ \bibinfo
  {author} {\bibfnamefont {C.}~\bibnamefont {Nayak}},\ }\bibfield  {title}
  {\bibinfo {title} {Breakdown of a topological phase: Quantum phase transition
  in a loop gas model with tension},\ }\href
  {https://doi.org/10.1103/PhysRevLett.98.070602} {\bibfield  {journal}
  {\bibinfo  {journal} {Phys. Rev. Lett.}\ }\textbf {\bibinfo {volume} {98}},\
  \bibinfo {pages} {070602} (\bibinfo {year} {2007})}\BibitemShut {NoStop}%
\bibitem [{\citenamefont {Vidal}\ \emph {et~al.}(2009)\citenamefont {Vidal},
  \citenamefont {Dusuel},\ and\ \citenamefont {Schmidt}}]{Vidal2009low-energy}%
  \BibitemOpen
  \bibfield  {author} {\bibinfo {author} {\bibfnamefont {J.}~\bibnamefont
  {Vidal}}, \bibinfo {author} {\bibfnamefont {S.}~\bibnamefont {Dusuel}},\ and\
  \bibinfo {author} {\bibfnamefont {K.~P.}\ \bibnamefont {Schmidt}},\
  }\bibfield  {title} {\bibinfo {title} {Low-energy effective theory of the
  toric code model in a parallel magnetic field},\ }\href
  {https://doi.org/10.1103/PhysRevB.79.033109} {\bibfield  {journal} {\bibinfo
  {journal} {Phys. Rev. B}\ }\textbf {\bibinfo {volume} {79}},\ \bibinfo
  {pages} {033109} (\bibinfo {year} {2009})}\BibitemShut {NoStop}%
\bibitem [{\citenamefont {Wu}\ \emph {et~al.}(2012)\citenamefont {Wu},
  \citenamefont {Deng},\ and\ \citenamefont {Prokof'ev}}]{Wu2012phase}%
  \BibitemOpen
  \bibfield  {author} {\bibinfo {author} {\bibfnamefont {F.}~\bibnamefont
  {Wu}}, \bibinfo {author} {\bibfnamefont {Y.}~\bibnamefont {Deng}},\ and\
  \bibinfo {author} {\bibfnamefont {N.}~\bibnamefont {Prokof'ev}},\ }\bibfield
  {title} {\bibinfo {title} {Phase diagram of the toric code model in a
  parallel magnetic field},\ }\href
  {https://doi.org/10.1103/PhysRevB.85.195104} {\bibfield  {journal} {\bibinfo
  {journal} {Phys. Rev. B}\ }\textbf {\bibinfo {volume} {85}},\ \bibinfo
  {pages} {195104} (\bibinfo {year} {2012})}\BibitemShut {NoStop}%
\bibitem [{SM()}]{SM}%
  \BibitemOpen
  \href@noop {} {}\bibinfo {howpublished} {See Supplemental Material for
  details on the numerical TDVP algorithm, on how to derive the effective
  models at strong electric coupling and on the string dissipation as the Higgs
  phase is approached.}\BibitemShut {Stop}%
\bibitem [{vid()}]{videos}%
  \BibitemOpen
  \href@noop {} {}\bibinfo {howpublished}
  {\href{https://youtube.com/playlist?list=PLoUsb3eaKix6dip3vE-ri_j7ESA_YI2nJ&si=fY131G58iKyr7FSq}{Video
  playlist} for quench dynamics and breaking of flux strings in a $2+1$D
  $\mathbb{Z}_2$ lattice gauge theory.}\BibitemShut {Stop}%
\bibitem [{\citenamefont {MATHIEU}\ \emph {et~al.}(2009)\citenamefont
  {MATHIEU}, \citenamefont {KOCHELEV},\ and\ \citenamefont
  {VENTO}}]{Mathieu2009glueballs}%
  \BibitemOpen
  \bibfield  {author} {\bibinfo {author} {\bibfnamefont {V.}~\bibnamefont
  {MATHIEU}}, \bibinfo {author} {\bibfnamefont {N.}~\bibnamefont {KOCHELEV}},\
  and\ \bibinfo {author} {\bibfnamefont {V.}~\bibnamefont {VENTO}},\ }\bibfield
   {title} {\bibinfo {title} {The physics of glueballs},\ }\href
  {https://doi.org/10.1142/S0218301309012124} {\bibfield  {journal} {\bibinfo
  {journal} {International Journal of Modern Physics E}\ }\textbf {\bibinfo
  {volume} {18}},\ \bibinfo {pages} {1} (\bibinfo {year} {2009})},\ \Eprint
  {https://arxiv.org/abs/https://doi.org/10.1142/S0218301309012124}
  {https://doi.org/10.1142/S0218301309012124} \BibitemShut {NoStop}%
\end{thebibliography}%

\clearpage
\pagebreak
\newpage

\onecolumngrid

\setcounter{equation}{0}
\setcounter{figure}{0}
\setcounter{table}{0}
\setcounter{page}{1}

\renewcommand{\theequation}{S\arabic{equation}}
\renewcommand{\thefigure}{S\arabic{figure}}
\renewcommand{\thetable}{S\arabic{table}}

\begin{center}
\textbf{\large Supplemental Material:\\String Breaking Dynamics and Glueball Formation in a $2+1$D Lattice Gauge Theory}
\end{center}

\section{Computational Method}
\begin{figure}
    \centering
    \includegraphics[width=0.5\linewidth]{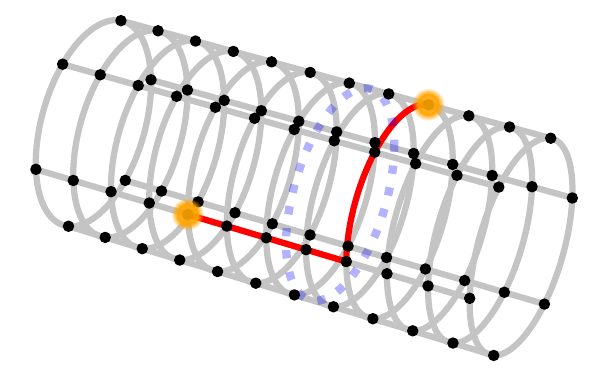}
    \caption{Schematics of the geometry used in this study. The red solid line, yellow dots, and blue dashed line represent the initial L-shaped string, the static charges, and the cut used to compute the entanglement entropy respectively.}
    \label{fig:MPS}
\end{figure}

\begin{figure}
    \centering
    \includegraphics[width=0.8\linewidth]{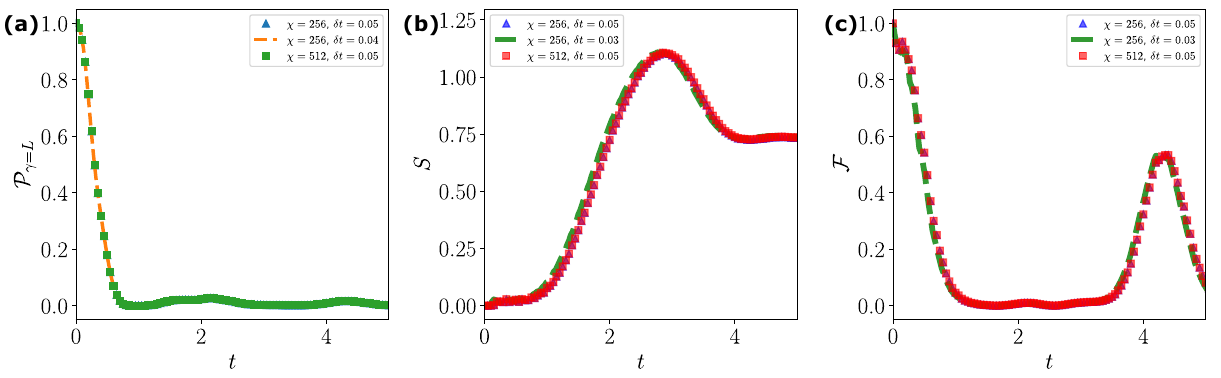}
    \caption{Convergence tests showing different observables for several values of the time step $\delta t$ and bond dimension $\chi$ in different regimes. (a) L string fidelity for $J_s=2$, $J_p=1$, $h_x=4$, and $h_z=1$, i.e., at the first order resonance condition. (b) Entanglement entropy as a function of time for a minimal L-shape string at $J_s=15$, $J_p=1$, $h_x=12$, and $h_z=1$, i.e. deeply in the confined phase, off-resonance. (c) Fidelity for non-minimal snake string as a function of time at $J_s=15$, $J_p=1$, $h_x=12$, and $h_z=1$.}
    \label{fig:conv}
\end{figure}

To perform time evolution, we employ a two-site update TDVP algorithm. As it is standard for the application of 1d algorithms such as TDVP and DMRG to two-dimensional systems, we use a cylindrical geometry, with circumference of size $L_y$ and open boundary condition in the horizontal direction $L_x$, as shown in \cref{fig:MPS}. The bond dimension grows dynamically with each time step, and the simulations in the main draft employ a $6\times 6$ lattice site with $72$ links and 30 matter vertices. After careful convergence tests with respect to bond dimension $\chi$ and time step size $dt$, we establish that the values $\chi=256$ and $\delta t = 0.05$ guarantee sufficiently accurate results in all the regimes that we consider. To this end, we note that the dynamics in the confined phase is mostly confined within the patch spanned by the two static charges, and as a consequence the entanglement growth is also limited. This allows to perform efficient simulations with relatively low values of the bond dimension $\chi$. Convergence tests are shown in \cref{fig:conv} both for the resonance regime and in the large mass and string tension limits considered in the main text.

\section{Effective model for the perturbative dynamics at large electric coupling}

We describe here a general perturbative method that can be used to obtain the effective Hamiltonian acting on a relevant degenerate subspace as $h_x\rightarrow \infty$. Here, the only energetically viable transitions are the ones between states containing strings of equal length, which are separated for the other subsectors of the Hilbert space by a large gap $\Delta E \approx 2h_x$. The case of first order mesonic resonances $h_x=2J_s$ will also be considered at the end of this section.
Length-preserving transitions involve acting with $B_{\rr*}$ on plaquettes where exactly two electric field operators out of four are flipped ($\langle\hat{\sigma}^x\rangle=-1$), which can occur either on corners or on parallel links. These imply that the dynamics is restricted to the patch spanned by the two static charges, since no plaquette acting outside of the boundary can satisfy these conditions. Given an initial state, one can keep track of all possible configurations that can be reached by applying iteratively the plaquette operator whenever allowed, and construct a basis for the relevant subspace. The number of basis states is exponentially smaller than the full Hilbert space, and allows to obtain exact diagonalization results up to patch sizes of $6\times4$ (58 spins) with moderate computational resources. The \cref{tab:config_basis_sizes} shows Hilbert space sizes for a variety of initial configurations and patch sizes.

\subsection{Free fermionic case}
The method described above can be readily applied to an initial configuration consisting of a minimal length string, where the only possible transitions involve flipping corners. As discussed in \cite{borla2025stringbreaking21dmathbbz2}, these transitions can be put in one to one correspondence with the nearest neighbor hoppings of fermions (or hardcore bosons) on a one dimensional chain of the same length as the string at filling $N/L$, with $N$ given by the vertical displacement of its endpoints. For example, an initial L-shaped string corresponds to the initial fermionic state $|000011\rangle$ and the application of a plaquette operator at the corner connects it to the state $|000101\rangle$. 

As opposed to the general case described above, which requires the numerical diagonalization of a complicated effective Hamiltonian, the time evolution of such fermionic states under
\begin{equation}
    H_f =-t \sum_{i=1}^{L-1} c^{\dagger}_i c_{i+1} + \text{h.c.}
    \label{eq:ff_ham}
\end{equation}
can be determined analytically. Even in the absence of translational invariance the single particle eigenstates are readily obtained by diagonalizing the tridiagonal Hamiltonian
\begin{equation}
    H_{i,i+1} = H_{i+1,i} = 1, \quad H_{ij}=0 \,\,\text{otherwise}, \quad i=1\dots L-1.
\end{equation}
The eigenstates are 
\begin{equation}
    \phi_j^{(n)}= \sqrt{\frac{2}{L+1}} \sin(\frac{n \pi j}{L+1})
\end{equation}
with energies 
\begin{equation}
    \epsilon^{(n)}=-2 \cos(\frac{n\pi}{L+1}).
\end{equation}

\begin{figure*}[t!]
\captionsetup[subfigure]{labelformat=empty}
    \subfloat[\label{subfig:pert_res_fidelity}]{}
    \subfloat[\label{subfig:pert_res_breaking}]{}
    \centering
    \includegraphics[width=0.9\linewidth]{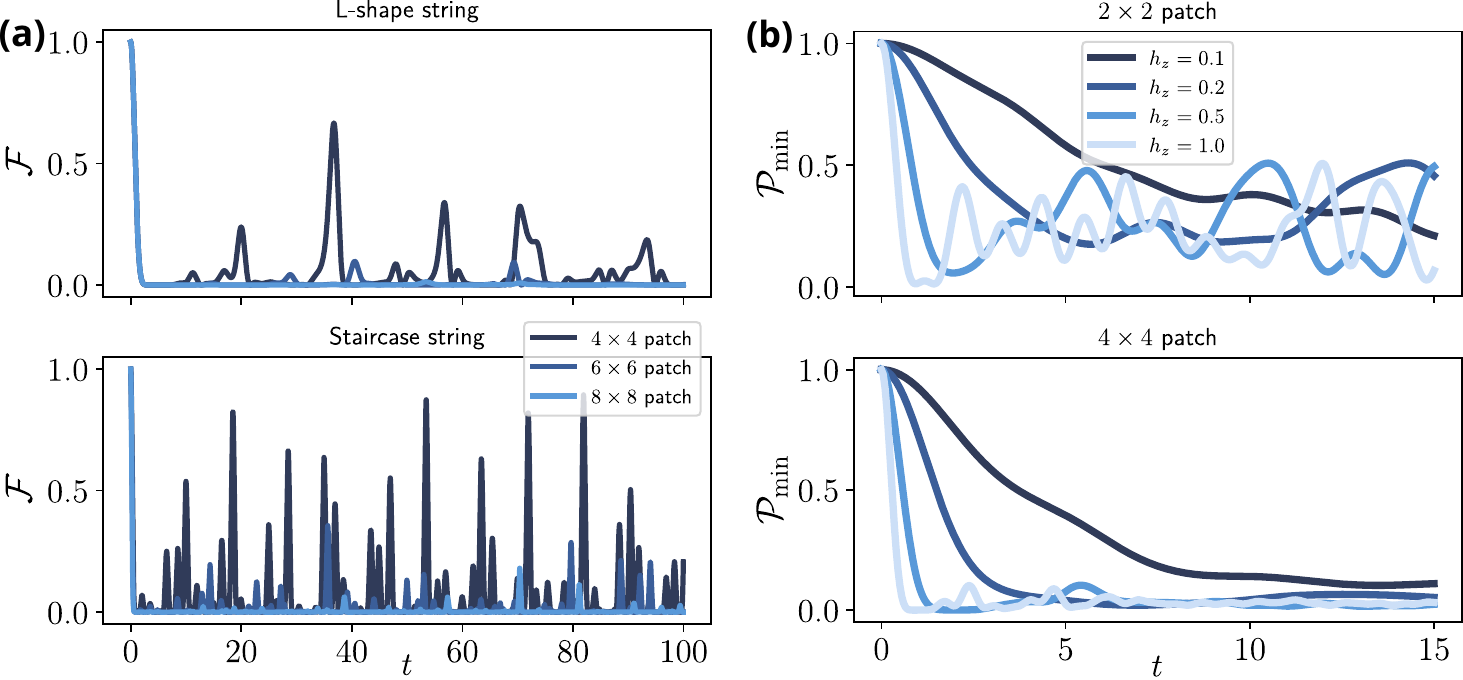}
    \caption{Perturbative results for minimal string dynamics. (a) Off-resonance fidelity for L-shaped strings (top) and staircases (bottom) for different sizes of the patch. As expected, the amplitude of the revivals decreases for larger system sizes. In the case of the staircase string, however, they are sizable and observable even at large timescales. (b) String breaking, measured by $\mathcal{P}_{\text{min}}$, as a function of the matter coupling $h_z$, at $J_p=1$. For the small $2\times 2$ patch, we see frequent revivals of the string due to the low number of string broken configurations available. On the $4 \times 4$ patch, one can observe proper string breaking, with $\mathcal{P}_{\text{min}}$ settling to very small values after a timescale $t_c \approx h_z^{-1}$.}
    \label{supfig:pert_results}
\end{figure*}

The Hamiltonian \cref{eq:ff_ham} becomes diagonal with respect to the operators
\begin{equation}
    b_n = \sum_{j=1}^L \phi_j^{n} c_j
\end{equation}
which evolve in time as 
\begin{equation}
    b_n(t) = b_n e^{-i \epsilon_n t}.
\end{equation}
From this we get the time evolution of the original $c$ operators
\begin{equation}
    c_j(t) = \sum_{n=1}^{L} \phi_j^{(n)} b_n(t) = \sum_{n=1}^{L} \phi_j^{(n)} b_n e^{-i \varepsilon_n t} = \sum_{k=1}^{L} \left[ \sum_{n=1}^{L} \phi_j^{(n)} \phi_k^{(n)} e^{-i \varepsilon_n t} \right] c_k = \sum_{k=1}^L U_{jk}(t)c_k
\end{equation}
and similarly
\begin{equation}
    c^\dagger_j(t) = \sum_{k=1}^L U^*_{jk}(t)c_k^\dagger.
\end{equation}
This can be applied promptly to the time evolution of any initial state. For example, for a two-particle state we have
\begin{equation}
    |\psi(t)\rangle = c^\dagger_a(t) c^\dagger_b(t) |0 \rangle.
\end{equation}
Overlaps of such states with arbitrary basis states can be conveniently expressed in terms of determinants, exploiting the antisymmetry of fermionic wavefunctions. In \cref{subfig:pert_res_fidelity} we show how these exact results can be used to compute the long-time dynamics of minimal strings for system sizes significantly larger than those typically tractable with exact diagonalization.

\subsection{General case and ``snake'' strings}

\begin{figure}[t!]
    \centering
    \includegraphics[width=0.80\linewidth]{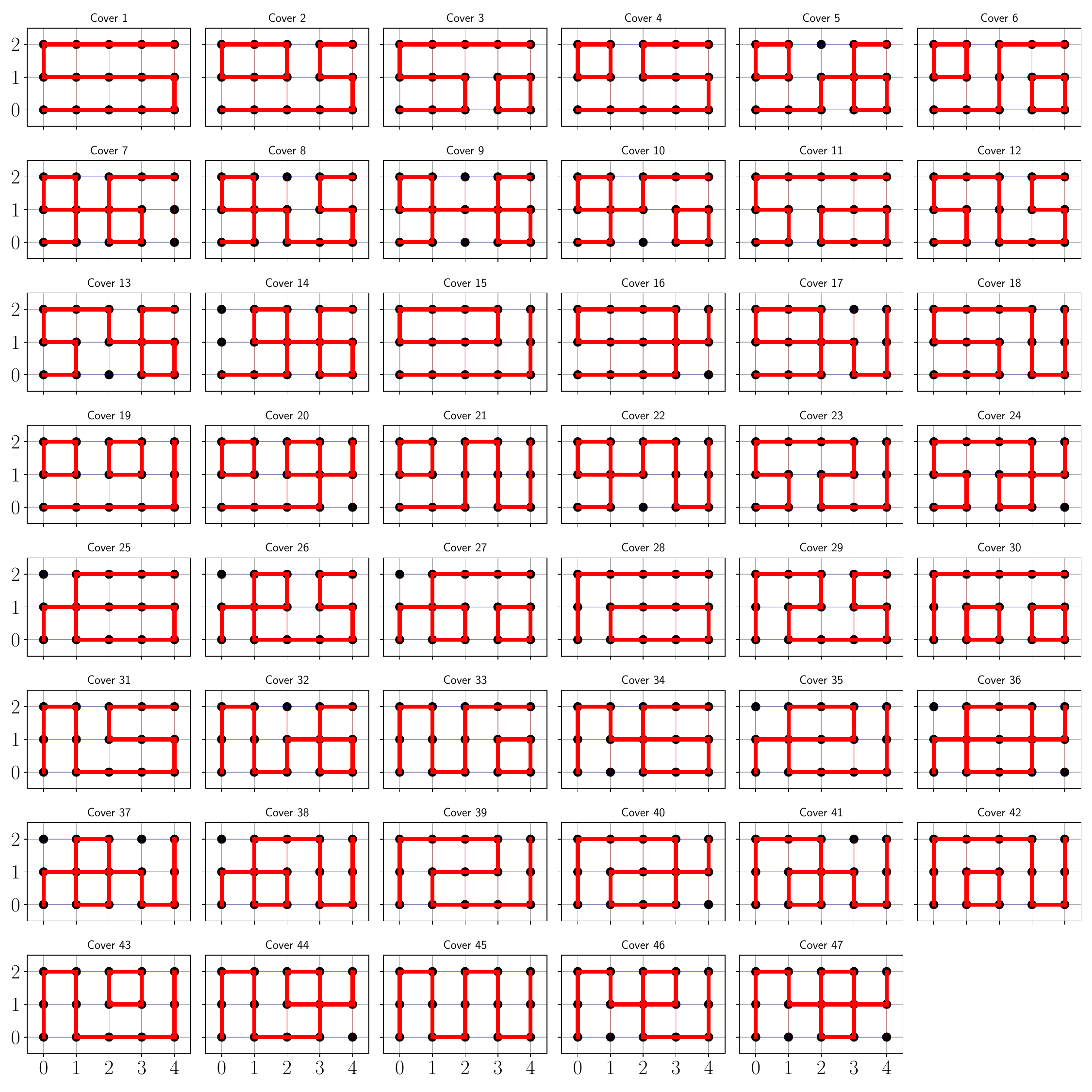}
    \caption{All valid configurations of $L=14$ strings which are completely contained within the patch. Note that covers 7 and 14 do not have any plaquettes with exactly two active electric fields, and therefore cannot be connected to any of the others.}
    \label{fig:path}
\end{figure}

We now give and example of the general procedure through a specific example involving an initial string of maximal length $L=14$, which winds within a $4\times2$ patch. This is the string  considered for \cref{fig:snake_probs} of the main text. The relevant degenerate subspace is the set of $L=14$ string configurations which are entirely contained inside the patch. There are $47$ of them, shown in \cref{fig:path}. Interestingly, we notice that two of them ($|7\rangle$ and $|14\rangle$), have no matrix elements with all the others and are thus immobile. We expect such frozen states to become more and more present as the size of the patch is increased. The iterative method described above does not detect such states, as it only generates connected configurations which form a complete basis for the $45\times 45$ effective Hamiltonian. Among these, there are several which contain not only strings but also one or more disconnected loops of various sizes. At any given time an initial long connected string can reduce its size by nucleating an electric loop.

\subsection{Resonant case and string breaking}
The scheme illustrated above can be readily adapted to the resonant case, where the degenerate subspace includes not only open and closed strings of equal total length, but also all the states where a single link of the string is replaced by two dynamical $\Zt$ charges at its ends. This is done in practice by identifying ``flippable'' links in each electric-basis configuration as those for which the neighboring star operators have the same sign, with suitable modifications on the two links connecting to the static charges. While this dramatically increases the size of the basis, the dynamics is still entirely contained within the patch and can be extracted with exact diagonalization for relatively large patches at least in the case of minimal length strings.

\begin{table}[h]
\centering
\begin{tabular}{|c|c|c|c|}
\hline
\textbf{Initial state} & \textbf{System Size} & \textbf{Basis (no mesons)} & \textbf{Basis (with mesons)} \\
\hline
L-shape & \(4 \times 2 \text{ (22 spins)}\) & 15 & 315 \\
L-shape & \(4 \times 4 \text{ (40 spins)}\) & 70 & 3850 \\
L-shape & \(6 \times 4 \text{ (58 spins)}\) & 210 & n.d. \\
Snake & \(4 \times 2 \text{ (22 spins)}\) & 45 & 504928 $\approx 2^{19}$ \\
Snake & \(4 \times 4 \text{ (40 spins)}\) & 8102 $\approx 2^{13}$ & n.d. \\
Snake & \(6 \times 4 \text{ (58 spins)}\) & 1545178 $\approx 2^{20}$ & n.d. \\

\hline
\end{tabular}
\caption{Examples of dimensions of effective Hamiltonians for different initial configurations, with or without the inclusion of first-order mesonic resonances. When considering mesonic resonances, the Hilbert space grows considerably faster and quickly saturates the sizes which can be treated with exact diagonalization.}
\label{tab:config_basis_sizes}
\end{table}

\section{Approaching the Higgs phase}

\begin{figure}[t]
    \centering
    \includegraphics[width=0.8\linewidth]{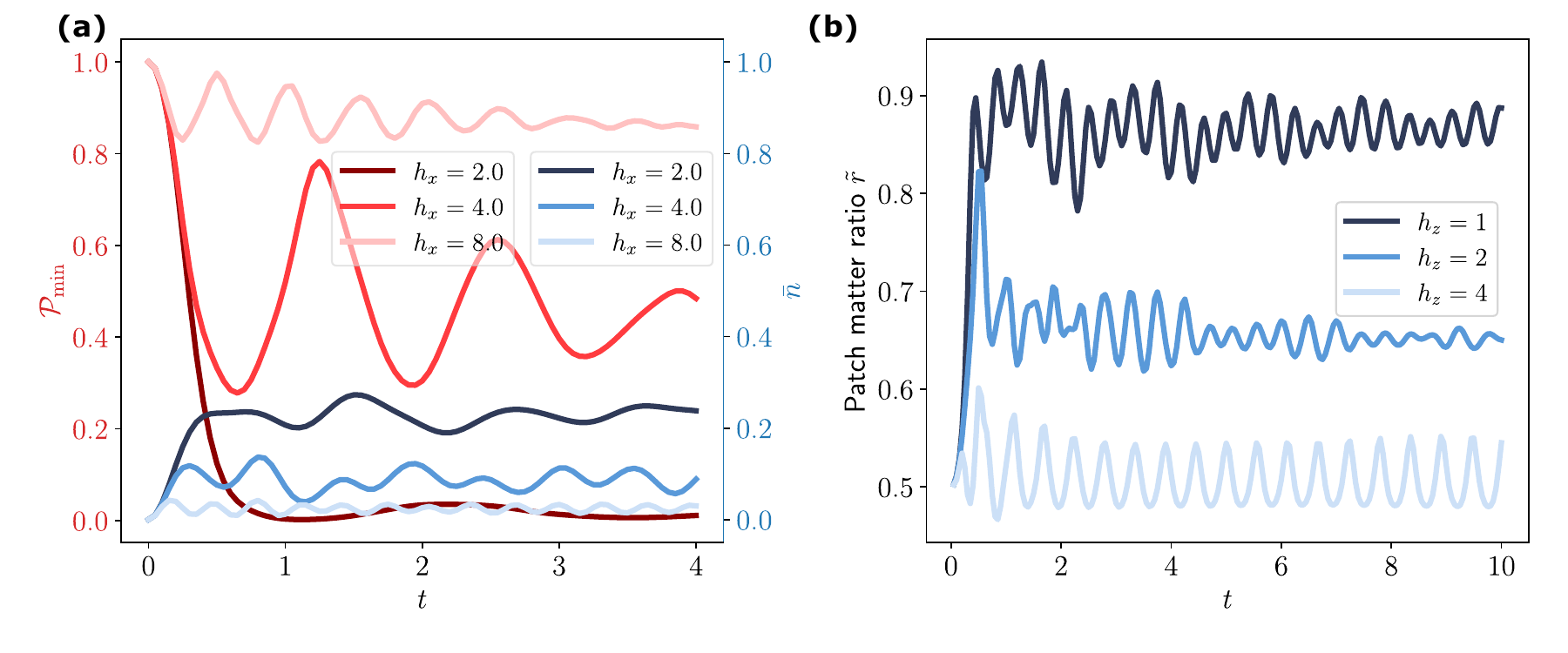}
    \caption{String dynamics as the Higgs phase is approached Left: string probabilities by varying string tension $h_x$ with fixed $h_z=1$, $J_s=1$ and $J_p=1$. By decreasing $h_x$, one can observe that the rapid dissipation of the string and stronger charge fluctuations. Right: ratio \cref{supeq:ratio} between matter density within the patch and in the whole system. We start deep inside the confined phase at the resonant quench parameters $J_s=2$, $J_p=1$, $h_x=4$, $h_z=1$, and then increase $h_z$. The plot shows how particles are progressively more delocalized, and they proliferate uniformly across the system as the Higgs phase is approached.}
    \label{fig:critical_hx}
\end{figure}

The main text is focused on the string dynamics deep in the confined phase, where even far-from-equilibrium strings are well-defined objects with a large energy gap separating them from non-stringy states. Charge fluctuation here plays a role only when certain resonance conditions are met, leading to string breaking. Deep in the Higgs phase of the model, on the other hand, charge fluctuations are strong and can lead to string dissipation caused by the proliferation of charges. We considered the off-resonant (string broken) and resonant scenarios separately. 

For generic off-resonance values of the quench parameters, we start deep in the confined phase with high string tension $h_x=8$. From \cref{fig:critical_hx}, one observes that the probability of finding a string within the patch stays close to $1$, while decreasing $h_x$ causes them to rapidly go to zero, accompanied by matter creation both where static charges are located and elsewhere on the lattice, with comparable magnitude, saturating around $\langle \hat{n}\rangle \approx 0.25 $ per site. 

We now consider the case where the quench parameters are set to resonant values, so that the dynamics deep in the confined phase exhibits string breaking. As charge fluctuations are enhanced by increasing the matter coupling $h_z$, the system is driven from the string broken towards the Higgs phase. As opposed to the off-resonant case, where matter creation is almost entirely suppressed in the confined phase, here mesons pop up within the patch as a result of string breaking. These exhibit restricted mobility, as they can resonate with string configurations but their hopping is heavily suppressed. To quantify this, we define the observable

\begin{equation}
    \tilde{r} = \frac{\sum_{i\in {\rm patch}}\langle\hat{n}_{i}\rangle}{\sum_{i\in {\rm all ~sites}}\langle\hat{n}_{i}\rangle},
    \label{supeq:ratio}
\end{equation}

which measures the fraction of matter created within the patch compared to the total matter created across the system. As shown in Fig \cref{fig:critical_hx}, in the confined regime matter creation is highly localized, leading to $\tilde{r}\approx 1$. In contrast, as the system approaches the Higgs phase, charge fluctuations delocalize, and matter is more uniformly distributed throughout the lattice.

\section{Snapshots of system configurations}
\begin{figure*}[t!]
    \centering
    \includegraphics[width=0.8\linewidth]{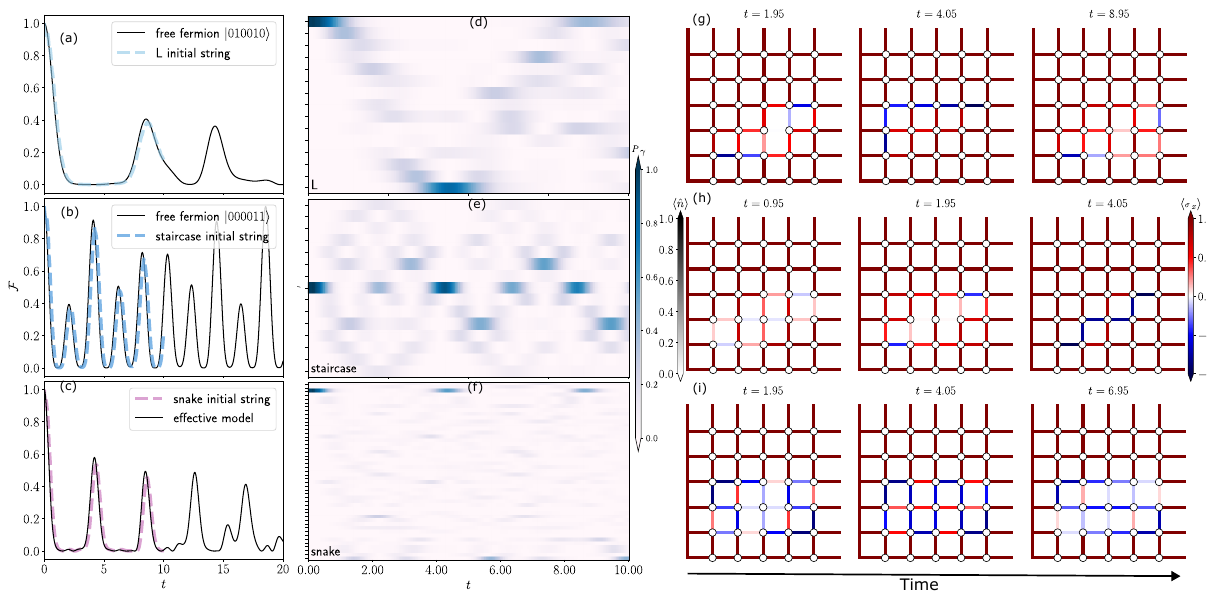}
    \caption{(a)-(c) Comparison of free fermion evolution and quench of different initial strings, where we can see that for minimal covering (a) L string and (b) diagonal string, the free fermion picture matches quantitatively well up to Trotter error. (d)-(f) String-configuration-resolved probability distributions and (g)-(i) some snapshots of intermediate string configurations after quenching from L (g), diagonal (h), and snake (i) initial strings respectively. We can see that after evolving into the superposition of the diagonal covering state, around $t=4$ it turns into the ``opposite" L covering of the rectangle patch, which corresponds to the zero $\mathcal{F}$ plateau in (a) and the large value in (d). For all the quenches in this figure, the parameters of the Hamiltonian are $J_p=1$, $J_s=15$, $h_x=12$, and $h_z=1$.}
    \label{fig:revival}
\end{figure*}

\begin{figure*}[t!]
    \centering
    \includegraphics[width=0.8\linewidth]{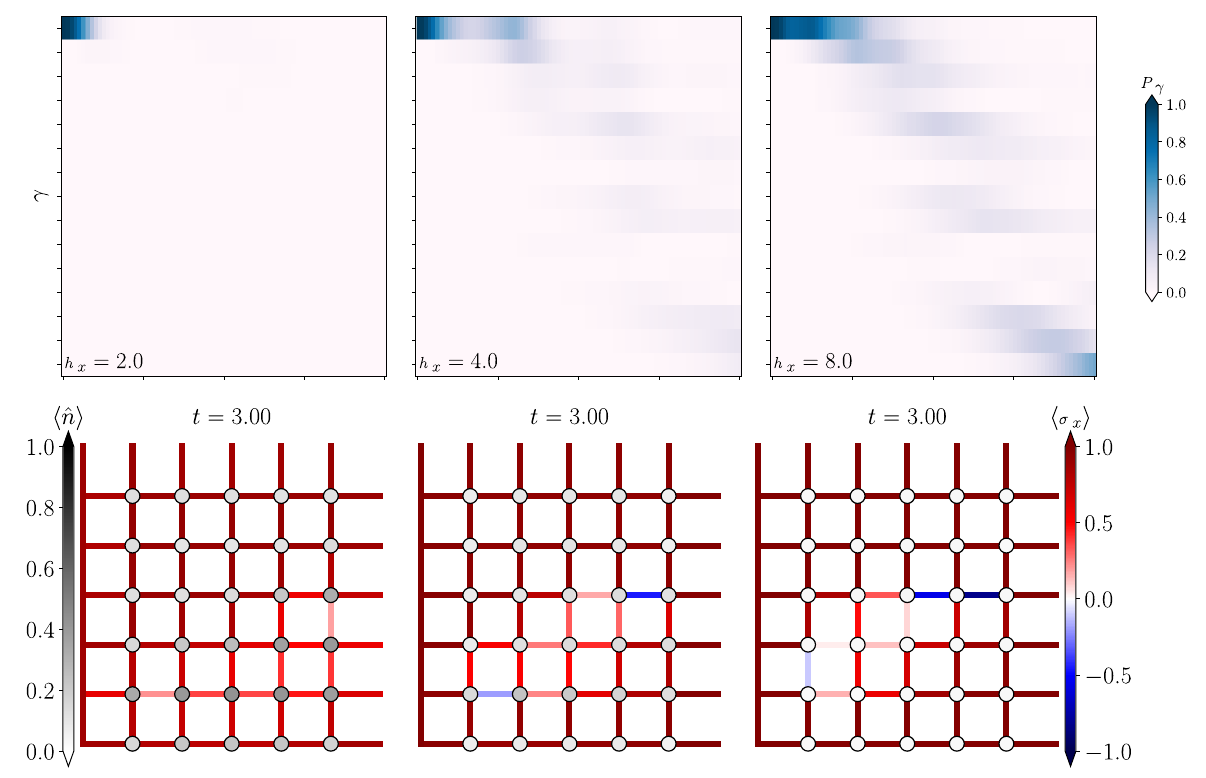}
    \caption{Configuration-resolved string probabilities distribution varying $h_x$ with $J_s=1$, $J_p=1$, and $h_z=1$ and corresponding string and charge snapshots at intermediate timescale.}
    \label{fig:higgs_snap}
\end{figure*}

\begin{figure*}
    \centering
    \includegraphics[width=0.8\linewidth]{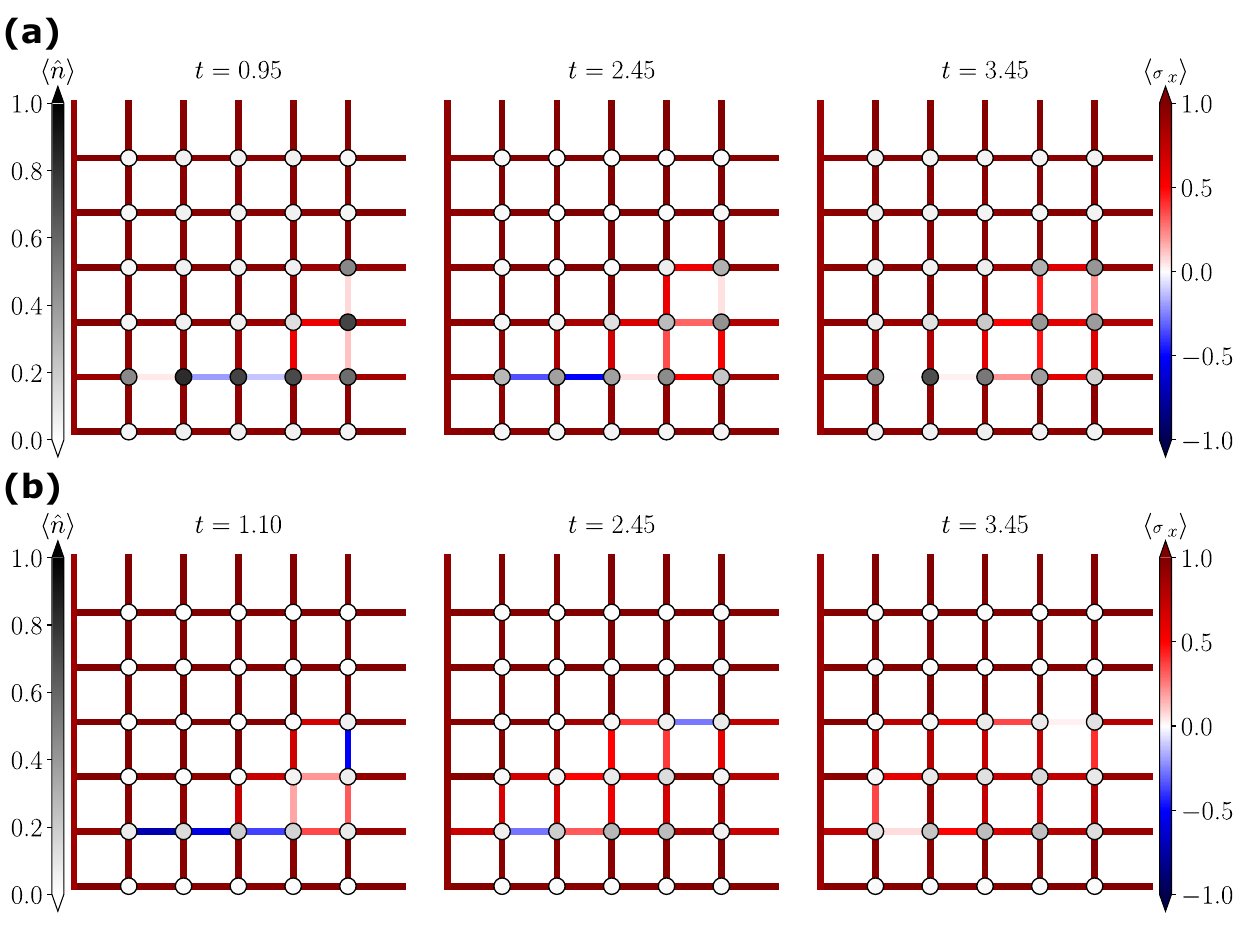}
    \caption{Electric field (links) and matter (sites) configurations snapshots for the first-order (a) and second-order (b) resonance with $J_s=2$, $J_p=1$, $h_x=4$, and $h_z=1$ and $J_s=4$, $J_p=1$, $h_x=4$, and $h_z=1$, respectively. Matter creation within the patch as a result of string breaking is clearly visible.}
    \label{fig:1st}
\end{figure*}

In \cref{fig:revival}, \cref{fig:higgs_snap} and \cref{fig:1st} we show snapshots of the electric field and matter configurations over the whole lattice. These show how the initial string configuration evolves into broken and unbroken states.

\end{document}